\documentclass[twocolumn]{aastex631}
\usepackage{amsmath}

\newcommand{\hi}{H\textsc{i}}

\shorttitle{A Galactic Eclipse}
\shortauthors{Murray et al.}

\begin{document}

\title{A Galactic Eclipse: The Small Magellanic Cloud is Forming Stars in Two, Superimposed Systems}

\correspondingauthor{Claire Murray}
\email{cmurray1@stsci.edu}

\author[0000-0002-7743-8129]{Claire E. Murray}
\affil{Space Telescope Science Institute, 
3700 San Martin Drive, 
Baltimore, MD, 21218}
\affil{Department of Physics \& Astronomy, 
Johns Hopkins University,
3400 N. Charles Street, 
Baltimore, MD 21218}

\author[0000-0001-5388-0994]{Sten Hasselquist}
\affil{Space Telescope Science Institute, 
3700 San Martin Drive, 
Baltimore, MD, 21218}

\author[0000-0003-4797-7030]{Joshua E. G. Peek}
\affil{Space Telescope Science Institute, 
3700 San Martin Drive, 
Baltimore, MD, 21218}
\affil{Department of Physics \& Astronomy, 
Johns Hopkins University,
3400 N. Charles Street, 
Baltimore, MD 21218}

\author[0000-0003-0588-7360]{Christina Willecke Lindberg}
\affil{Department of Physics \& Astronomy, 
Johns Hopkins University,
3400 N. Charles Street, 
Baltimore, MD 21218}
\affil{Space Telescope Science Institute, 
3700 San Martin Drive, 
Baltimore, MD, 21218}

\author[0009-0000-0733-2479]{Andres Almeida}
\affiliation{Department of Astronomy, University of Virginia, Charlottesville, VA, 22904, USA}

\author[0000-0003-1680-1884]{Yumi Choi}
\affiliation{NOIRLab, 950 N. Cherry Ave, Tucson, AZ 85719, USA}

\author[0000-0002-7661-4856]{Jessica E. M. Craig}
\affil{Lennard-Jones Laboratories, Keele University, ST5 5BG, UK}

\author[0000-0002-9214-8613]{Helga D\'enes}
\affiliation{School of Physical Sciences and Nanotechnology, Yachay Tech University, Hacienda San Jos\'e S/N, 100119, Urcuqu\'i, Ecuador }

\author[0000-0002-6300-7459]{John M. Dickey}
\affiliation{School of Natural Sciences, Private Bag 37, University of Tasmania, Hobart, TAS, Australia}

\author[0000-0003-4019-0673]{Enrico M.\ Di Teodoro}
\affil{Dipartimento di Fisica e Astronomia, Università degli Studi di Firenze, 50019 Sesto Fiorentino, Italy}

\author[0000-0002-0706-2306]{Christoph Federrath}
\affil{Research School of Astronomy and Astrophysics, Australian National University, Canberra, ACT~2611, Australia}
\affil{Australian Research Council Centre of Excellence in All Sky Astrophysics (ASTRO3D), Canberra, ACT~2611, Australia}

\author[0000-0002-1995-6198]{Isabella. A. Gerrard}
\affil{Research School of Astronomy and Astrophysics, Australian National University, Canberra, ACT~2611, Australia}

\author[0000-0002-1495-760X]{Steven J. Gibson}
\affil{Department of Physics \& Astronomy, Western Kentucky University, 1906 College Heights Blvd., Bowling Green, KY 42101, USA}

\author[0000-0002-4814-958X]{Denis Leahy}
\affil{Department of Physics \& Astronomy, 
University of Calgary,
2500 University Dr. NW, 
Calgary, AB, Canada T2N 1N4}

\author[0000-0002-9888-0784]{Min-Young Lee}
\affil{Korea Astronomy and Space Science Institute, 
776 Daedeok-daero, Daejeon 34055, Republic of Korea}
\affil{Department of Astronomy and Space Science, 
University of Science and Technology, 
217 Gajeong-ro, Daejeon 34113, Republic of Korea}

\author[0000-0001-6846-5347]{Callum Lynn}
\affil{Research School of Astronomy and Astrophysics, Australian National University, Canberra, ACT~2611, Australia}

\author[0000-0003-0742-2006]{Yik Ki Ma}
\affil{Research School of Astronomy and Astrophysics, Australian National University, Canberra, ACT~2611, Australia}

\author[0000-0002-5501-232X]{Antoine Marchal}
\affil{Research School of Astronomy and Astrophysics, Australian National University, Canberra, ACT~2611, Australia}

\author[0000-0003-2730-957X]{N.~M.\ McClure-Griffiths}
\affil{Research School of Astronomy and Astrophysics, Australian National University, Canberra, ACT~2611, Australia}

\author[0000-0002-1793-3689]{David Nidever}
\affil{Department of Physics, Montana State University, P.O. Box 173840, Bozeman, MT 59717-3840}

\author[0000-0002-2712-4156]{Hiep Nguyen}
\affil{Research School of Astronomy and Astrophysics, Australian National University, Canberra, ACT~2611, Australia}

\author[0000-0001-9504-7386]{Nickolas~M.~Pingel}
\affiliation{Department of Astronomy, The University of Wisconsin--Madison, 475 N. Charter Street, Madison, WI 53706, USA}

\author[0000-0003-1356-1096]{Elizabeth Tarantino}
\affil{Space Telescope Science Institute, 
3700 San Martin Drive, 
Baltimore, MD, 21218}

\author[0000-0002-2082-1370]{Lucero Uscanga}
\affiliation{Departamento de Astronom\'ia, Universidad de Guanajuato, A.P. 144, 36000 Guanajuato, Gto., Mexico}

\author[0000-0002-1272-3017]{Jacco Th. van Loon}
\affil{Lennard-Jones Laboratories, Keele University, ST5 5BG, UK}

\begin{abstract}

The structure and dynamics of the star-forming disk of the Small Magellanic Cloud (SMC) have long confounded us. The SMC is widely used as a prototype for galactic physics at low-metallicity, and yet we fundamentally lack an understanding of the structure of its interstellar medium (ISM). In this work, we present a new model for the SMC by comparing the kinematics of young, massive stars with the structure of the ISM traced by high-resolution observations of neutral atomic hydrogen (\hi) from the Galactic Australian Square Kilometer Array Pathfinder survey (GASKAP-\hi). Specifically, we identify thousands of young, massive stars with precise radial velocity constraints from the \emph{Gaia} and APOGEE surveys and match these stars to the ISM structures in which they likely formed. By comparing the average dust extinction towards these stars, we find evidence that the SMC is composed of two structures with distinct stellar and gaseous chemical compositions. We construct a simple model that successfully reproduces the observations and shows that the ISM of the SMC is arranged into two, superimposed, star-forming systems with similar gas mass separated by $\sim5\rm\,kpc$ along the line of sight. 

\end{abstract}


\section{Introduction}
\label{sec:intro}

As one of our nearest neighbors, the Small Magellanic Cloud (SMC) is among the most highly scrutinized galaxies in the Universe. Located at a distance of $\sim 62$ kpc \citep[see compliation of historical estimates in][]{degrijs2015}, the SMC is close enough that individual stars and interstellar medium (ISM) structures can be resolved. 

What makes the SMC especially interesting is its markedly different interstellar conditions to the Milky Way. In particular, the SMC has a low metallicity \citep[$\sim 20\%$ Solar;][]{russell1992}, making it an excellent laboratory for deciphering the physics of the ISM at higher redshift (e.g., cosmic noon), where similar pristine conditions are expected to be prevalent.

However, the line of sight structure of the SMC is not well constrained, due in large part to the fact that different observational tracers paint disparate pictures for its geometry. For example, the oldest stellar populations in the SMC are distributed spherically within a radius of $7-12\rm\,kpc$ \citep{subramanian2012}, and do not show significant signs of rotation  \citep{harris2006, gaiacollaboration2018, niederhofer2018, zivick2018, niederhofer2021}, except for in the very central region \citep[within $\pm1\rm\,kpc$ from the SMC optical center;][]{zivick2021}. Stars with measurable distances (red clump stars, Cepheid variables, RR Lyrae) are greatly dispersed along the line of sight \citep[$\sim20-30\rm\,kpc$ depth;][]{mathewson1988, nidever2013, scowcroft2016, jacyszyn2016, ripepi2017}. In contrast, young main sequence and red giant branch (RGB) stars display a radial velocity gradient indicative of rotation \citep{evans2008, dobbie2014, elyoussoufi2023}. In addition, the velocity distribution of stars in the central SMC exhibits signs of tidal disruption at the hands of the Large Magellanic Cloud (LMC) \citep{niederhofer2018, zivick2019, deleo2020,zivick2021, niederhofer2021}. There is mounting evidence for the presence of distinct substructures along the line of sight \citep[e.g.,][]{hatzidimitriou1989, martinez2019, cullinane2023}. In addition, there is a well-known ``bimodality" in distance observed in red clump stars on the Eastern side of the system \citep[e.g.,][]{nidever2013} and recently traced throughout the SMC \citep{subramanian2017, tatton2021, omkumar2021, almeida2023}. Furthermore, recent chemical analysis of RGB stars in the SMC exhibit complex distributions in both radial velocity and metallicity ([Fe/H]), suggesting that chemical enrichment in the system ``has not been uniform", pointing to either strong influence of star formation (SF) bursts \citep{hasselquist2021, massana2022} or chemically distinct substructures \citep{mucciarelli2023}.

\begin{figure*}[ht!]
\centering
\includegraphics[width=0.65\textwidth]{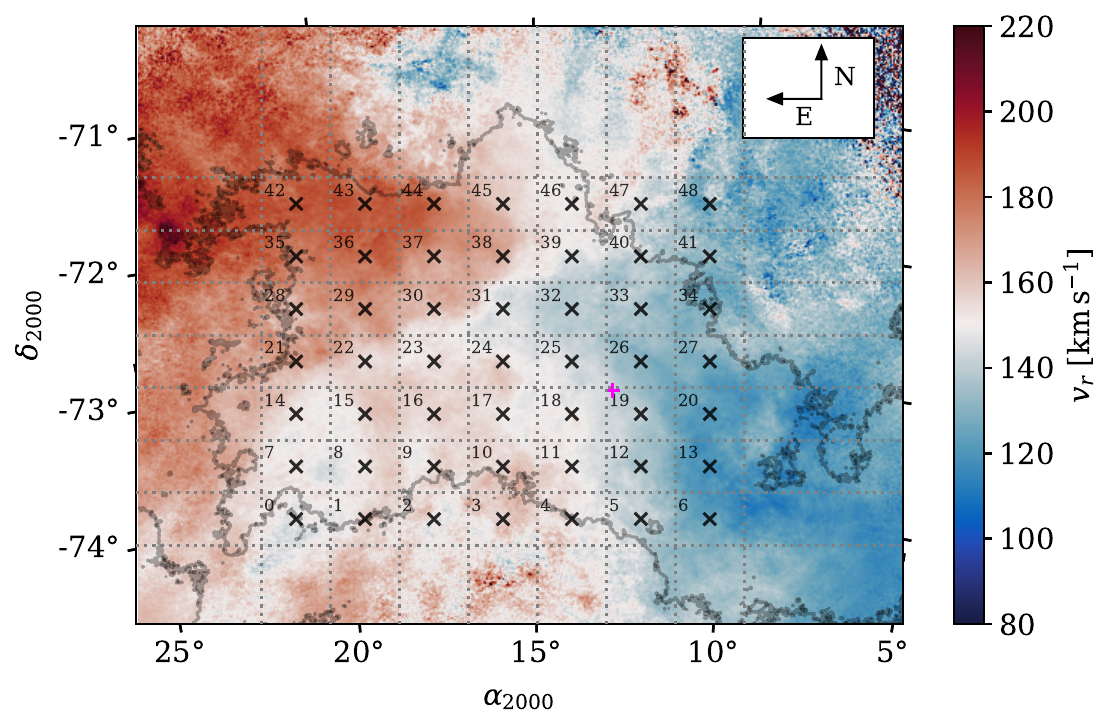}
\includegraphics[width=0.95\textwidth]{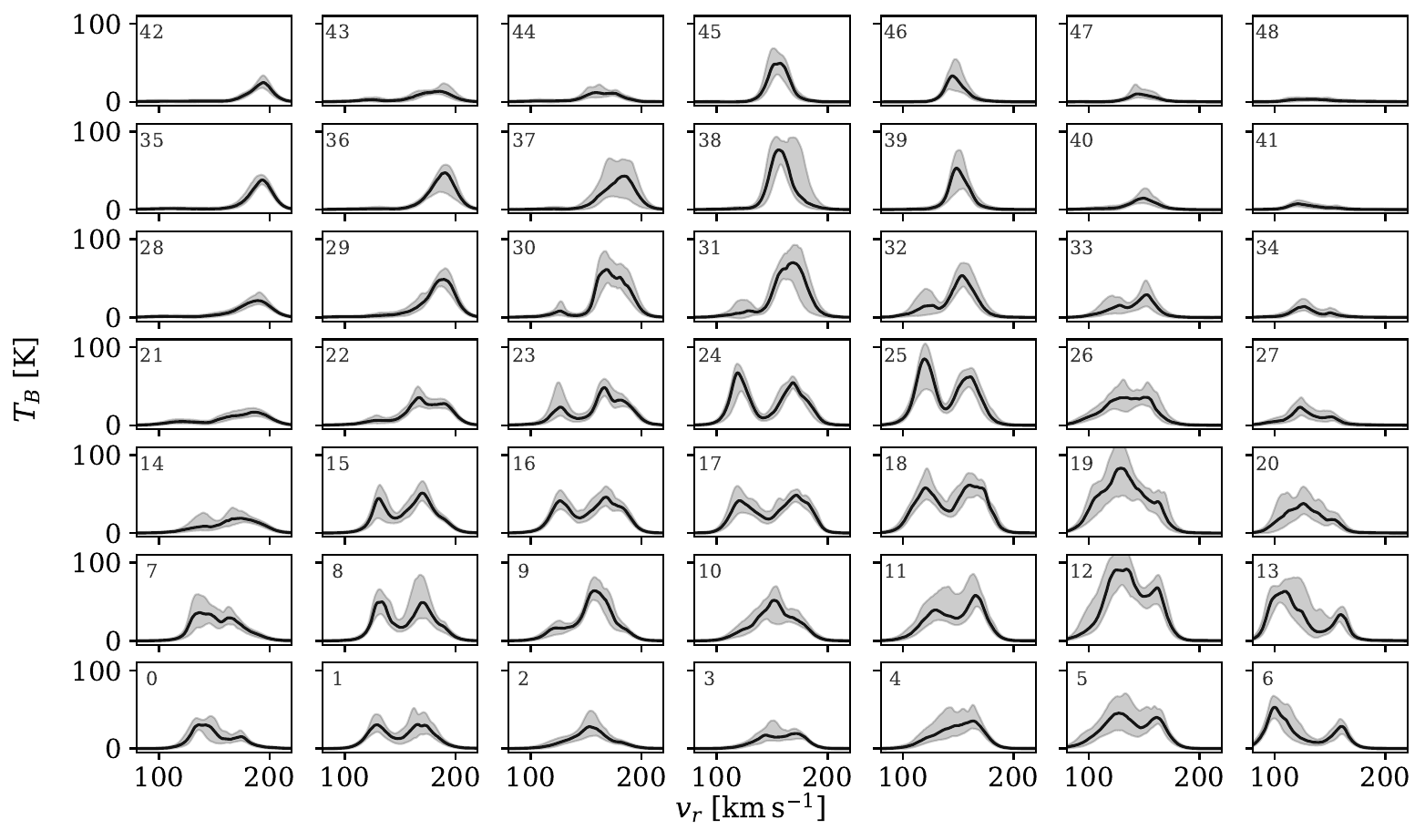}
\caption{Top: Intensity-weighted mean velocity map of the SMC from \citet{pingel2022}. The single contour indicates an \hi\ column density of $15\times 10^{20}\rm\,cm^{-2}$. The center of the SMC used in this work \citep[defined based on the stellar populations observed by \emph{Gaia}][]{zivick2021} is marked with a magenta cross. Spatial bins are overlaid and identified with numbers. Bottom: The average  \hi\ brightness temperature spectra ($T_b(v_r)$) profiles from spatial bins marked at left (black). Each panel includes shaded grey envelopes denoting the $16^{\rm th}$ through $84^{\rm th}$ percentile of the $T_b(v_r)$ within each spatial bin. We observe that the radial velocity structure of \hi\ emission features multiple, distinct velocity peaks (and typically two, dominant components).
\label{fig:hi_complex}}
\end{figure*}

In addition to its diversely structured stellar populations, the ISM of the SMC is highly disturbed \citep{stanimirovic1999}. Neutral hydrogen (\hi) emission in the SMC is organized into multiple, distinct peaks in radial velocity. This structure was originally interpreted as the presence of two, distinct sub-systems separated in space \citep{kerr1954, johnson1961, hindman1964}. \citet{mathewson1984} further showed that the velocities of stars, HII regions and planetary nebulae matched that of the \hi, and concluded that the SMC has been ``badly torn" by its interactions with the LMC, in agreement with the leading theoretical models at that time \citep{murai1980}. The widely held interpretation is that the low velocity gas must be in front \citep{hindman1964, mathewson1984}, given that optical absorption is typically associated with the low-velocity \hi\ peak \citep[e.g.,][]{danforth2002, welty2012}. But the interpretation of these results has not always necessitated the presence of two distinct components. For example, a recent study of gaseous filaments in the SMC establishes that the filaments in the low and high velocity components show a preference for being aligned with each other, which suggests that they must somehow be physically related \citep{ma2023}. Another alternative interpretation is that the system is a series of overlapping supershells \citep{hindman1967, staveleysmith1997}.

Despite this complexity, the line-of-sight integrated \hi\ velocity field of the SMC is broadly consistent with a rotating disk \citep{stanimirovic2004, diteodoro2019}. This disk model has underpinned our concept of the SMC system and has long-defined its total dynamical mass -- a key ingredient to numerical simulations. However, the velocity gradient observed in \hi\ emission is perpendicular to that observed in the radial velocities of young stars \citep{evans2008, dobbie2014}. In addition, the distance gradient observed in Cepheids from the North East to South West is oriented $\sim90$ degrees away from the minor axis of the gas disk \citep{scowcroft2016}. Finally, the 3D kinematics of young, massive stars throughout the SMC, which are young enough to trace the kinematics of their birth clouds in the ISM, are inconsistent with the disk rotation model \citep{murray2019}. 

These observational inconsistencies have traditionally been explained away by the fact that the SMC has been severely disrupted by its recent interactions with the nearby LMC. Although precise observations of stellar proper motions in the LMC and SMC \citep{kallivayalil2006a, kallivayalil2006b, besla2007, piatek2008, zivick2018} indicate that the LMC and SMC are likely on their first infall into the Milky Way (MW) halo \citep[although a second infall scenario is also possible;][]{vasiliev2024}, spatially-resolved star formation histories show that the two galaxies have been interacting with each other repeatedly for the past several Gyr \citep[e.g.,][]{massana2022}. These interactions have produced a wealth of debris in the outskirts of both Clouds \citep{besla2016, mackey2016, belokurov2016, carrera2017, pieres2017, choi2018a, martinez2019}. In particular, the Clouds likely made a direct collision $\sim150$ Myr ago \citep{zivick2021, choi2022}. These interactions have produced the massive, gaseous debris fields known as the Leading Arm, Bridge and Stream \citep{wannier1972, mathewson1974, putman1998, nidever2010, for2014, for2016}. 

As the more massive of the two galaxies by a factor of $\sim10$ \citep{penarrubia2016, erkal2019}, the LMC has survived this eventful interaction history as a nearly face-on rotating disk \citep{vandermarel2014, gaia2021}. As a result, theoretical efforts to reconstruct the formation history of the broader Magellanic System (MS) largely focus on the interaction between the LMC and MW \citep[e.g.,][]{besla2012, pardy2018, lucchini2021}. However, constraining the precise orbital histories (and futures) of the MCs and their satellites relies on the influence of the SMC \citep{patel2020}. It is thus critical to develop a model that reconciles these diverse observational constraints. 

In this work, we present a new model for gas in the SMC, and show that it is composed of not one, but two structures located at distinct distances along the line of sight, distributed across the inner $\pm4^{\circ}$ of the SMC. In Section~\ref{sec:data}, we discuss the data products used in our analysis. In Section~\ref{sec:analysis}, we present our analysis procedure, in which we determine the order of ISM components along the line of sight. In Section~\ref{sec:model}, we present a toy model to describe our observational results. In Section~\ref{sec:discussion}, we discuss this new model in the context of the literature and the history of the MS. Finally, we conclude in Section~\ref{sec:concl}.

\section{Data}
\label{sec:data}

In this section, we describe the data products used in our analysis. Specifically, we will combine information from the following sources:

\begin{itemize}
    \item High-resolution observations of the SMC's neutral gas emission at $21\rm\,cm$ from the GASKAP-\hi\ survey \citep{pingel2022}, which trace the radial velocity of the bulk gas population in the system.
    \item Line of sight extinction and radial velocity measurements of bright, young ($10-100$ Myr) stars in the SMC from the \emph{Gaia} mission DR3 \citep{gaia2016, gaia2023, babusiaux2023} and APOGEE DR17 \citep{majewski2017, zasowski2017, abdurrouf2022}.
\end{itemize}

We also use supplementary information from the APOGEE spectral fits, as well as large-area molecular gas observations of the system to assess its chemical state. 

\subsection{Neutral gas radial velocities}

The ISM of the SMC is heavily dominated by atomic gas \citep{leroy2007, jameson2016}, and therefore to trace the bulk distribution of gas, we use the recent survey of 21 cm emission in the SMC from the Galactic Australian Square Kilometer Array Pathfinder (GASKAP-\hi) collaboration \citep{pingel2022}. This dataset was constructed from 20.9 hours of integration time by ASKAP \citep{hotan2021} on the SMC ($\rm RA = 00h58m43.280s, Dec = -72d31m49.03s$) in December 2019. ASKAP is composed of 36 antennas, each equipped with a phased array feed receiver which combine to give the telescope an instantaneous field of view of $\sim 25$ square degrees \citep{hotan2021}. The resulting data products were combined with single-dish observations from the GASS survey \citep{mcg2009, kalberla2010, kalberla2015} to produce an \hi\ emission data cube with an angular resolution of $30^{\prime \prime}$ ($\sim 10$ pc at a distance of $62\rm\,kpc$), sensitivity to \hi\ brightness temperature of $1.1\rm\,K$ and a per-channel velocity resolution of $0.98\rm\,km\,s^{-1}$ -- by far the highest-resolution, fully-resolved \hi\ survey of the SMC to date \citep{pingel2022}. 

\begin{figure*}[ht!]
\centering
\includegraphics[width=0.99\textwidth]{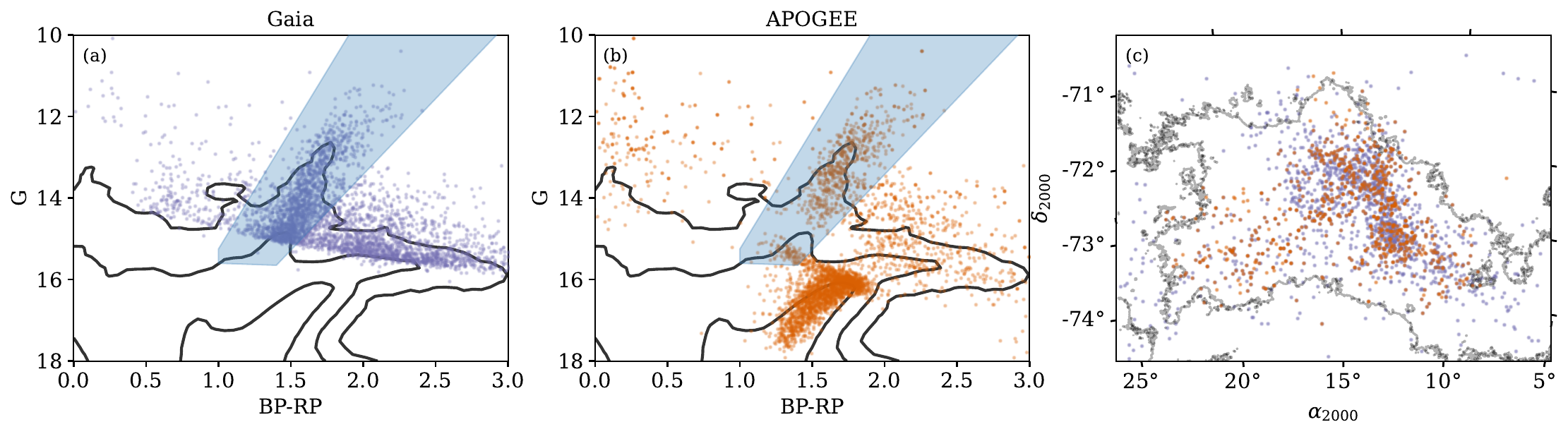}
\caption{CMDs of the \textit{Gaia} (a) and APOGEE (b) samples (before RSG selection). Black contours in each panel denote the $1$, $2$ and $3\sigma$ limits of the total \emph{Gaia} SMC sample. Red supergiants (RSGs) are indicated by the blue highlighted region in each panel. (c) Spatial distribution of selected RSG stars for the \textit{Gaia} and APOGEE samples. The \hi\ distribution is highlighted by a single gray contour at $15\times10^{20}\rm\,cm^{-2}$ (see Figure~\ref{fig:hi_complex}). 
\label{fig:cmd_compare} }
\end{figure*}

In Figure~\ref{fig:hi_complex}, we display the intensity-weighted mean radial velocity of the GASKAP-\hi\ cube (a.k.a., the first moment) for visualization of the velocity structure of gas in the SMC. There is a clear gradient in radial velocity across the main body of the SMC, which seems to resemble the signature of a rotating disk \citep[e.g.,][]{diteodoro2019, gerrard2023}. In the right-hand panel of Figure~\ref{fig:hi_complex}, we display example brightness temperature spectra ($T_B(v)$) drawn from a regular grid of positions overlaid on the first moment map. We observe that these example spectra display complex, multi-peaked structure. In particular, across the main body of the SMC, where we see strong emission, there are often two distinct spectral components, one at high velocity ($\sim 170\rm\,km\,s^{-1}$) and one at low velocity ($\sim 130\rm\,km\,s^{-1}$).

\subsection{Line-of-sight extinction and radial velocity of embedded stars}
\label{sec:stars}

Although 21 cm emission provides a high-resolution probe of the position-position-velocity structure of the gas, it does not directly probe the line of sight distance axis. To sample this third spatial dimension, we use measurements of extinction by dust toward individual stars. As extinction provides a cumulative measurement of the effect of dust along the line of sight, we can compare extinction values towards stars at different radial velocities in order to determine the relative spatial order of these stars along the line of sight (i.e., ``front" and ``behind"). 

As our primary goal is to connect the line of sight structure of the stars with the radial velocity structure of the gas, we make the assumption that the gas, the dust, and the young stars coexist spatially and share bulk kinematics on the same spatial scales probed by each tracer. With these assumptions, we can compare total extinction measured to samples of stars with different radial velocities in order to determine the relative order along the line of sight. Although the assumption that dust and gas are well-mixed is widely accepted, the assumption that the stars trace the same 3D motions as the gas relies on targeting only the youngest stellar populations, whose lives are short enough that their motions are coupled to the motions of their parent gas clouds \citep[e.g.,][]{grossschedl2021}. The precise definition of ``young" to satisfy this requirement is not well-defined, especially in the low-metallicity, dynamically complex environment of the SMC. In general, stellar populations will lose memory of their birthplaces on a dynamical timescale (i.e., a crossing timescale of about $100\rm\, Myr$ for the SMC). As we will describe below, we endeavor to select stars young enough (ages $\sim 10$s of Myr) such that our assumption is valid when averaged across large spatial areas, rather than in individual cases. 

For this study, we assemble SMC stars observed by the APOGEE survey and the \emph{Gaia} mission. Both products include precise radial velocity measurements for thousands of stars, and their selection is described below.

\subsubsection{Gaia}

To construct an initial sample of SMC stars from \emph{Gaia} DR3 \citep{gaia2016, gaia2023, babusiaux2023}, we follow the strategy described by \citet{gaia2018}. First, to construct a proper motion filter for SMC stars, we select all stars from DR3 within $3^{\circ}$ of $(\alpha_{2000}, \delta_{2000})=(13.15833^{\circ}, -72.8002^{\circ})$ with low parallax over error ($\varpi/\sigma_{\varpi}<10$) to minimize foreground contamination. Following Equation 2 of \citet{gaia2018}, we compute the orthographic projections of the source coordinates, and further select all stars with mean G-band magnitude $G<19\rm\,mag$. Next, we compute the median proper motions of these stars in $x$ and $y$ ($\mu_x$ and $\mu_y$), and exclude all stars with values beyond $4\times$ the robust scatter estimate \citep{gaia2018} from the median in both axes. Then, we compute the covariance matrix of $\mu_x,\,\mu_y$ ($\sigma$) and ultimately define a filter on proper motions defined as $\mu^{T}\sigma^{-1}\mu<9.21$. We acknowledge that genuine SMC stars may be excluded by this cut, as stars outside the inner $3^{\circ}$ may have significantly different proper motions. However, we estimate this to be a small effect on our analysis as our focus is on young stars with kinematics consistent with the bulk SMC population.

To select a complete sample of SMC stars, we select all stars within $8^{\circ}$ of the SMC center (although this is larger than the GASKAP-\hi\ footprint, here we simply follow the previous analysis of \citet{gaia2018}) with $\varpi/\sigma_{\varpi}<10$ and $G<20$, and apply the proper motion filter defined on the central regions of the SMC above to produce a catalog of 2,006,326 SMC stars. As a last step, we restrict this catalog to the footprint of the GASKAP-\hi\ SMC cube, yielding 1,687,195 stars. 

Since our primary interest is to use stars to probe the velocity structure of gas in the system, we further select only stars with radial velocity measurements from \emph{Gaia}. This removes the majority of stars in our initial SMC/GASKAP-\hi\ sample, as radial velocities are only measurable at the SMC distance for the brightest stars. In total, we have 3,783 stars with radial velocities. From these, we select only those within the SMC velocity range ($80<v_r<250\rm\,km\,s^{-1}$; 3,707 stars). Finally, we convert the \emph{Gaia} radial velocities from their native barycentric reference frame to the Local Standard of Rest frame to be consistent with the \hi. %

\begin{figure}[ht!]
\centering
\includegraphics[width=0.45\textwidth]{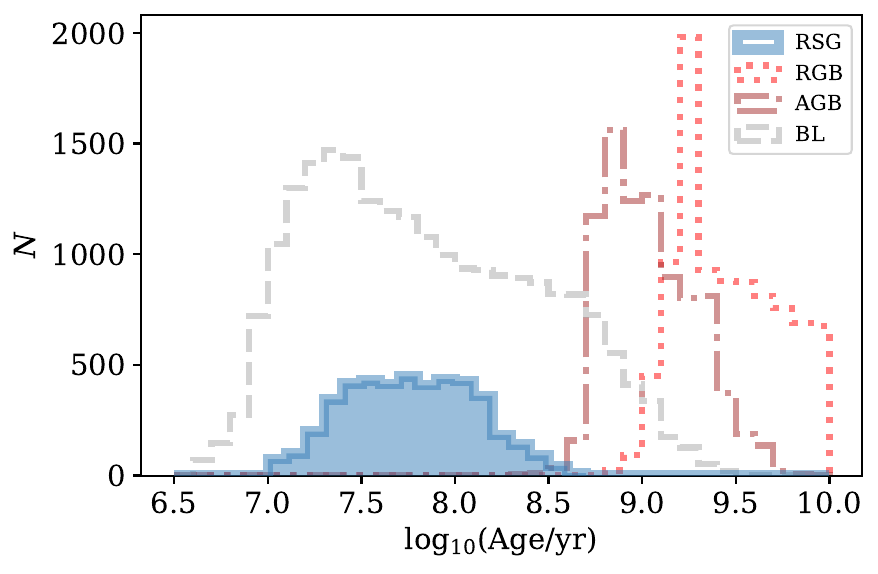}
\caption{Ages of synthetic stellar populations with a range of SMC-like metallicities. We compare synthetic stars passing our RSG selection criteria (Figure~\ref{fig:cmd_compare}) with synthetic stars in the RGB, AGB and ``blue loop" (BL) \citep[the selection of these populations in color-magnitude space is defined in Section 2.3.2 of][]{gaia2021}. We note that the blue loop sample (BL) overlaps with the RSG selection in Figure~\ref{fig:hi_complex}. 
\label{fig:ages} }
\end{figure}

\subsubsection{APOGEE}

In addition to the \emph{Gaia} stars, we select stars from the second iteration of the Apache Point Observatory Galactic Evolution Experiment (APOGEE-2, \citealt{majewski2017}), which was a part of the fourth iteration of the Sloan Digital Sky Survey (SDSS-IV, \citealt{blanton2017}). APOGEE-2 consisted of two $H$-band spectrographs \citep{wilson2019}, one on the SDSS 2.5m telescope at Apache Point Observatory \citep{gunn2006} and one on the 2.5m du Pont Telescope at Las Campanas Observatory \citep{bowen1973}, the latter of which allows for observations of the SMC due to its position in the Southern hemisphere. SMC stars were targeted through a combination of main survey fields and special fields from ``Contributed Programs'', as described and detailed in \citet{zasowski2017,nidever2020,santana2021}. 

In this work, we make use of APOGEE radial velocities and metallicities from the 17th Data Release (DR17) of SDSS-IV \citep{abdurrouf2022}. APOGEE measures radial velocities by cross-correlating the observed spectra with synthetic spectra, using methodology described in detail in \citet{nidever2015} and updates described in \citet{abdurrouf2022}. \citet{jonsson2020} found that the APOGEE DR16 radial velocities agreed with Gaia DR2 to within $1-2\rm\,km\,s^{-1}$ depending on the apparent magnitude of the star. We find similar agreement. For the 548 stars in common between APOGEE and \emph{Gaia}, the RVs agree to within $0.03^{+2.2}_{-1.6}\rm\,km\,s^{-1}$ (mean and $\pm 1 \sigma$ errors), which is comparable to the average uncertainty in radial velocity in the \emph{Gaia} sample of $1.5\rm\,km\,s^{-1}$. APOGEE metallicities are measured using the APOGEE Stellar Parameters and Chemical Abundance Pipeline (ASPCAP, \citealt{garcia-perez2016}), which uses the FERRE code (\citealt{AllendePrieto2006}) to match the normalized, RV-corrected, and sky-subtracted spectra to a library of synthetic stellar spectra, the generation of which is described in \citet{Zamora2015}. Updates to this spectral grid are described in \citet{abdurrouf2022}, which include spectral grids generated using the SYNSPEC code \citep{Hubeny2021} with non-local thermodynamic equilibrium corrections from \citet{Osorio2020}.

\begin{figure}[ht!]
\centering
\includegraphics[width=0.45\textwidth]{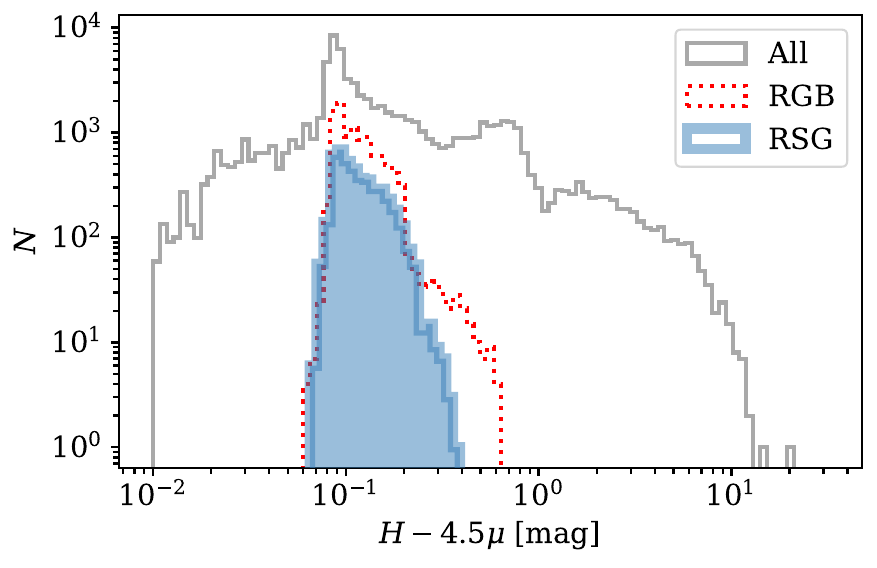}
\caption{Synthetic IR color histograms for three subsets of the simulated stellar population: all (black), those passing our \emph{Gaia} color selection for RSGs (blue), and those passing the \emph{Gaia} color selection for RGBs defined in \citep{gaia2021} (red). We observe that the simulated RSGs span an even narrower $H-4.5\mu$m distribution than the RGBs. 
\label{fig:rjce_test} }
\end{figure}

\begin{figure}[ht!]
\centering
\includegraphics[width=0.45\textwidth]{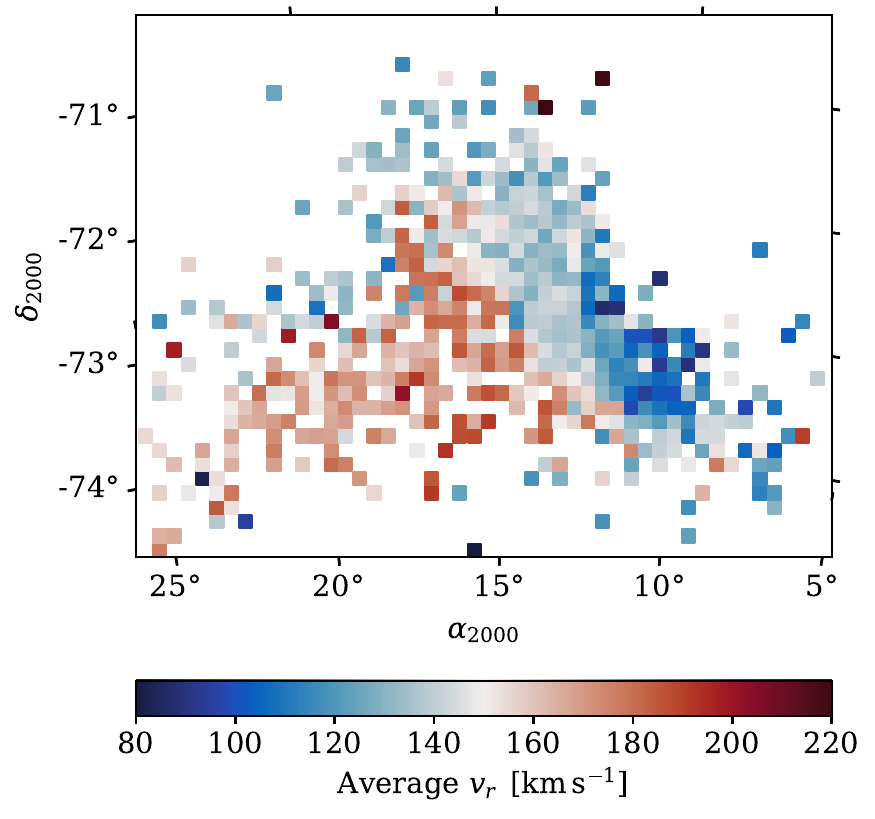}
\caption{Average radial velocity of the selected stellar catalog from \emph{Gaia} and APOGEE ($v_r$, LSR) (binned spatially according to the pixel sizes shown). The stars display a radial velocity gradient across the main body of the SMC with similar magnitude as the gas (Figure~\ref{fig:hi_complex}).
\label{fig:hi_complex_gaia1} }
\end{figure}

We select all APOGEE stars within a radius of $9^{\circ}$ from the SMC center (5,938 stars). This search radius is significantly larger than the GASAKP-\hi\ footprint, and therefore is aribitrarily selected. We then compute the proper motions in $x$ and $y$ in the same way as for the Gaia sample and exclude all stars with values beyond $4\times$ the robust scatter estimate \citep{gaia2018} from the median of the \emph{Gaia} sample in both axes. We then convert the Heliocentric velocities from APOGEE to the LSR frame and select stars with SMC velocities ($80<v_r<250\rm\,km\,s^{-1}$) residing within the GASKAP-\hi\ footprint (2,407 stars). 

\begin{figure*}[ht!]
\centering
\includegraphics[width=0.95\textwidth]{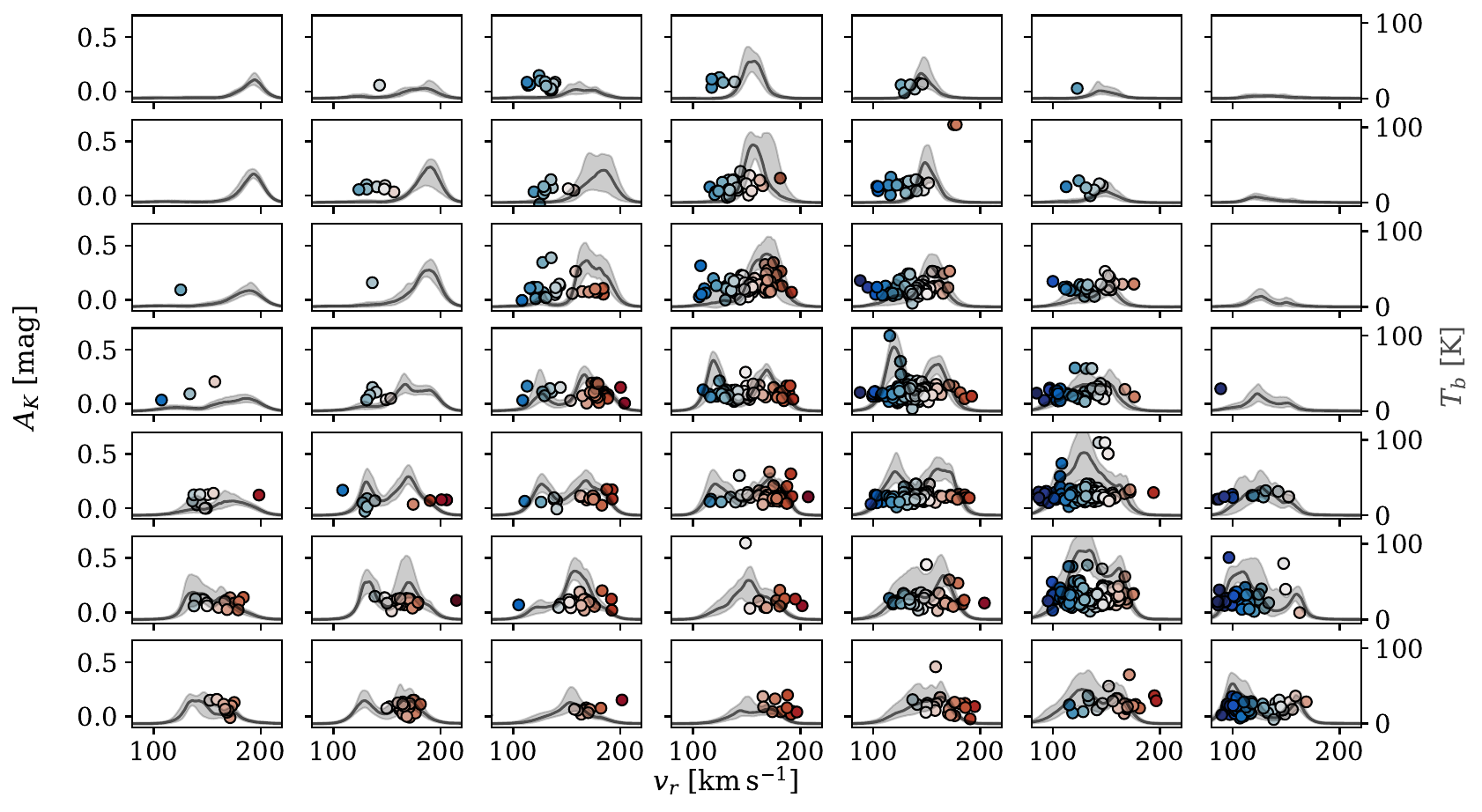}
\caption{$A_K$ versus velocity for stars within the same regularly-spaced spatial bins as in Figure~\ref{fig:hi_complex}. The stars are colored by the radial velocity according to the same colorbar as in Figure~\ref{fig:hi_complex}. The average \hi\ $T_b(v_r)$ profiles are overlaid in black, and include shaded grey envelopes denoting the $16^{\rm th}$ through $84^{\rm th}$ percentiles of the $T_b(v_r)$ within each spatial bin. In general, the stellar radial velocities span the same range as the \hi\ velocities across the SMC.
\label{fig:hi_complex_gaia2} }
\end{figure*}

\subsubsection{Young star selection}
\label{sec:young_star_selection}

In Figure~\ref{fig:cmd_compare}, we display color-magnitude diagrams (CMDs) for the two samples. We use the \emph{Gaia} observed magnitudes, $BP$-$RP$ (a blue-red color index) and $G$. 

In the case of the \emph{Gaia} sample, the stars with measured radial velocities are all bright ($G \lesssim 16$). In both samples, the stars span a wide range of stellar types, including pulsating variables, red supergiants, and asymptotic giant branch stars \citep{gaia2021}.

The APOGEE stars span a wider range in magnitude, as this survey included both shallow and deep observations. As a result there is a gap between the faintest and brightest RSGs around $G\sim15$. We will discuss the effect of the magnitude distribution of this sample on our results in Section~\ref{sec:analysis}. 

From these stars, we need to sub-select according to two conditions: (1) stars that are young enough to support our assumption that their motions trace the bulk motions of their parent gas clouds; (2) stars of a spectral type that is well-understood enough for the line of sight extinction towards them to be accurately inferred with available data. To satisfy these criteria, we select all stars within the region highlighted in Figure~\ref{fig:cmd_compare}, as this region of the CMD is dominated by RSGs. As a test of our assumption, we generate a set of synthetic stellar isochrone tables for a range of SMC metallicities ([M/H] = $-1.0$ to $-0.65$) from PARSEC \citep{bressan2012, tang2014, chen2014, chen2015, marigo2017, pastorelli2019, pastorelli2020}. In Figure~\ref{fig:ages}, we plot histograms of the ages of stars in our RSG selection (blue), as well as the ages of a range of stellar populations identified in the SMC by \citet{gaia2021}, including red giant branch stars (RGB), asymptotic giant branch stars (AGB), and blue loop stars (BL). The RSG are of intermediate age ($\sim10-100$ Myr), but are among the youngest stars available in the two surveys. In addition, RSGs in the LMC have been shown to closely mimic the kinematics of the \hi\ \citep{olsen2015}.

Following the RSG selection, we combine the \emph{Gaia} and APOGEE samples by cross-matching the catalogs and removing overlapping stars from the \emph{Gaia} catalog, yielding a total of 1,947 stars (782 APOGEE, 1,165 \emph{Gaia}).
In Figure~\ref{fig:cmd_compare}c, we plot the spatial distribution of these samples. Because the APOGEE survey of the SMC was limited to individual pointings, these stars are not uniformly spread across the galaxy as the \emph{Gaia} stars. We will discuss the effect of the difference in spatial sampling in Section~\ref{sec:analysis}.

\subsubsection{Estimating Extinction}

To estimate the extinction towards each source in the sample in the same manner, we use the ``Rayleigh Jeans Color Excess" (RJCE) method \citep{majewski2011}. RJCE was developed to quantify the amount of reddening towards individual stars based on a single observed color, $H-4.5\rm \,\mu m$. These wavelengths sample the Rayleigh-Jeans tail of the stellar SED where red clump and red giant stars have the same intrinsic color. As a result, variations in their observed colors can be attributed to the effect of dust extinction \citep{majewski2011}.

We note that the RJCE method has not been explicitly validated on the stellar types selected here (RSG). To test the validity of the method, we use the same set of synthetic isochrones from Figure~\ref{fig:ages} and plot histograms of the $H-4.5\mu$m colors for the RSG sample, compared with the RGB sample \citep[where the color-magnitude selection for SMC RGB is defined in][]{gaia2021} in Figure~\ref{fig:rjce_test}. We observe here that our RSG selected stars span an even narrower $H-4.5\mu$m color range than the RGB stars, granting us confidence that these stars are likely good candidates for the RJCE method. 

To compute extinction in $K_s$ band, $A_K$, we cross-match our sample with the catalog of stars from the \emph{Spitzer} SAGE-SMC survey \citep{gordon2011}, specifically the ``more reliable" catalog of IRAC Epoch 0, 1, and 2 catalog. This catalog includes IRAC $4.5\mu\rm m$ photometry, as well as $H$-band photometry from 2MASS \citep{skrutskie2006}. We find a match within a search radius of $3^{\prime \prime}$ for 1,934 of our stars. We then compute the extinction in the $K_s$ band, $A_K$, using RJCE via \citet{majewski2011} Equation 1, reproduced here,

\begin{equation}
A_K = 0.918\, (H - [4.5 \mu] - 0.08).
\end{equation}

\noindent We remove stars with NaN-values (117 stars) or extraneously red $3.6-4.5\mu\rm m$ colors (defined as $[3.6-4.5]>0.15$; 10 stars).  Our final catalog has 1,801 stars.

\begin{figure}[ht!]
\includegraphics[width=0.5\textwidth]{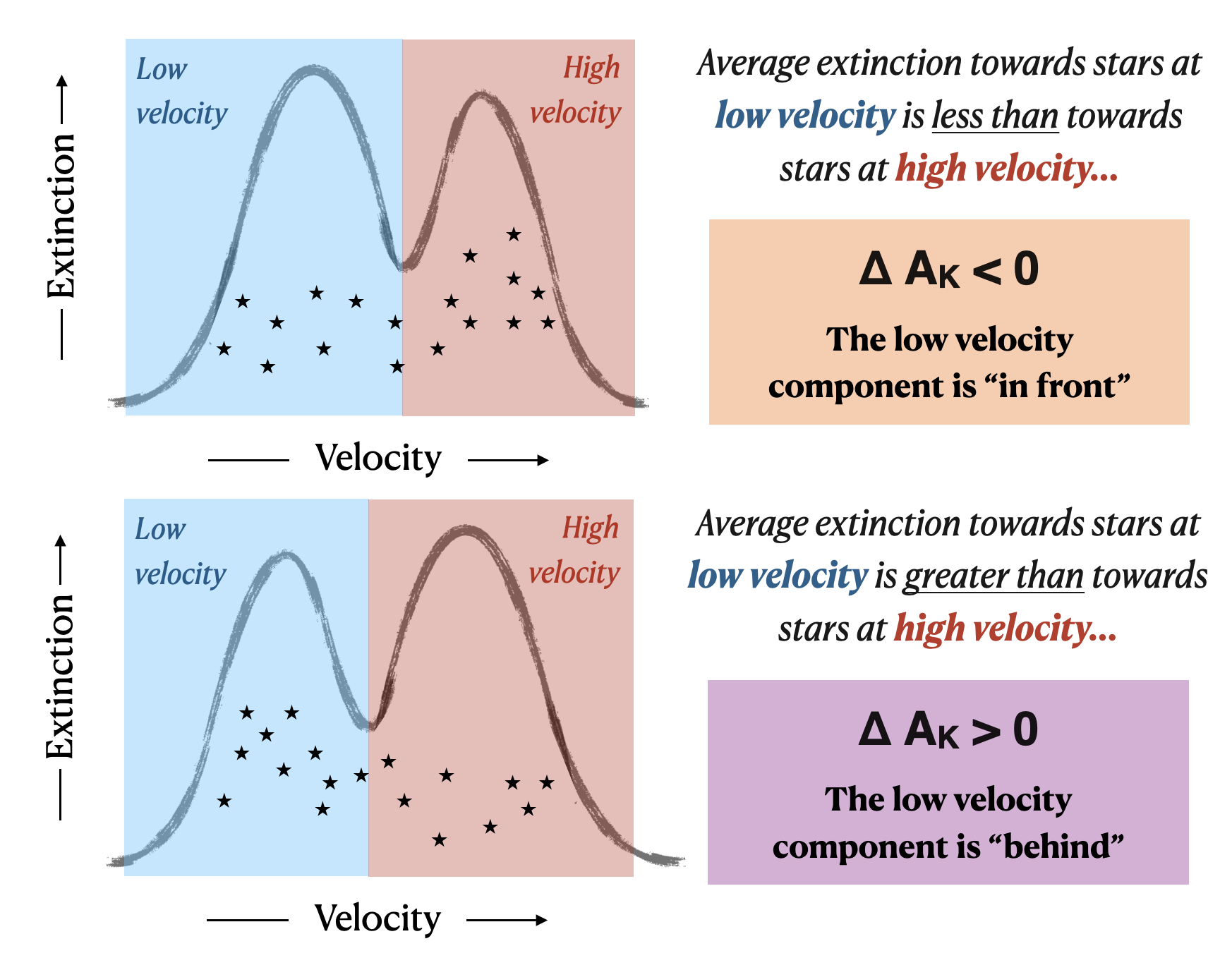}
\caption{Cartoon schematic of the differential extinction estimation process. Top: The average extinction towards stars with ``low velocity" ($v_r<M_1$) is less than the average extinction towards stars with ``high velocity" ($v_r>M_1$) -- therefore, $\Delta A_K<0$ and the ``low velocity" gas component is in front of the high velocity component in this direction. Bottom: The average extinction towards stars with ``low velocity" ($v_r<M_1$) is greater than the average extinction towards stars with ``high velocity" ($v_r>M_1$)-- therefore, $\Delta A_K>0$ and the ``low velocity" gas component is behind the high velocity component in this direction. 
\label{fig:schematic}}
\end{figure}

\begin{figure}[ht!]
\includegraphics[width=0.5\textwidth]{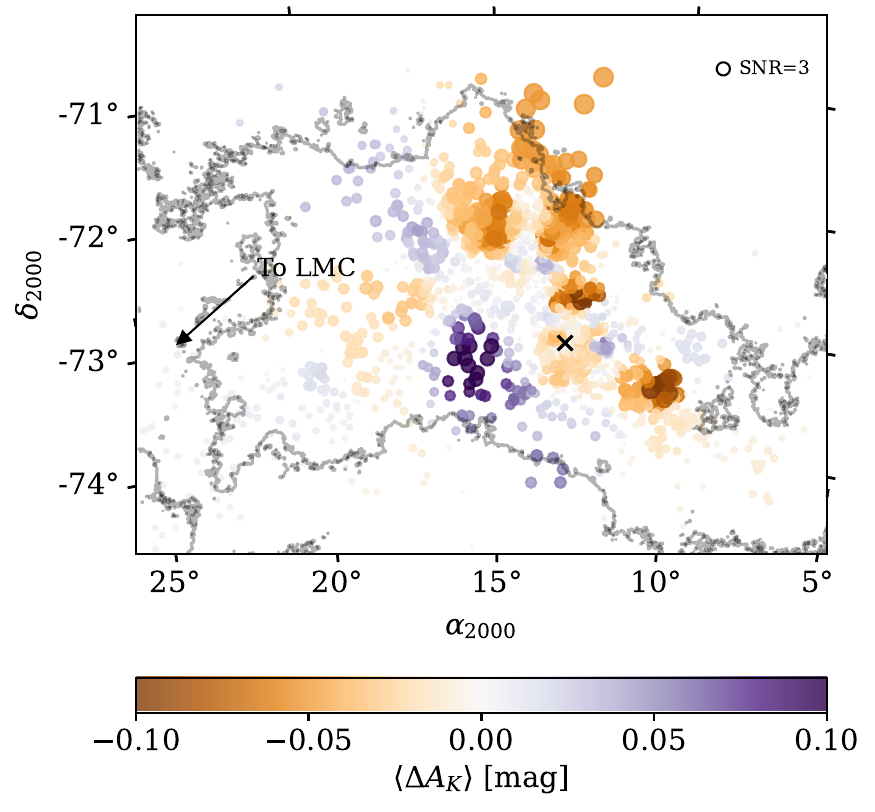}
\caption{The difference in extinction towards low velocity gas and high velocity gas ($\langle \Delta A_K \rangle$), averaged across binning parameters at the location of each source. The symbol sizes are scaled by the signal to noise in $\langle \Delta A_K \rangle$ (larger symbols = more confidence). The assumed center of the SMC is marked with an ``x" \citep[$\alpha_{0}, \delta_{0} = 13.15833, \, -72.80028$][]{zivick2021} and the direction towards the LMC is indicated with an arrow. 
\label{fig:results}}
\end{figure}

To further validate our measurements, we compare $A_K$ with the results of \citet{zhang2023}, who estimated stellar parameters and extinction towards the $220$ million stars with XP spectra from \emph{Gaia} DR3 using a forward-modeling approach trained on the LAMOST spectral library. We find that the values of $A_K$ are positively correlated with the extinction parameter, $E$ (i.e., a scalar proportional to total extinction), from \citet{zhang2023} with high significance (Pearson $R=0.52$) and the null hypothesis that the two samples are uncorrelated can be rejected with $p\ll 0.001$.\footnote{We note that we do not use the \citet{zhang2023} values in our analysis directly given the uncertainty in validating stellar models trained for Milky Way conditions on the SMC. However, as a test we repeated our full analysis procedure using $E$ instead of $A_K$ and found statistically indistinguishable results. This lends confidence that we are not significantly biased by our choice of extinction measurement.}

\begin{figure*}[ht!]
\includegraphics[width=\textwidth]{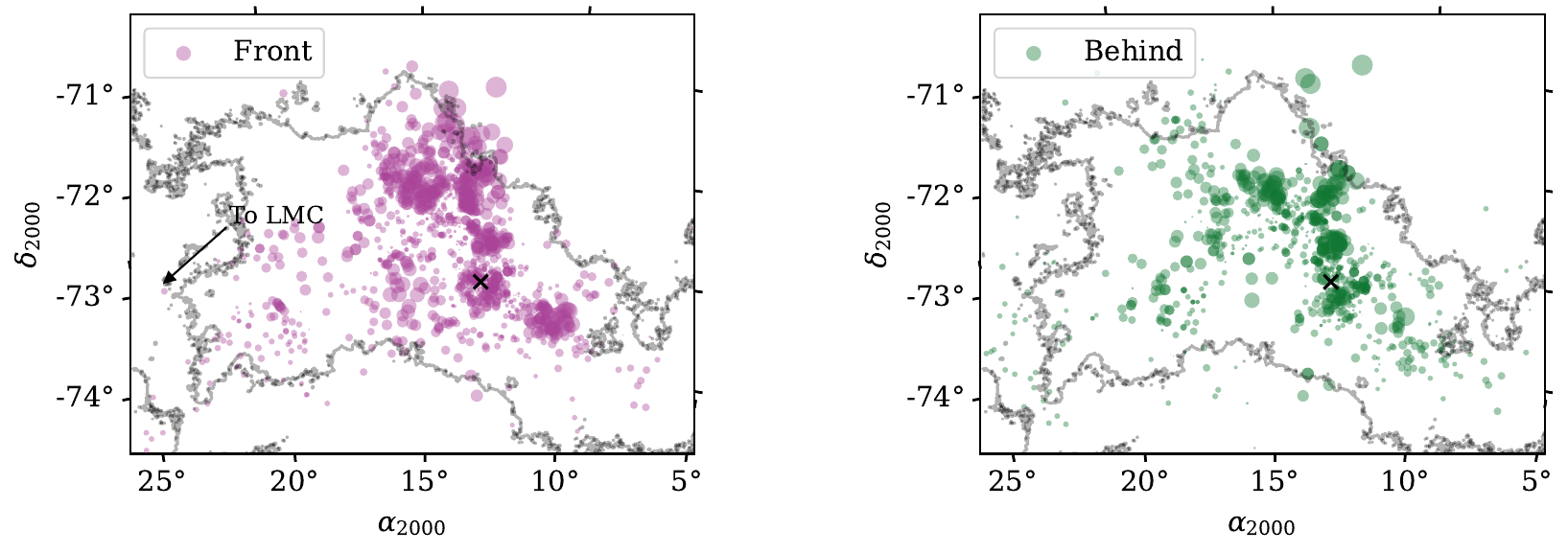}
\caption{Spatial distribution of stars in the ``front" structure ($\langle \Delta A_K \rangle <0$ at low velocity and $\langle \Delta A_K \rangle >0$ at high velocity) and the ``behind" structure ($\langle \Delta A_K \rangle >0$ at low velocity and $\langle \Delta A_K \rangle <0$ at high velocity). As in Figure~\ref{fig:results}, the points are scaled by the significance in the $\langle \Delta A_K \rangle$ measurement.
\label{fig:fb_compare}}
\end{figure*}

We emphasize that, as will be elaborated on in the ensuing analysis, our key metric for this work is a differential measurement in extinction, and therefore the precise zeropoint calibration to the RJCE method are not needed for our analysis. In addition, the correction for Milky Way foreground extinction is not needed, as we assume that the foreground extinction does not vary significantly between spatial bins considered here (see Section~\ref{sec:analysis}). 

In Figure~\ref{fig:hi_complex_gaia1}, we plot the spatially-binned average radial velocities of the stars in our sample. The stars are distributed according to the regions of highest \hi\ column density, and display an East-West gradient in $v_r$ with the same magnitude as seen in $T_b(v_r)$ shown in Figure~\ref{fig:hi_complex}. In Figure~\ref{fig:hi_complex_gaia2}, we plot $A_K$ vs.~$v_r$ for stars in the same example bins shown, and observe that the stars follow the same radial velocity range of the \hi\ profiles. Of the 1,801 total stars in the sample, 995 ($55\%$) are at low velocity (i.e., $v_r<M_1$) and 806 are at high velocity. 

\section{Analysis}
\label{sec:analysis}

The GASKAP-\hi\ data cube and our selected sample of stars provide two high-resolution probes of the 3D structure of the SMC system. In the following analysis, we match the stars with peaks in $T_b(v_r)$ based on their $v_r$ (Figure~\ref{fig:hi_complex}) and compare the average extinction towards the \hi\ velocity structures observed across the SMC. This differential measurement -- the average difference in extinction towards the ``low" and ``high" velocity peaks observed in Figures~\ref{fig:hi_complex} -- determines the order of these components along the line of sight. 

\subsection{Matching Stars with Gas}

The process of matching the stars with \hi\ peaks requires a series of selection parameters, which we describe below.

\subsubsection{Number of \hi\ components}

To identify the number of \hi\ components and their properties, we follow \citet{murray2019}, who used the first numerical derivative of the brightness temperature spectrum in each pixel of the \hi\ data cube from \citet{mcg2018} to identify peaks in radial velocity. In Figure 2 from \citet{murray2019}, they conclude that most of the SMC main body features at least two components. However, in detail, this process requires choosing two free parameters: the minimum $T_B$ threshold for a component to be considered ``significant" ($T_{B,\rm min}$), and a Gaussian smoothing kernel width ($\sigma_G$) for suppressing the effect of spectral noise on the numerical derivative, which is necessary to determine the number of components along the line of sight. We find that varying these selections ($1<T_{b, \rm \, min}<10\rm\,K$, $2<\sigma_G<10$) does not have a significant effect on our results, as the majority of the stars are located along high-$T_b$, complex LOS. For the ensuing analysis, we select $T_{b,\rm\,min}=5\rm\,K$ and $\sigma_G=4$ channels.

\begin{figure*}[ht!]
\includegraphics[width=\textwidth]{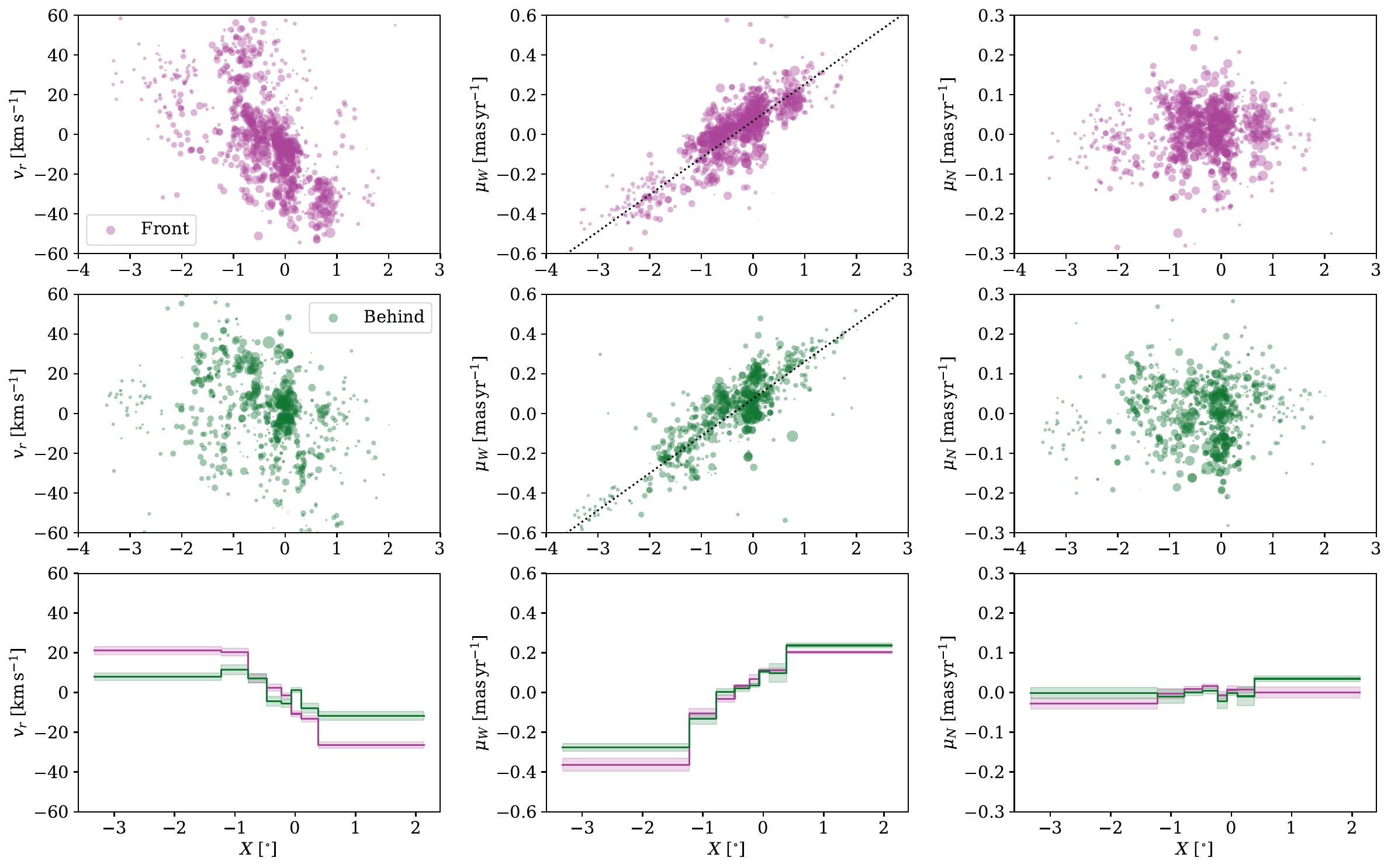}
\caption{Comparing the 3D kinematics ($v_r$, $\mu_W$ and $\mu_N$; left, middle and right, respectively) as a function of Cartesian East-West ($X$) position for stars in the front and behind structures, after subtracting the systemic motions of the SMC. The bottom row of panels compares the binned average values for each velocity component in 10 bins of equal size for the front and behind structures, highlighting in particular the difference in $v_r$ as a function of $X$. 
\label{fig:kinematics}}
\end{figure*}

\subsubsection{Spatial binning of stars}

Given that the stellar sample is unevenly distributed across the SMC, with most stars lying in the central regions and far fewer in the outskirts (Figure~\ref{fig:cmd_compare}c), we use Voronoi binning \citep{cappellari2003}, an adaptive binning scheme that allows us to ensure consistent source counts across spatial bins. This process also requires several binning parameters. First is the degree of regular-spaced spatial binning before applying the Voronoi binning algorithm (as in practice, using all $\sim 4$ million individual \hi\ spaxels is prohibitively slow), parameterized by the number of pixels per side per bin ($N_{{\rm pix}, V}$). The second free parameter is the signal-to-noise ratio per bin ($\rm(SNR)_V$). We select $N_{{\rm pix},V}=30,\,40,\,50,\,60$ and $\rm(SNR)_V=3,\,4,\,5,\,6$ and repeat our analysis (below) for each unique combination of these parameters. This allows us to account for the effect of binning choice, while ensuring that there are a uniform number of stars per spatial bin.

\subsection{Identifying structures via differential extinction}

For each combination of selection parameters, in each bin we identify stars at ``low" velocity ($v_r<M_1$) and at ``high" velocity ($v_r>M_1$) where $M_1$ is the intensity-weighted average radial velocity of the gas along the line of sight. If there is at least one source in the ``low" and ``high" velocity samples, we compute the average extinction towards the ``low" and ``high" samples ($\langle A_K \rangle_{\rm low}$, $\langle A_K \rangle_{\rm high}$). Finally, we compute the difference between these values, defined as ``low" minus ``high":

\begin{equation}
    \Delta A_K = \langle A_K \rangle_{\rm low} - \langle A_K \rangle_{\rm high},
\end{equation}

\noindent and assign this value of $\Delta A_K$ to all stars within the bin. In Figure~\ref{fig:schematic}, we illustrate how this process works with simple cartoons for two cases. When $\Delta A_K <0$, there is less extinction towards the ``low" velocity component than the ``high" velocity component, which implies that the low velocity component is in front of the high velocity component. Conversely, when $ \Delta A_K >0$, there is more extinction towards the low velocity component, which implies that it is behind the high velocity component. 

In Figure~\ref{fig:results}, we plot the average of $\Delta A_K$ over all 16 unique combinations of binning parameters for each star ($\langle \Delta A_K \rangle$). In each trial, we find that there is a roughly equal number of stars at low and high velocity within each spatial bin, with a total number per bin that ranges between $\sim6$ and $\sim 40$ stars. In addition, the value of $\langle \Delta A_K \rangle$ for each star is derived from a non-uniform (Voronoi) binning scheme which ensures that a different set of stars is selected in each trial. This choice allows us to minimize the impact of single outliers on the results.

We observe clear trends in $\langle \Delta A_K \rangle$ across the face of the SMC. The points in Figure~\ref{fig:results} have sizes that are scaled by the signal to noise in the measurement of $\langle \Delta A_K \rangle$. Given the fact that the dust content of the SMC is diffuse (outside of the major star-forming regions), the magnitude of the extinction difference is small ($\lesssim 0.1 \rm \, mag$). However, the fact that the sign of $\langle \Delta A_K \rangle$ is constant across large areas ($100 \rm s$ of pc, much larger than the spatial bin sizes) indicates that the signal detected at each location is significant. 
 
\subsection{Selecting ``front" and ``behind" samples}

Building from the spatial trends observed in Figure~\ref{fig:results}, we select all stars in ``front" (i.e., negative $\langle \Delta A_K \rangle$ at low velocity and positive $\langle \Delta A_K \rangle$ at high velocity) and ``behind (positive $\langle \Delta A_K \rangle$ at low velocity and negative $\langle \Delta A_K \rangle$ at high velocity). 

In Figure~\ref{fig:fb_compare}, we plot the spatial distributions of the ``front" and ``behind" samples. We find 1,014 stars in the front sample and 787 in the behind sample. Of these, $35\%$ and $42\%$ are from APOGEE in the front and behind samples respectively, highlighting the fact that both samples have similar representation from the \emph{Gaia}-only and APOGEE coverage.

\subsection{Kinematics}

To analyze the kinematics of the front and behind \hi\ samples, we first correct for the systemic motion of the SMC. We estimate the mean proper motion of the both the front and behind samples to be $(\mu_{\alpha,\rm sys},\,\mu_{\delta,\rm sys}) = (0.77, \,-1.25)\,\rm mas \,yr^{-1}$, which is in good agreement with previous estimates of the proper motion of the SMC \citep[e.g.,][]{Jacyszyn-Dobrzeniecka2017, zivick2018}. We estimate the systemic radial velocity to be the median of the GASKAP-\hi\ first moment map, equal to $v_{r, \rm sys}=151\rm\,km\,s^{-1}$. We subtract $v_{r,\rm sys}$, $\mu_{\alpha,\rm sys}$, $\mu_{\delta,\rm sys}$ from $v_r$, $\mu_{\alpha}$ and $\mu_{\delta}$ in the sample.

For convenience, we next convert the observed coordinates ($\alpha$, $\delta$) coordinates to a Cartesian frame in the plane of the sky ($X$, $Y$) following \citet{vandeven2006} Equation 1, reproduced here,

\begin{equation} 
\begin{aligned}
X &= - r_0 \, \cos{\delta}\,\sin{\Delta \alpha}, \\
Y &= r_0 \,(\sin{\delta}\cos{\delta_0} - \cos{\delta}\sin{\delta_0}\cos{\Delta \alpha}),
\end{aligned}
\label{eq:cartesian}
\end{equation}

\noindent where $\Delta \alpha=\alpha-\alpha_0$, $\Delta \delta = \delta - \delta_0$ and ($\alpha_0$, $\delta_0$) is the assumed center of the SMC \citep[here, from][]{zivick2021}. In this frame, positive ``$X$" pointing West and positive ``$Y$" pointing North. We convert $\mu_{\alpha}$ and $\mu_{\delta}$ to the Cartesian frame ($\mu_W$ and $\mu_N$) using the orthographic projection from Equation 2 of \citet{gaia2018}, with positive $\mu_W$ pointing West and positive $\mu_N$ pointing North. 

Another important observational effect is ``perspective rotation", or the apparent kinematic signature imposed by the systemic motion of the objects with large angular extent \citet[e.g.,][]{vandeven2006}. Correcting for this effect requires an assumed distance to the star along the line of sight, which in this case is not known. Therefore, we incorporate this correction in when we model our results in Section~\ref{sec:model}.

In Figure~\ref{fig:kinematics}, we plot the three residual (i.e., systemic-corrected) velocities ($v_r$, $\mu_W$ and $\mu_N$) for the stars in each component. In the bottom row, we compare the average values as a function of East-West coordinate ($X$), binned so that there is an equal number of stars per bin. 

To compare the velocities between the front and behind samples, we conducted an Anderson-Darling test to compare the distributions of $v_r$, $\mu_W$ and $\mu_N$. The null hypothesis that the two samples are drawn from the same parent population can be rejected with a p-value of $<0.001$ for $v_r$, and $0.02$ for $\mu_W$ and $\mu_N$. This is confirmed visually in Figure~\ref{fig:kinematics}, it is clear that the slope of the two distributions of $v_r$ are significantly different. 

In addition, both front and behind display a strong gradient in $\mu_W$. We fit a line to this gradient (Figure~\ref{fig:kinematics}) and find the slope to be $0.185$ ($0.187$) $\rm mas/(yr\, deg)$, with an intercept of $0.068$ ($0.075$) $\rm km\,s^{-1}$, in the front and behind components respectively. This gradient is consistent with being caused by the tidal influence of the LMC on the SMC, which we will discuss further in Section~\ref{sec:model}.

\begin{figure}[ht!]
\includegraphics[width=0.45\textwidth]{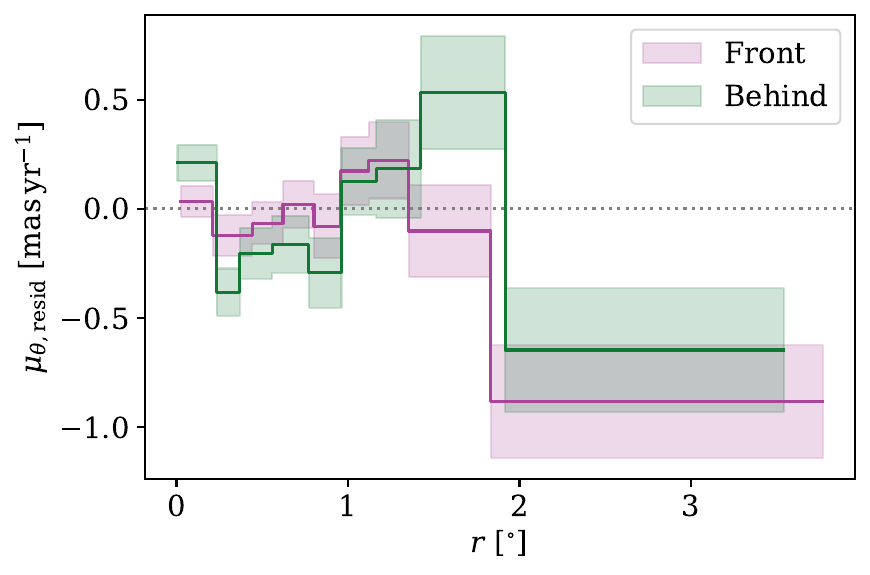}
\caption{The residual tangential proper motion ($\mu_{\theta,\rm resid}$) as a function of radius in the front and behind components, after subtracting the gradient in $\mu_w$ (Figure~\ref{fig:kinematics}). Here we plot the average (line) and standard error (shading) in bins of equal numbers of stars (6 bins per sample). 
\label{fig:pmtan}}
\end{figure}

To further explore the residual proper motions in the system, we subtract the gradient in $\mu_W$ and look for signatures of rotation in the residuals. To quantify this, we convert $\mu_{W, \rm resid}$ and $\mu_N$ to radial and tangential proper motions ($\mu_{r, \rm resid}$ and $\mu_{\theta, \rm resid}$). In Figure~\ref{fig:pmtan}, we plot $\mu_{\theta,\rm resid}$ as a function of radius in both components. Coherent rotation should appear as a nonzero $\mu_{\theta,\rm resid}$ as a function of radius, with positive or negative values depending on the direction of rotation \citep[e.g.,][]{zivick2021}. We bin the stars so that there is an equal number of stars in each of 10 radial bins ($\sim100$ stars), and compute the mean and standard error.

In both components, the residual tangential proper motions are consistent with zero in the innermost radii ($r\lesssim 1.5^{\circ}$). At larger radii in the behind component, $\mu_{\theta,\rm resid}$ switches from positive to negative, whereas in the front component $\mu_{\theta,\rm resid}$ remains negative. We interpret these results as there being tentative evidence for rotation in the front component, however the uncertainties are large given that this analysis relies on an assumed galaxy center and orientation of rotation. 

\begin{figure}[ht!]
\includegraphics[width=0.5\textwidth]{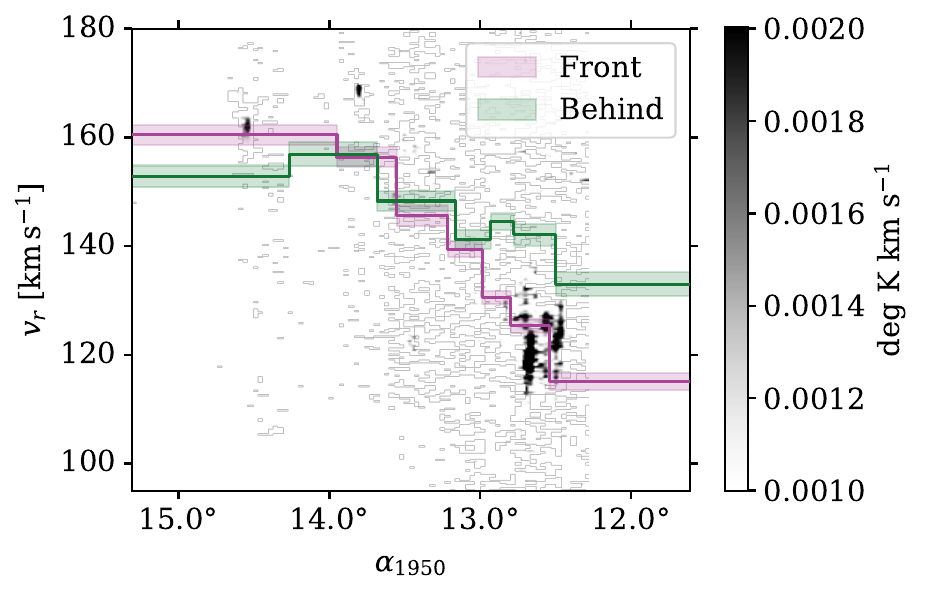}
\caption{Comparing right ascension ($\alpha_{1950}$) vs.~radial velocity ($v_r$) in bins of equal number of stars ($\sim100$ per bin) in the front and behind components (mean and standard error) with observed structure in molecular gas \citep[NANTEN;][]{mizuno2001}. The colorbar indicates the integrated $^{12}$CO(1-0) flux density.
\label{fig:co_compare}}
\end{figure}

\subsection{Where is the molecular gas?}

Resolving the molecular gas content of the SMC is important for gauging the amount of fuel available for star-formation in the system. In addition, the ratio of mass in the atomic and molecular gas phases is sensitive to the influence of physical mechanisms in the system (e.g., radiation fields, shocks, turbulence), and traces the chemical state of the ISM. 

The best available tracer for the cold, dense molecular gas phase in the SMC, which represents the conditions necessary for star formation, is carbon monoxide (CO) emission at millimeter wavelengths. Existing CO surveys of the SMC have established that there is very little CO relative to the rich atomic gas reservoir \citep{leroy2007, jameson2016}. In addition, CO in the SMC displays a much simpler radial velocity distribution than the \hi -- appearing as a single component with a velocity consistent with one of the ``low" or ``high" velocity components seen in \hi\ emission but not the other \citep{rubio1991, israel1993, rubio1996, mizuno2001, muller2010, tokuda2021, saldano2022}. In addition, high-resolution observations with the ALMA observatory have revealed a wealth of clouds with weak CO emission and extremely small angular size \citep[$<1^{\prime \prime}$; e.g.,][]{tokuda2021}.

\begin{figure*}[ht!]
\includegraphics[width=0.45\textwidth]{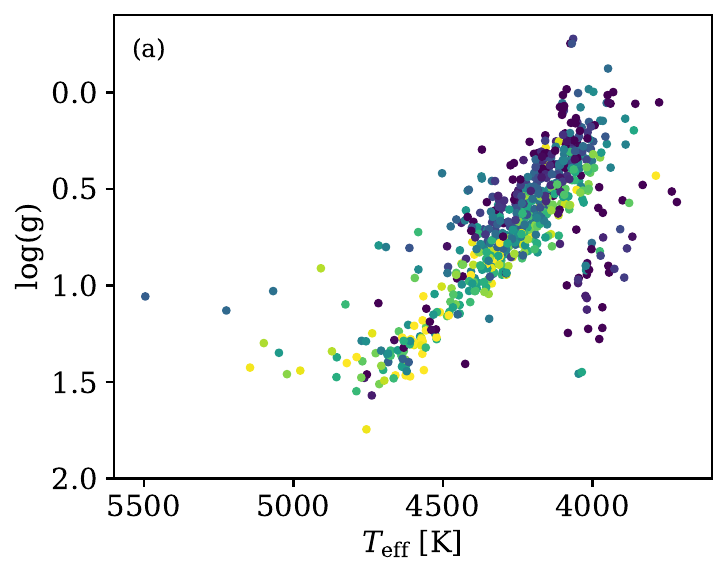}
\includegraphics[width=0.55\textwidth]{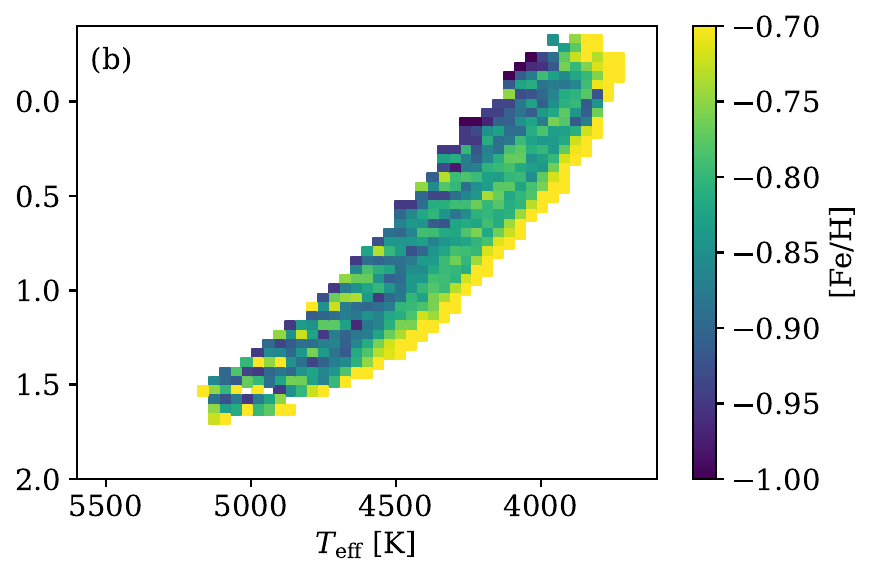}
\caption{(a): APOGEE log(g) vs.~Teff for the RSGs analyzed in this work, colored by metallicity ([Fe/H]). (b): Binned average log(g) vs.~Teff for synthetic stellar isochrones with the same RSG selection. We note that the clump of stars at $T_{\rm eff} \sim 4000\rm\,K$ in panel (a) are all bright, dusty RSGs, whose effect we discuss in the text. }
\label{fig:logg_teff}
\end{figure*}

To explore the molecular gas content of the front and behind components, we compare our results with observations from \citet{mizuno2001}, who mapped the $^{12}$CO(1-0) transition across the SMC with the NANTEN telescope. Although the resolution is coarse ($2.6^{\prime}$, $\sim45\rm\,pc$ at the distance of the SMC), it is the survey with the widest and most contiguous spatial coverage. Due to the large angular size of the SMC, most studies are limited to individual star-forming regions. We compute a $0^{\rm th}$ moment map along the declination axis of the NANTEN SMC data cube, and compare the structure in right ascension ($\alpha_{1950}$) vs.~radial velocity (LSR; $v_r$) with the stars in the front and behind components in Figure~\ref{fig:co_compare}.

Although there is overlap between the radial velocities in the front and behind components across the middle of the SMC, we observe in Figure~\ref{fig:co_compare} that the front component matches the RA vs.~radial velocity structure of the $^{12}$CO emission better than the behind component. The correspondence between CO velocities and the front component is particularly true on the Eastern side ($\alpha_{1950}\sim 14.5^{\circ}$) and in the central region ($\alpha_{1950}\sim 12.5^{\circ}$). 

However, the interpretation of these results becomes complicated if the dependence of CO emission strength on gas-phase metallicity is considered. The ratio between CO and H$_2$ is known to decrease steeply as a function of metallicity \citep{schruba2012, accurso2017}. So, if the behind component has significantly lower gas-phase metallicity, CO emission will be much more difficult to detect despite the possible presence of molecular gas. In addition, CO structures at larger distances would also be more difficult to detect due to the effects of beam dilution.

\begin{figure}[ht!]
\includegraphics[width=0.45\textwidth]{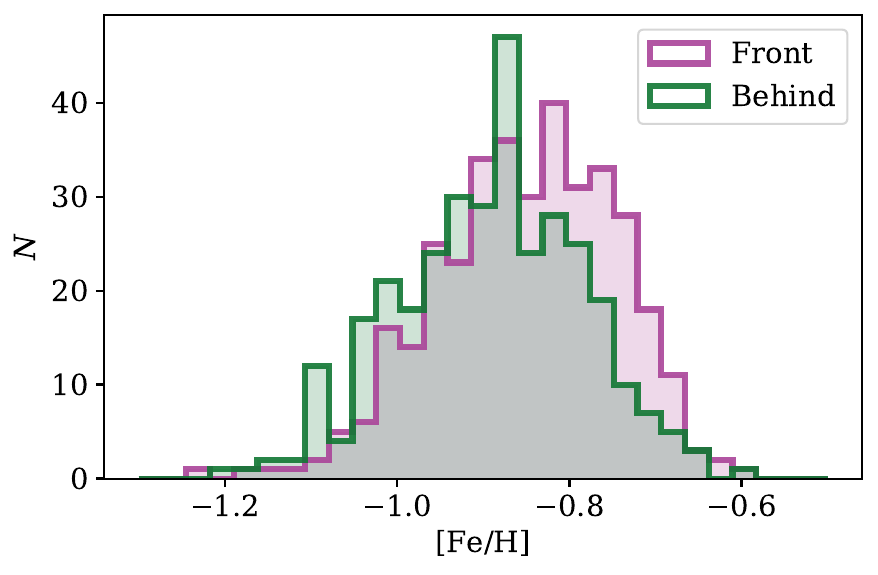}
\caption{Histograms of [Fe/H] for APOGEE stars in the front and behind components.
\label{fig:feh}}
\end{figure}

\begin{figure*}[ht!]
\includegraphics[width=0.95\textwidth]{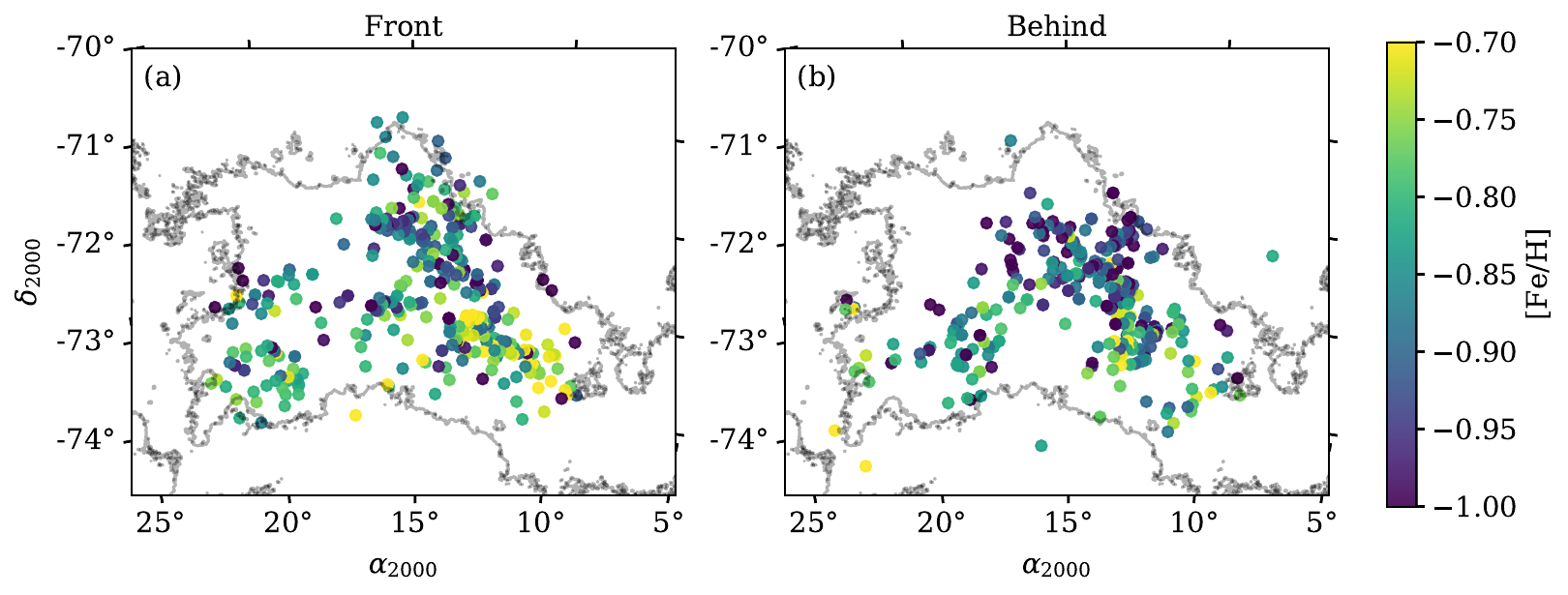}
\caption{Spatial distribution of stars in the front (left) and behind (right) systems, colored by metallicity, [Fe/H] (for those with estimates from APOGEE). The single contour denotes an \hi\ column density of $15\times 10^{20}\rm\,cm^{-2}$.
\label{fig:feh_maps}}
\end{figure*}

\subsection{Stellar Metallicity}

To explore differences in metallicity, we analyze the APOGEE stars in front and behind components. In addition to radial velocities, APOGEE provides stellar parameter measurements, including stellar metallicity, [Fe/H]. The APOGEE survey provides accurate and precise metallicity measurements for RGBs across the MW (see e.g., \citealt{jonsson2020}), but there have been no assessments of the accuracy of metallicities derived from RSG spectra. While a detailed comparison of APOGEE RSG metallicities with optical studies is beyond the scope of this work, we use isochrones to determine if the relative metallicities of red supergiants are informative. Figure \ref{fig:logg_teff} shows a Kiel diagram colored by metallicity for the APOGEE RSGs (a) and the RSG isochrones described and used in Section \ref{sec:young_star_selection} (b). APOGEE finds that the more metal-poor stars are generally cooler and that the more metal-rich stars are generally hotter at a fixed luminosity. Overall, the APOGEE stars show the same qualitative trend of metal-rich stars having higher temperatures at fixed log(g), suggesting that the APOGEE metallicities are at least useful in a relative sense.

We note that the subset of RSGs with low log(g) sitting below the distribution in Figure~\ref{fig:logg_teff}a are all bright ($G\lesssim13 \rm \,mag$) and ``dusty" ($A_K>0.2$). These stars are present in both the front and behind components, and excluding them does not affect the analysis presented here (i.e., all observed signals persist). 

Histograms of the metallicities derived by APOGEE are shown in Figure~\ref{fig:feh}. Although there is significant overlap, there is evidence that stars in the front component have higher metallicity. A Kolmogorov-Smirnoff test between the two samples rejects the null hypothesis that they were drawn from the same parent population with $p=10^{-8}$. 

To check whether the observed difference in metallicity is due to a bias in observed magnitudes (e.g., is the behind component dominated by the brightest, intrinsically lowest-metallicity sources?), we compare the CMDs of the two components. We find that both components have stars spanning the full range of observed magnitudes in the RSG selection (Figure~\ref{fig:cmd_compare}). The completeness in our sample at these bright magnitudes is expected to be close to 1, and therefore the difference in metallicity is not likely to be a result of observational selection.

In Figure~\ref{fig:feh_maps} we plot the spatial distribution of RSGs in the front and behind components, colored by metallicity. The majority of the highest metallicity sources are located in the SW Bar, where the majority of the dust and molecular gas content resides \citep{jameson2016}. Interestingly, we observe that the lowest metallicity sources are in the behind component and are distributed in the Northern Bar. As a possible explanation, there are known outflows of cold \hi\ from this region \citep[e.g.,][]{mcg2018}, and thus stars which form within outflowing gas may be more metal poor than those forming with the Bar, which is a possibility for both the front and behind components. 

\section{Simple Model for Component Distances}
\label{sec:model}

In the following section, we construct a simple toy model to constrain the distances to the front and behind components based on their observed kinematics. Although the history of the LMC-SMC system is extremely complex to model, even in the latest hydrodynamical N-body simulations \citep{besla2012}, our goal is to determine whether a simple, physically-meaningful model can explain our results in broad strokes.

We treat the two components (front and behind) as independent structures experiencing tidal effects from the LMC (evident by the gradient in $\mu_W$; Figure~\ref{fig:kinematics}) as well as rotation (evident from residual $\mu_{\theta}$; Figure~\ref{fig:pmtan}). We do not consider the bulk motion of the LMC and SMC in the model, we model all stars in the reference frame of the LMC-SMC system, and calculate the tidal force at a fixed distance to the LMC over time.

For each component, we initialize a population of stars with positions distributed normally with $\sigma=1\rm\,kpc$ in each of the $x$, $y$ and $z$ dimensions. This value was chosen to roughly model the observed distribution of stars in $X$ and $Y$, and we assume for simplicity that the systems are symmetric in $Z$. The stars are given velocities with random Gaussian amplitude and $\sigma=20\rm\,km\,s^{-1}$ in the $x$, $y$, and $z$ directions. This value was chosen based on the average standard deviation in the values of $\mu_N$ for both samples, converted from $\rm mas/yr$ to $\rm km/s$ assuming a distance of $62\rm\,kpc$. In the plane of the sky, the LMC is located at a distance of $d_{\rm LMC,\,sky} = 20\rm\,kpc$, and has a mass of $M_{\rm LMC} =1.4\times10^{11}\rm\,M_{\odot}$ \citep{erkal2019}.

\begin{figure*}[ht!]
\includegraphics[width=\textwidth]{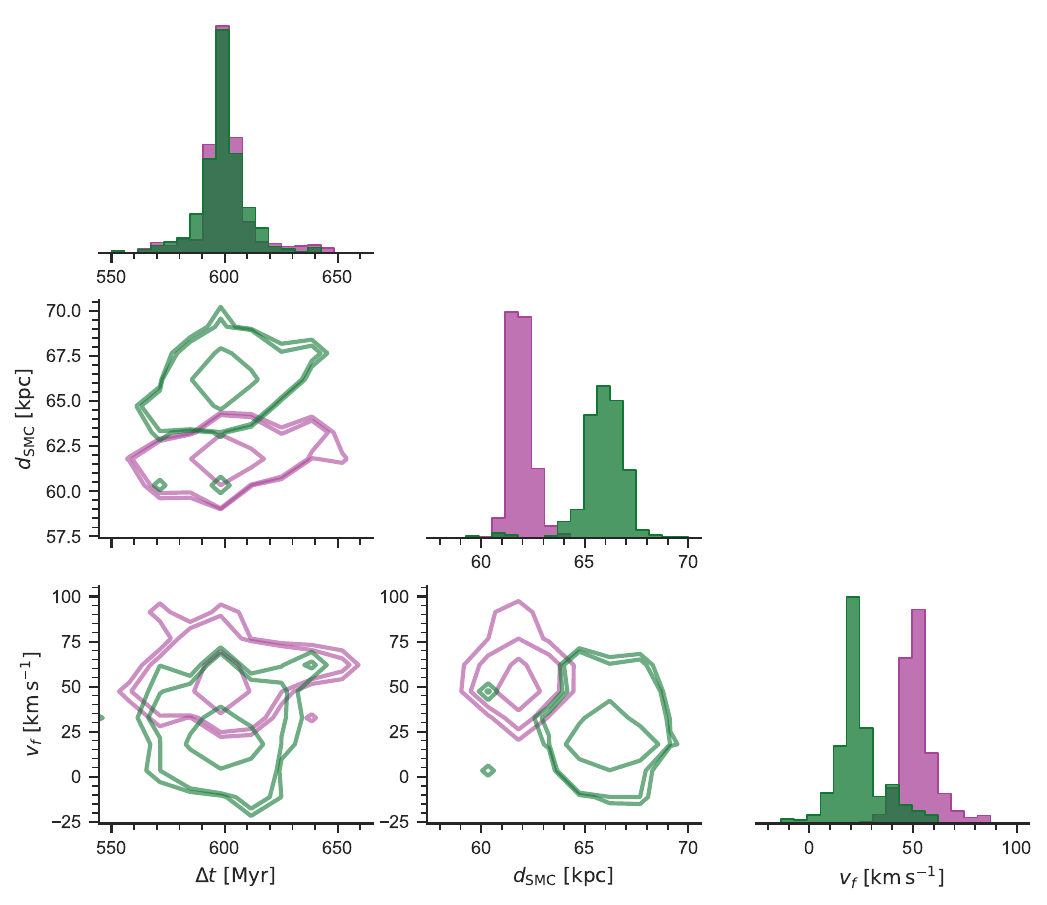}
\caption{Results of the MCMC simple model fit procedure to the front (purple) and behind (green) structures. 2D histograms display the $1$, $2$ and $3\sigma$ limits of each posterior distribution. Histograms display 1D marginalized posterior PDFs of each parameter.
\label{fig:toy_model_details}}
\end{figure*}

Next, we construct a simple prescription for the differential tidal force on the stars due to the LMC. Considering the large uncertainties in the SMC geometry, we only consider a simplistic tidal prescription (i.e., omitting the influence of the Milky Way and angular momentum effects). By integrating the tidal acceleration on the stars over a timescale $\Delta t$, and assuming the distance from the center of the system to the the LMC ($r_{\rm LMC}$) is much larger than the distance from each star to the center of mass ($r$), we can compute the velocity in the LMC direction of each star as a result of tides via:

\begin{equation}
v_{t,\rm LMC}(r) \approx \frac{2 \,G\, M_{\rm LMC} \, r^2}{r_{\rm LMC}^3} \, \Delta t.
\end{equation}

In addition to the tidal velocity, we also include a simple prescription for a rotational component ($v_{\rm rot}$) as a function of the observed coordinates, 

\begin{equation}
v_{\rm rot}(X) = \frac{2}{\pi}\, v_f \arctan \left ( \frac{X}{X_f}\right),
\end{equation}

\noindent following \citet{diteodoro2019} for asymptotic velocity $v_f$ and scale radius $X_f$ (we assume $X_f=1\rm\,kpc$). Of course, this prescription assumes a particular orientation of the rotation in each system, as we do not take inclination or higher-order kinematic effects into account. We emphasize that this is out of scope for this work.

Given this simple model, we can project the resulting velocities into the three observable components, $v_r$, $\mu_W$ and $\mu_N$ based on the values of the three free parameters, $d_{\rm SMC}$, $\Delta t$ and $v_f$ and compare them with the results of Figure~\ref{fig:kinematics}. Because we declined to correct the observations for perspective rotation, we add this effect to the model (i.e., for which we know the distance to each source), using Equation 6 of \citet{vandeven2006}, reproduced below for clarity:

\begin{equation} 
\begin{aligned}
\mu_X^{\rm pr} &= -6.1363\times 10^{-5} \, X \, v_{r,\,\rm sys} / d_{\rm SMC} \\
\mu_Y^{\rm pr} &= -6.1363\times 10^{-5} \, Y \, v_{r,\,\rm sys} / d_{\rm SMC} \\
v_r^{\rm pr} &= 1.3790\times 10^{-3} \, (X \,\mu_{X,\rm sys} +  Y\, \mu_{Y, \rm sys} ) / d_{\rm SMC},
\end{aligned}
\end{equation}

\noindent where $d_{\rm SMC}$ is the line of sight distance to the center of the system in our observed reference frame, and is computed from $r_{\rm LMC}$ via,

\begin{equation}
d_{\rm SMC} = 50\, + \sqrt{r_{\rm LMC}^2 + d_{\rm LMC, \, sky}^2} \rm\, kpc.
\end{equation}

\begin{figure*}[ht!]
\includegraphics[width=\textwidth]{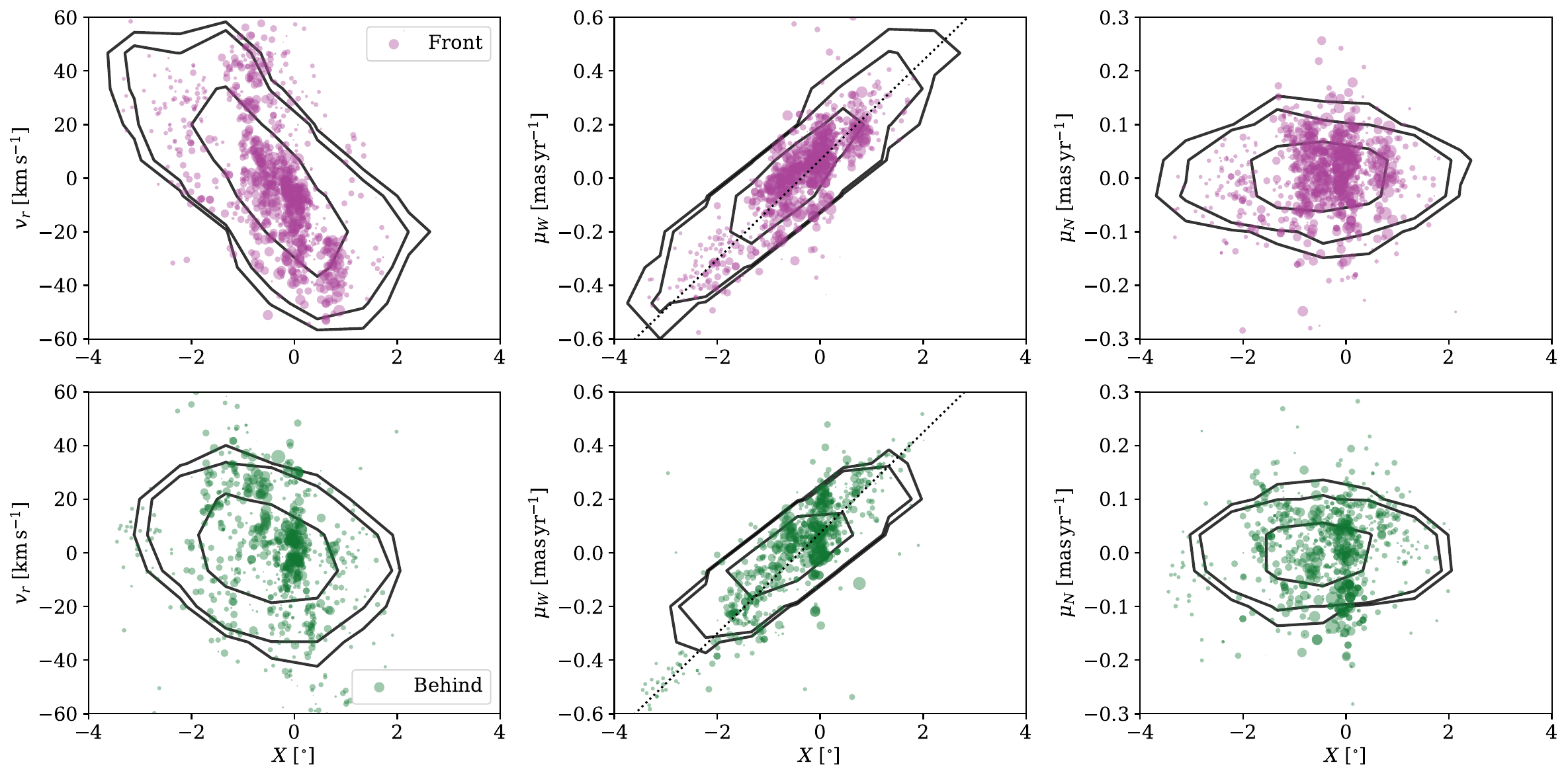}
\caption{The best-fitting toy model results (black contours, denoting $1$, $2$ and $3\sigma$ limits of the model distributions) overlaid on the observed velocity components of the front and behind structures (Figure~\ref{fig:kinematics}).
\label{fig:toy_model}}
\end{figure*}

To perform the fit, we use Markov Chain Monte Carlo sampling implemented in the \texttt{emcee} Python package \citep{foremanmackey2013}. Following Bayes' rule, the posterior probability that our observations $\mathbf{x}$ (where $\mathbf{x}\equiv v_r,\,\mu_W$) are consistent with our model, $m(\mathbf{\theta})$ (where $\mathbf{\theta}\equiv d_{\rm SMC},\,\Delta t,\, v_f)$ is proportional to the product of the likelihood ${\rm P}(\mathbf{x} | \theta)$ and the prior ${\rm P}(\theta)$,

\begin{equation}
    {\rm P}(\mathbf{\theta} | \mathbf{x}) \propto {\rm P}( \mathbf{x} | \mathbf{\theta}) \, \,
    {\rm P}(\mathbf{\theta}).
\end{equation}

\noindent Here we assume the likelihood is roughly Gaussian and independent for each of the observed quantities so that, 

\begin{equation}
{\rm P}( \mathbf{x} | \mathbf{\theta}) = \prod_x \frac{1}{\sqrt{2\pi} \sigma_x} \exp{\left [ - \frac{1}{2} \frac{m(\mathbf{\theta}) - \mathbf{x}}{\sigma_x^2} \right ] }
\end{equation}

\noindent where the subscript $x$ runs between the values of $\mathbf{x}$ (i.e., $v_r$ and $\mu_W$). We assume a flat prior on each of the three free parameters, and allow them to have values in the range of $50<d_{\rm SMC}<80\rm\,kpc$ (used to compute $r_{\rm LMC}$), $\Delta t = 0-1000\rm\,Myr$ and $-200<v_f<200\rm\,km\,s^{-1}$. We initialize the model at the same position for both samples ($d_{\rm smc},\,\Delta t, \,v_f)=(62\rm\,kpc,\,600\rm\,Myr,\,30\rm\,km\,s^{-1})$, and sample the posterior with 1,000 walkers and 10,000 steps per parameter, excluding a burn-in phase of 1,000 steps. 

In Figure~\ref{fig:toy_model_details} we display a ``corner" plot of the 2D distributions of the posterior, as well as marginalized 1D PDFs. We compute the $50^{\rm th}$ percentile of each 1D PDF and compute the uncertainties based on the $16^{\rm th}$ and $84^{\rm th}$ percentiles. These results are summarized in Table~\ref{tab:fit_results}.

\begin{deluxetable*}{c l | cc}
\tablecaption{Simple Model Fit Results \label{tab:fit_results}}
\tablehead{\colhead{Parameter} & \colhead{Description} & \colhead{Front} & \colhead{Behind}}  
\startdata
$d_{\rm SMC}$ [kpc] & Line of sight distance  & $61^ {+2}_{-1}$ & $66^{+2}_{-2}$ \\
$\Delta t$ [Myr] & Time since tidal interaction with LMC began & $600^{+9}_{-8}$ & $600^{+7}_{-6}$ \\
$v_f$ [$\rm km\,s^{-1}$] & Asymptotic rotation velocity & $50^{+9}_{-5}$ & $22^{+12}_{-6}$ \\
\hline
$M_{\rm HI}$ [$M_{\odot}$] & Total \hi\ mass &  $2.6\times10^{8} \left  ( \frac{D}{61\rm \,kpc}  \right )^2$ & $ 2.8\times10^{8} \left  ( \frac{D}{66\rm \,kpc}  \right )^2$ \\
\enddata
\tablecomments{Parameters derived from the MCMC fit of our simple toy model to the observed velocities of RSGs in the front and behind systems. }
\end{deluxetable*}

In Figure~\ref{fig:toy_model}, we overlay the contours from one realization of the toy model. We observe that there is good agreement between the observations and the model contours in all panels, which is encouraging given the simplicity of the model (i.e., we did not take into account effects of the components on each other, the Milky Way, or any external effects beyond tidal acceleration from a simplistic point-mass LMC). The shape of the $\mu_W(X)$ distribution (middle column; Figure~\ref{fig:toy_model}) is constrained by the $d_{\rm SMC}$ and $\Delta t$ parameters, and the shape of the $v_r(X)$ distributions (left column; Figure~\ref{fig:toy_model}) is also constrained by $v_f$. We emphasize again that the structures were modeled independently, without prior information about their order along the line of sight (i.e., flat prior on all model parameters). 

It is highly encouraging that we recover the fact that the front component is closer along the line of sight ($61^{+2}_{-1}\rm\,kpc$ versus $\sim 66^{+2}_{-2}\rm\,kpc$). In addition, our results agree with previous constraints for the distance structure to the young, star-forming components of the SMC. In particular, we agree with the analysis of \citet{yanchulova2017}, who inferred the properties of dust along the line of sight by modeling the detailed structure of the red giant branch and red clump traced by UV-IR photometry of the Southwest Bar from the \emph{HST} SMIDGE survey. They concluded that the dust in that region is localized to a thin layer at $\sim 62\rm\,kpc$ within a broad stellar distribution spread $\sim10\rm\,kpc$ along the line of sight. The dust in this region is spatially coincident with molecular gas traced by CO emission, which we have shown is consistent with being located in the front component. We find consistent results for the distance to the dust in the front component of $\sim 61\rm\,kpc$.

The other two free parameters in the model, $\Delta t$ and $v_f$ are also consistent with expectations. The two components have statistically indistinguishable $\Delta t$, indicating that they have undergone tidal influence of the LMC for the same amount of time, which is reasonable. In the front component, $v_f$ is larger than in the behind component. This agrees in broad strokes with observations of $\mu_{\theta}$, which is more consistently negative in the front component than the behind. We emphasize that $v_f$ is extremely uncertain, as it relies on an assumed rotation axis and structure (i.e., higher-order effects such as inclination are not taken into account). Properly modeling rotation in each system requires a much more sophisticated approach, which we consider to be beyond the scope of this work. We emphasize again that this model was designed to constrain $d_{\rm SMC}$ in particular, which influences all kinematic components (rather than just $v_r$ in the case of $v_f$). 

\subsection{Estimating total \hi\ mass}

Based on the results of this simple model, we estimate the total \hi\ masses of the two structures. The first step is to split the GASKAP-\hi\ cube into the front and behind structures, which requires an estimate for $\langle \Delta A_K \rangle$ across the full survey footprint. To estimate this, we compute the binned average of $\langle \Delta A_K \rangle$ in square bins 10 pixels on a side, and for pixels without a measurement, we use the average of the nearest 10 pixels with a measurement.

We then model all spectra in the SMC \hi\ cube with Gaussian functions using GaussPy+ \citep{lindner2015, riener2019}, an extremely efficient, reproducible method for spectral line decomposition. We use the simplest form of GaussPy+, which leverages the Autonomous Gaussian Decomposition algorithm to identify the positions, widths, and amplitudes of spectral lines based on their numerical derivatives, controlled by only three free parameters: smoothing scales for detecting narrow and broad components ($\alpha_1$ and $\alpha_2$) and the desired signal to noise ratio (S/N). The optimal values for $\alpha_1$ and $\alpha_2$ are chosen by running the algorithm on a training set of known decomposition, and in this case we use the helper functions from GaussPy+ to construct a training set selected from the SMC cube itself, and decompose them using an independent iterative method \citep[for more details see][]{riener2019}. From this process, we find that the optimal decomposition parameters are: $\alpha_1 =1.08$ and $\alpha_2=4.95$ and S/N$=3$. 

\begin{figure*}[ht!]
\includegraphics[width=\textwidth]{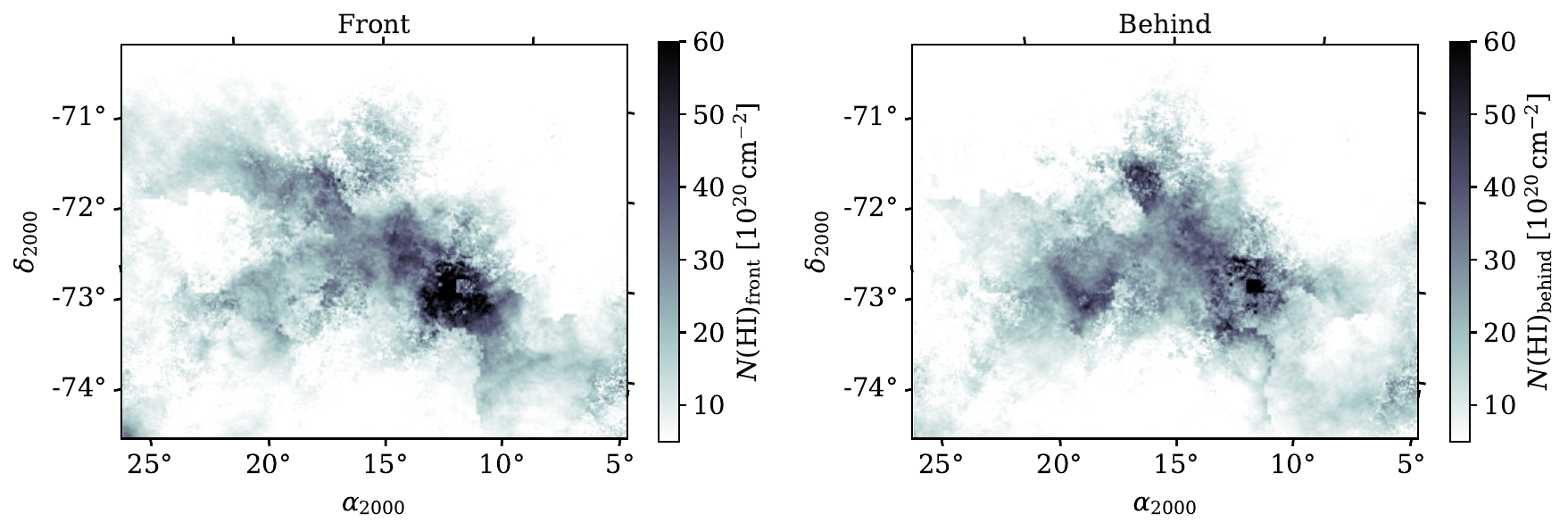}
\caption{\hi\ column density maps of the front (left) and behind (right) components. 
\label{fig:hi_components}}
\end{figure*}

Next, for each pixel, if $\langle \Delta A_K \rangle <0$,  we assign the emission for all fitted Gaussian components with mean velocities less than the first moment to the front component and the rest to the behind component (and vice-versa for $\langle \Delta A_K\rangle >0$). We assume the average distances to the front and behind structures from the simple model, $61\rm\,kpc$ and $66\rm\,kpc$ respectively, and report the \hi\ mass in Table~\ref{tab:fit_results}. 

Overall, the structures appear to have roughly the same \hi\ mass. This is not surprising, given the fact that the amplitudes of the \hi\ features at low and high velocities are similar throughout the system. Our estimates generally agree with previous estimates for the \hi\ mass of the SMC, which range from $\sim3-6\times10^8\rm\,M_{\odot}$ \citep{hindman1967, bajaja1982, stanimirovic1999, diteodoro2019}. 

In Figure~\ref{fig:hi_components} we display \hi\ column density maps, computed in the optically-thin limit, for the front and behind components.

\section{Discussion}
\label{sec:discussion}

Taken together, our results paint a picture in which the SMC is composed of two, distinct, star-forming systems which overlap along the line of sight. 

At face value, this result is consistent with previous conclusions that multi-peaked \hi\ velocity profiles of the SMC arise due to the presence of separate sub-systems at different line of sight distances \citep{kerr1954, johnson1961, mathewson1984}. The key difference is that we have shown that the two systems cannot be simply classified as having low or high velocity, as in fact the order of the velocity components along the line of sight changes as you move across the face of the SMC. 

Another important difference between this work and previous analysis of the line of sight structure of the ISM is that we probe a wide range of SMC environments. For example, optical absorption lines detected towards SMC stars are typically associated with \hi\ at low velocities \citep{danforth2002, welty2012}, prompting the conclusion that gas at low velocity must be in front along the line of sight across the system. However, these studies are limited to handfuls of detections, biased towards the regions which feature the highest densities and highest optical depths. Our method allows us to probe the low-density diffuse medium of the SMC, and the smoothness of $\langle \Delta A_K \rangle$ (Figure~\ref{fig:results}) provides supporting evidence that we are sensitive to real changes in the column density structure in different regions.

In addition to our agreement with the only constraint for the dust geometry of the SMC in the SW Bar \citep{yanchulova2017}, we have general agreement with the picture painted by older stellar populations in the system. Cepheid variables, stars for which precise distances can be estimated, are highly-elongated (up to $\sim20\rm\,kpc$) along the line of sight \citep{scowcroft2016}. Although they do not show the same spatial distribution as inferred for the star-forming ISM in this work (as they are much more elongated and are generally uniformly distributed), there is clear evidence for the presence of distinct sub-structures with similar average positions along the line of sight, i.e., $\sim60$ and $\sim66\rm\,kpc$ \citep{ripepi2017, Jacyszyn-Dobrzeniecka2017}. These substructures are angled across the field of view such that they overlap along the line of sight when viewed in the plane of sight, making it difficult to conduct detailed comparisons with the morphologies of the front and behind components identified here (which also overlap along the line of sight). In addition, the bimodality in distance observed in red clump stars in the East \citep{nidever2013, subramanian2017, tatton2021, omkumar2021, almeida2023} underscores the potential for multiple components to the SMC system. The separation between the red clump structures is larger than what we infer here ($\sim 10\rm\,kpc$), and we consider it beyond the scope of this work to determine if they are directly related (as this will require more detailed numerical modeling of the system than performed here). However, overall, we consider it to be encouraging that there is a wealth of ancillary evidence for significant substructure along the line of sight to the system.

In addition to the fact that the reservoir of cold, dense star-forming material, also displays evidence for significant substructure along the line of sight, we find that both systems identified here are actively forming stars. Taking a conservative estimate for the lifetimes of these stars to be $100\rm\,Myr$ (most are much younger; Figure~\ref{fig:ages}), and a rough estimate for the relative speed of the stars with respect to the gas of $\sim 10-100\rm\,km\,s^{-1}$, the stars will travel $\sim 0.1-1\rm\,kpc$ in their lifetimes. This is only $\lesssim20\%$ of the inferred separation of the two systems along the line of sight ($\sim 5\rm\,kpc$), which suggests the RSGs likely formed within the system they are located in today. 

One possible explanation for our results is that the two systems are remnants of distinct galaxies. The difference in observed stellar metallicities, as well as the distinct difference between the molecular gas features in the two structures, suggests that they may have evolved in different ways. Although the chemical enrichment history appears similar across the SMC \citep{almeida2023}, there is evidence for ``non-uniform" enrichment in the outskirts of the system \citep{mucciarelli2023}.

Alternatively, the front structure is the main galaxy \citep[the ``SMC Remnant";][]{mathewson1984} and the behind structure is tidal debris resulting from the interaction with the LMC. The complex nature of the debris field surrounding the SMC, including a bifurcated bridge towards the LMC \citep{muller2004}, and the bifurcated, trailing Stream \citep{putman1998, nidever2010}, support this picture. Furthermore, although in this work we have made the choice to focus on the properties of the two dominant \hi\ velocity features, it is clear that in some regions of the SMC there are more than two components. Although these additional components tend to have significantly smaller column density than the two main peaks, their presence highlights the possibility that additional subsystems may be identified. 

One particular hypothesis of note is that the behind component is a ``counter-bridge" of material pulled from the SMC as a result of its interactions with the LMC \citep{toomre1972, besla2012, diaz2012}. According to numerical simulations of the MS, this structure is expected to sit directly behind the main body of the SMC, and if pulled from the outskirts of the system, may explain its lower observed metallicity. The caveat to this interpretation is that the two systems have roughly equal \hi\ mass, which is not expected in the counter-bridge model. In addition, the spatial distribution of \hi\ in the system indicates that the ``loop" in the North East, posited to be part of the counter bridge \citep{muller2007} is actually at the near SMC distance. However, the 3D positions and kinematics of clusters in the SMC periphery have consistent distances with the front and behind components found here, as well as being consistent with the numerical predictions of the counter bridge \citep{dias_2022}.

Ultimately, reconciling these results within the broad landscape of observational constraints for the structure of the SMC requires further work. This includes direct distance constraints to the ISM components, as well as numerical studies of the structure of the SMC in the context of the history of the full MS. However, the implications for our understanding of star formation in low-metallicity dwarf galaxies are already significant. The SMC is used as a template for the evolution of the dust-to-gas ratio with metallicity \citep[e.g., see ][]{clark2023}. Accurately associating the diffuse, atomic gas to the metal-rich ISM is crucial for inferring this parameter, which underlies our understanding of the chemical enrichment of galaxies.

\section{Summary}
\label{sec:concl}

In this study we investigate the 3D structure of the star-forming ISM in the SMC. We compile a sample of young, massive stars from \emph{Gaia} DR3 and APOGEE whose kinematics are consistent with distinct \hi\ velocity components observed by the high-resolution GASKAP-\hi\ survey. By comparing the line of sight extinction to these stars across the SMC, we find that the order of the low and high \hi\ velocity components changes between SMC environments. We separate the stellar sample into sub-systems, one in front along the line of sight and one behind. We find that these systems have distinct metallicity distributions, and there is evidence that the molecular gas in the SMC resides mostly in the front component. We construct a simple toy model to explain the  kinematics of the two components undergoing the tidal influence of the LMC as well as rotation. We find that the two components lie $5\rm\,kpc$ from each other along the line of sight (average distances of $61$ and $66\rm\,kpc$ respectively). These results are broadly consistent with previous estimates of the line of sight structure of the SMC from diverse tracers.

\section{acknowledgements}
The authors acknowledge the anonymous referee, whose comments improved the clarity of the manuscript. C.E.M. acknowledges insightful conversations with Christopher Clark, Karl Gordon, Peter Scicluna, Lara Cullinane and David French during the development of this project. The authors acknowledge Interstellar Institute’s program ``II6” and the Paris-Saclay University’s Institut Pascal for hosting discussions that nourished the development of the ideas behind this work. J.E.M.C acknowledges STFC studentship. E.D.T. was supported by the European Research Council (ERC) under grant agreement no. 101040751. C.F.~acknowledges funding provided by the Australian Research Council (Discovery Project DP230102280), and the Australia-Germany Joint Research Cooperation Scheme (UA-DAAD).

The Australian SKA Pathfinder is part of the Australia Telescope National
Facility which is managed by CSIRO. Operation of ASKAP is funded by
the Australian Government with support from the National Collaborative
Research Infrastructure Strategy. ASKAP uses the resources of the Pawsey
Supercomputing Centre. Establishment of ASKAP, the Murchison Radioastronomy Observatory and the Pawsey Supercomputing Centre are initiatives of the Australian Government, with support from the Government of Western Australia and the Science and Industry Endowment Fund. We acknowledge the Wajarri Yamatji people as the traditional owners of the Observatory site. 

This work has made use of data from the European Space Agency (ESA) mission {\it Gaia} (\url{https://www.cosmos.esa.int/gaia}), processed by the {\it Gaia} Data Processing and Analysis Consortium (DPAC,
\url{https://www.cosmos.esa.int/web/gaia/dpac/consortium}). Funding for the DPAC has been provided by national institutions, in particular the institutions participating in the {\it Gaia} Multilateral Agreement.

Funding for the Sloan Digital Sky Survey IV has been provided by the Alfred P. Sloan Foundation, the U.S. Department of Energy Office of Science, and the Participating Institutions. SDSS acknowledges support and resources from the Center for High-Performance Computing at the University of Utah. The SDSS web site is www.sdss4.org.

SDSS is managed by the Astrophysical Research Consortium for the Participating Institutions of the SDSS Collaboration including the Brazilian Participation Group, the Carnegie Institution for Science, Carnegie Mellon University, Center for Astrophysics | Harvard \& Smithsonian (CfA), the Chilean Participation Group, the French Participation Group, Instituto de Astrofísica de Canarias, The Johns Hopkins University, Kavli Institute for the Physics and Mathematics of the Universe (IPMU) / University of Tokyo, the Korean Participation Group, Lawrence Berkeley National Laboratory, Leibniz Institut für Astrophysik Potsdam (AIP), Max-Planck-Institut für Astronomie (MPIA Heidelberg), Max-Planck-Institut für Astrophysik (MPA Garching), Max-Planck-Institut für Extraterrestrische Physik (MPE), National Astronomical Observatories of China, New Mexico State University, New York University, University of Notre Dame, Observatório Nacional / MCTI, The Ohio State University, Pennsylvania State University, Shanghai Astronomical Observatory, United Kingdom Participation Group, Universidad Nacional Autónoma de México, University of Arizona, University of Colorado Boulder, University of Oxford, University of Portsmouth, University of Utah, University of Virginia, University of Washington, University of Wisconsin, Vanderbilt University, and Yale University.

\facilities{ASKAP, Gaia, SDSS}

\software{AstroPy \citep{astropy2013, astropy2018}, Matplotlib  \citep{Hunter:2007}, NumPy \citep{harris2020array}, SciPy  \citep{Virtanen_2020} , glue \citep{glue2015} (DOI 10.5281/zenodo.774845). }

\bibliography{ms}{}

\begin{thebibliography}{}
\expandafter\ifx\csname natexlab\endcsname\relax\def\natexlab#1{#1}\fi
\providecommand{\url}[1]{\href{#1}{#1}}
\providecommand{\dodoi}[1]{doi:~\href{http://doi.org/#1}{\nolinkurl{#1}}}
\providecommand{\doeprint}[1]{\href{http://ascl.net/#1}{\nolinkurl{http://ascl.net/#1}}}
\providecommand{\doarXiv}[1]{\href{https://arxiv.org/abs/#1}{\nolinkurl{https://arxiv.org/abs/#1}}}

\bibitem[{{Abdurro'uf} {et~al.}(2022){Abdurro'uf}, {Accetta}, {Aerts}, {Silva
  Aguirre}, {Ahumada}, {Ajgaonkar}, {Filiz Ak}, {Alam}, {Allende Prieto},
  {Almeida}, {Anders}, {Anderson}, {Andrews}, {Anguiano}, {Aquino-Ort{\'\i}z},
  {Arag{\'o}n-Salamanca}, {Argudo-Fern{\'a}ndez}, {Ata}, {Aubert},
  {Avila-Reese}, {Badenes}, {Barb{\'a}}, {Barger}, {Barrera-Ballesteros},
  {Beaton}, {Beers}, {Belfiore}, {Bender}, {Bernardi}, {Bershady}, {Beutler},
  {Bidin}, {Bird}, {Bizyaev}, {Blanc}, {Blanton}, {Boardman}, {Bolton},
  {Boquien}, {Borissova}, {Bovy}, {Brandt}, {Brown}, {Brownstein}, {Brusa},
  {Buchner}, {Bundy}, {Burchett}, {Bureau}, {Burgasser}, {Cabang}, {Campbell},
  {Cappellari}, {Carlberg}, {Wanderley}, {Carrera}, {Cash}, {Chen}, {Chen},
  {Cherinka}, {Chiappini}, {Choi}, {Chojnowski}, {Chung}, {Clerc}, {Cohen},
  {Comerford}, {Comparat}, {da Costa}, {Covey}, {Crane}, {Cruz-Gonzalez},
  {Culhane}, {Cunha}, {Dai}, {Damke}, {Darling}, {Davidson}, {Davies},
  {Dawson}, {De Lee}, {Diamond-Stanic}, {Cano-D{\'\i}az}, {S{\'a}nchez},
  {Donor}, {Duckworth}, {Dwelly}, {Eisenstein}, {Elsworth}, {Emsellem},
  {Eracleous}, {Escoffier}, {Fan}, {Farr}, {Feng}, {Fern{\'a}ndez-Trincado},
  {Feuillet}, {Filipp}, {Fillingham}, {Frinchaboy}, {Fromenteau}, {Galbany},
  {Garc{\'\i}a}, {Garc{\'\i}a-Hern{\'a}ndez}, {Ge}, {Geisler}, {Gelfand},
  {G{\'e}ron}, {Gibson}, {Goddy}, {Godoy-Rivera}, {Grabowski}, {Green},
  {Greener}, {Grier}, {Griffith}, {Guo}, {Guy}, {Hadjara}, {Harding},
  {Hasselquist}, {Hayes}, {Hearty}, {Hern{\'a}ndez}, {Hill}, {Hogg},
  {Holtzman}, {Horta}, {Hsieh}, {Hsu}, {Hsu}, {Huber}, {Huertas-Company},
  {Hutchinson}, {Hwang}, {Ibarra-Medel}, {Chitham}, {Ilha}, {Imig}, {Jaekle},
  {Jayasinghe}, {Ji}, {Johnson}, {Jones}, {J{\"o}nsson}, {Katkov}, {Khalatyan},
  {Kinemuchi}, {Kisku}, {Knapen}, {Kneib}, {Kollmeier}, {Kong}, {Kounkel},
  {Kreckel}, {Krishnarao}, {Lacerna}, {Lane}, {Langgin}, {Lavender}, {Law},
  {Lazarz}, {Leung}, {Leung}, {Lewis}, {Li}, {Li}, {Lian}, {Liang}, {Lin},
  {Lin}, {Lin}, {Lintott}, {Long}, {Longa-Pe{\~n}a}, {L{\'o}pez-Cob{\'a}},
  {Lu}, {Lundgren}, {Luo}, {Mackereth}, {de la Macorra}, {Mahadevan},
  {Majewski}, {Manchado}, {Mandeville}, {Maraston}, {Margalef-Bentabol},
  {Masseron}, {Masters}, {Mathur}, {McDermid}, {Mckay}, {Merloni},
  {Merrifield}, {Meszaros}, {Miglio}, {Di Mille}, {Minniti}, {Minsley},
  {Monachesi}, {Moon}, {Mosser}, {Mulchaey}, {Muna}, {Mu{\~n}oz}, {Myers},
  {Myers}, {Nadathur}, {Nair}, {Nandra}, {Neumann}, {Newman}, {Nidever},
  {Nikakhtar}, {Nitschelm}, {O'Connell}, {Garma-Oehmichen}, {Luan Souza de
  Oliveira}, {Olney}, {Oravetz}, {Ortigoza-Urdaneta}, {Osorio}, {Otter},
  {Pace}, {Padilla}, {Pan}, {Pan}, {Parikh}, {Parker}, {Peirani}, {Pe{\~n}a
  Ram{\'\i}rez}, {Penny}, {Percival}, {Perez-Fournon}, {Pinsonneault},
  {Poidevin}, {Poovelil}, {Price-Whelan}, {B{\'a}rbara de Andrade Queiroz},
  {Raddick}, {Ray}, {Rembold}, {Riddle}, {Riffel}, {Riffel}, {Rix}, {Robin},
  {Rodr{\'\i}guez-Puebla}, {Roman-Lopes}, {Rom{\'a}n-Z{\'u}{\~n}iga}, {Rose},
  {Ross}, {Rossi}, {Rubin}, {Salvato}, {S{\'a}nchez}, {S{\'a}nchez-Gallego},
  {Sanderson}, {Santana Rojas}, {Sarceno}, {Sarmiento}, {Sayres}, {Sazonova},
  {Schaefer}, {Schiavon}, {Schlegel}, {Schneider}, {Schultheis}, {Schwope},
  {Serenelli}, {Serna}, {Shao}, {Shapiro}, {Sharma}, {Shen}, {Shetrone}, {Shu},
  {Simon}, {Skrutskie}, {Smethurst}, {Smith}, {Sobeck}, {Spoo}, {Sprague},
  {Stark}, {Stassun}, {Steinmetz}, {Stello}, {Stone-Martinez},
  {Storchi-Bergmann}, {Stringfellow}, {Stutz}, {Su}, {Taghizadeh-Popp},
  {Talbot}, {Tayar}, {Telles}, {Teske}, {Thakar}, {Theissen}, {Tkachenko},
  {Thomas}, {Tojeiro}, {Hernandez Toledo}, {Troup}, {Trump}, {Trussler},
  {Turner}, {Tuttle}, {Unda-Sanzana}, {V{\'a}zquez-Mata}, {Valentini},
  {Valenzuela}, {Vargas-Gonz{\'a}lez}, {Vargas-Maga{\~n}a}, {Alfaro},
  {Villanova}, {Vincenzo}, {Wake}, {Warfield}, {Washington}, {Weaver},
  {Weijmans}, {Weinberg}, {Weiss}, {Westfall}, {Wild}, {Wilde}, {Wilson},
  {Wilson}, {Wilson}, {Wolf}, {Wood-Vasey}, {Yan}, {Zamora}, {Zasowski},
  {Zhang}, {Zhao}, {Zheng}, {Zheng}, \& {Zhu}}]{abdurrouf2022}
{Abdurro'uf}, {Accetta}, K., {Aerts}, C., {et~al.} 2022, \apjs, 259, 35,
  \dodoi{10.3847/1538-4365/ac4414}

\bibitem[{{Accurso} {et~al.}(2017){Accurso}, {Saintonge}, {Catinella},
  {Cortese}, {Dav{\'e}}, {Dunsheath}, {Genzel}, {Gracia-Carpio}, {Heckman},
  {Jimmy}, {Kramer}, {Li}, {Lutz}, {Schiminovich}, {Schuster}, {Sternberg},
  {Sturm}, {Tacconi}, {Tran}, \& {Wang}}]{accurso2017}
{Accurso}, G., {Saintonge}, A., {Catinella}, B., {et~al.} 2017, \mnras, 470,
  4750, \dodoi{10.1093/mnras/stx1556}

\bibitem[{{Allende Prieto} {et~al.}(2006){Allende Prieto}, Beers, Wilhelm,
  Newberg, Rockosi, Yanny, \& Lee}]{AllendePrieto2006}
{Allende Prieto}, C., Beers, T.~C., Wilhelm, R., {et~al.} 2006, AJ, 636, 804,
  \dodoi{10.1086/498131}

\bibitem[{{Almeida} {et~al.}(2023){Almeida}, {Majewski}, {Nidever}, {Olsen},
  {Monachesi}, {Kallivayalil}, {Hasselquist}, {Choi}, {Povick}, {Wilson},
  {Geisler}, {Lane}, {Nitschelm}, {Sobeck}, \& {Stringfellow}}]{almeida2023}
{Almeida}, A., {Majewski}, S.~R., {Nidever}, D.~L., {et~al.} 2023, arXiv
  e-prints, arXiv:2308.13631, \dodoi{10.48550/arXiv.2308.13631}

\bibitem[{{Astropy Collaboration} {et~al.}(2013){Astropy Collaboration},
  {Robitaille}, {Tollerud}, {Greenfield}, {Droettboom}, {Bray}, {Aldcroft},
  {Davis}, {Ginsburg}, {Price-Whelan}, {Kerzendorf}, {Conley}, {Crighton},
  {Barbary}, {Muna}, {Ferguson}, {Grollier}, {Parikh}, {Nair}, {Unther},
  {Deil}, {Woillez}, {Conseil}, {Kramer}, {Turner}, {Singer}, {Fox}, {Weaver},
  {Zabalza}, {Edwards}, {Azalee Bostroem}, {Burke}, {Casey}, {Crawford},
  {Dencheva}, {Ely}, {Jenness}, {Labrie}, {Lim}, {Pierfederici}, {Pontzen},
  {Ptak}, {Refsdal}, {Servillat}, \& {Streicher}}]{astropy2013}
{Astropy Collaboration}, {Robitaille}, T.~P., {Tollerud}, E.~J., {et~al.} 2013,
  \aap, 558, A33, \dodoi{10.1051/0004-6361/201322068}

\bibitem[{{Astropy Collaboration} {et~al.}(2018){Astropy Collaboration},
  {Price-Whelan}, {Sip{\H{o}}cz}, {G{\"u}nther}, {Lim}, {Crawford}, {Conseil},
  {Shupe}, {Craig}, {Dencheva}, {Ginsburg}, {VanderPlas}, {Bradley},
  {P{\'e}rez-Su{\'a}rez}, {de Val-Borro}, {Aldcroft}, {Cruz}, {Robitaille},
  {Tollerud}, {Ardelean}, {Babej}, {Bach}, {Bachetti}, {Bakanov}, {Bamford},
  {Barentsen}, {Barmby}, {Baumbach}, {Berry}, {Biscani}, {Boquien}, {Bostroem},
  {Bouma}, {Brammer}, {Bray}, {Breytenbach}, {Buddelmeijer}, {Burke},
  {Calderone}, {Cano Rodr{\'\i}guez}, {Cara}, {Cardoso}, {Cheedella}, {Copin},
  {Corrales}, {Crichton}, {D'Avella}, {Deil}, {Depagne}, {Dietrich}, {Donath},
  {Droettboom}, {Earl}, {Erben}, {Fabbro}, {Ferreira}, {Finethy}, {Fox},
  {Garrison}, {Gibbons}, {Goldstein}, {Gommers}, {Greco}, {Greenfield},
  {Groener}, {Grollier}, {Hagen}, {Hirst}, {Homeier}, {Horton}, {Hosseinzadeh},
  {Hu}, {Hunkeler}, {Ivezi{\'c}}, {Jain}, {Jenness}, {Kanarek}, {Kendrew},
  {Kern}, {Kerzendorf}, {Khvalko}, {King}, {Kirkby}, {Kulkarni}, {Kumar},
  {Lee}, {Lenz}, {Littlefair}, {Ma}, {Macleod}, {Mastropietro}, {McCully},
  {Montagnac}, {Morris}, {Mueller}, {Mumford}, {Muna}, {Murphy}, {Nelson},
  {Nguyen}, {Ninan}, {N{\"o}the}, {Ogaz}, {Oh}, {Parejko}, {Parley}, {Pascual},
  {Patil}, {Patil}, {Plunkett}, {Prochaska}, {Rastogi}, {Reddy Janga},
  {Sabater}, {Sakurikar}, {Seifert}, {Sherbert}, {Sherwood-Taylor}, {Shih},
  {Sick}, {Silbiger}, {Singanamalla}, {Singer}, {Sladen}, {Sooley},
  {Sornarajah}, {Streicher}, {Teuben}, {Thomas}, {Tremblay}, {Turner},
  {Terr{\'o}n}, {van Kerkwijk}, {de la Vega}, {Watkins}, {Weaver}, {Whitmore},
  {Woillez}, {Zabalza}, \& {Astropy Contributors}}]{astropy2018}
{Astropy Collaboration}, {Price-Whelan}, A.~M., {Sip{\H{o}}cz}, B.~M., {et~al.}
  2018, \aj, 156, 123, \dodoi{10.3847/1538-3881/aabc4f}

\bibitem[{{Babusiaux} {et~al.}(2023){Babusiaux}, {Fabricius}, {Khanna},
  {Muraveva}, {Reyl{\'e}}, {Spoto}, {Vallenari}, {Luri}, {Arenou},
  {{\'A}lvarez}, {Anders}, {Antoja}, {Balbinot}, {Barache}, {Bauchet},
  {Bossini}, {Busonero}, {Cantat-Gaudin}, {Carrasco}, {Dafonte}, {Diakit{\'e}},
  {Figueras}, {Garcia-Gutierrez}, {Garofalo}, {Helmi}, {Jim{\'e}nez-Arranz},
  {Jordi}, {Kervella}, {Kostrzewa-Rutkowska}, {Leclerc}, {Licata}, {Manteiga},
  {Masip}, {Mongui{\'o}}, {Ramos}, {Robichon}, {Robin}, {Romero-G{\'o}mez},
  {S{\'a}ez}, {Santove{\~n}a}, {Spina}, {Torralba Elipe}, \&
  {Weiler}}]{babusiaux2023}
{Babusiaux}, C., {Fabricius}, C., {Khanna}, S., {et~al.} 2023, \aap, 674, A32,
  \dodoi{10.1051/0004-6361/202243790}

\bibitem[{{Bajaja} \& {Loiseau}(1982)}]{bajaja1982}
{Bajaja}, E., \& {Loiseau}, N. 1982, \aaps, 48, 71

\bibitem[{{Beaumont} {et~al.}(2015){Beaumont}, {Goodman}, \&
  {Greenfield}}]{glue2015}
{Beaumont}, C., {Goodman}, A., \& {Greenfield}, P. 2015, in Astronomical
  Society of the Pacific Conference Series, Vol. 495, Astronomical Data
  Analysis Software an Systems XXIV (ADASS XXIV), ed. A.~R. {Taylor} \&
  E.~{Rosolowsky}, 101

\bibitem[{{Belokurov} {et~al.}(2016){Belokurov}, {Erkal}, {Deason}, {Koposov},
  {De Angeli}, {Wyn Evans}, {Fraternali}, \& {Mackey}}]{belokurov2016}
{Belokurov}, V., {Erkal}, D., {Deason}, A.~J., {et~al.} 2016, \mnras,
  \dodoi{10.1093/mnras/stw3357}

\bibitem[{{Besla} {et~al.}(2007){Besla}, {Kallivayalil}, {Hernquist},
  {Robertson}, {Cox}, {van der Marel}, \& {Alcock}}]{besla2007}
{Besla}, G., {Kallivayalil}, N., {Hernquist}, L., {et~al.} 2007, \apj, 668,
  949, \dodoi{10.1086/521385}

\bibitem[{{Besla} {et~al.}(2012){Besla}, {Kallivayalil}, {Hernquist}, {van der
  Marel}, {Cox}, \& {Kere{\v s}}}]{besla2012}
---. 2012, \mnras, 421, 2109, \dodoi{10.1111/j.1365-2966.2012.20466.x}

\bibitem[{{Besla} {et~al.}(2016){Besla}, {Mart{\'\i}nez-Delgado}, {van der
  Marel}, {Beletsky}, {Seibert}, {Schlafly}, {Grebel}, \& {Neyer}}]{besla2016}
{Besla}, G., {Mart{\'\i}nez-Delgado}, D., {van der Marel}, R.~P., {et~al.}
  2016, \apj, 825, 20, \dodoi{10.3847/0004-637X/825/1/20}

\bibitem[{{Blanton} {et~al.}(2017){Blanton}, {Bershady}, {Abolfathi},
  {Albareti}, {Allende Prieto}, {Almeida}, {Alonso-Garc{\'\i}a}, {Anders},
  {Anderson}, {Andrews}, {Aquino-Ort{\'\i}z}, {Arag{\'o}n-Salamanca},
  {Argudo-Fern{\'a}ndez}, {Armengaud}, {Aubourg}, {Avila-Reese}, {Badenes},
  {Bailey}, {Barger}, {Barrera-Ballesteros}, {Bartosz}, {Bates}, {Baumgarten},
  {Bautista}, {Beaton}, {Beers}, {Belfiore}, {Bender}, {Berlind}, {Bernardi},
  {Beutler}, {Bird}, {Bizyaev}, {Blanc}, {Blomqvist}, {Bolton}, {Boquien},
  {Borissova}, {van den Bosch}, {Bovy}, {Brandt}, {Brinkmann}, {Brownstein},
  {Bundy}, {Burgasser}, {Burtin}, {Busca}, {Cappellari}, {Delgado Carigi},
  {Carlberg}, {Carnero Rosell}, {Carrera}, {Chanover}, {Cherinka}, {Cheung},
  {G{\'o}mez Maqueo Chew}, {Chiappini}, {Choi}, {Chojnowski}, {Chuang},
  {Chung}, {Cirolini}, {Clerc}, {Cohen}, {Comparat}, {da Costa}, {Cousinou},
  {Covey}, {Crane}, {Croft}, {Cruz-Gonzalez}, {Garrido Cuadra}, {Cunha},
  {Damke}, {Darling}, {Davies}, {Dawson}, {de la Macorra}, {Dell'Agli}, {De
  Lee}, {Delubac}, {Di Mille}, {Diamond-Stanic}, {Cano-D{\'\i}az}, {Donor},
  {Downes}, {Drory}, {du Mas des Bourboux}, {Duckworth}, {Dwelly}, {Dyer},
  {Ebelke}, {Eigenbrot}, {Eisenstein}, {Emsellem}, {Eracleous}, {Escoffier},
  {Evans}, {Fan}, {Fern{\'a}ndez-Alvar}, {Fernandez-Trincado}, {Feuillet},
  {Finoguenov}, {Fleming}, {Font-Ribera}, {Fredrickson}, {Freischlad},
  {Frinchaboy}, {Fuentes}, {Galbany}, {Garcia-Dias},
  {Garc{\'\i}a-Hern{\'a}ndez}, {Gaulme}, {Geisler}, {Gelfand},
  {Gil-Mar{\'\i}n}, {Gillespie}, {Goddard}, {Gonzalez-Perez}, {Grabowski},
  {Green}, {Grier}, {Gunn}, {Guo}, {Guy}, {Hagen}, {Hahn}, {Hall}, {Harding},
  {Hasselquist}, {Hawley}, {Hearty}, {Gonzalez Hern{\'a}ndez}, {Ho}, {Hogg},
  {Holley-Bockelmann}, {Holtzman}, {Holzer}, {Huehnerhoff}, {Hutchinson},
  {Hwang}, {Ibarra-Medel}, {da Silva Ilha}, {Ivans}, {Ivory}, {Jackson},
  {Jensen}, {Johnson}, {Jones}, {J{\"o}nsson}, {Jullo}, {Kamble}, {Kinemuchi},
  {Kirkby}, {Kitaura}, {Klaene}, {Knapp}, {Kneib}, {Kollmeier}, {Lacerna},
  {Lane}, {Lang}, {Law}, {Lazarz}, {Lee}, {Le Goff}, {Liang}, {Li}, {Li},
  {Lian}, {Lima}, {Lin}, {Lin}, {Bertran de Lis}, {Liu}, {de Icaza Lizaola},
  {Long}, {Lucatello}, {Lundgren}, {MacDonald}, {Deconto Machado}, {MacLeod},
  {Mahadevan}, {Geimba Maia}, {Maiolino}, {Majewski}, {Malanushenko},
  {Malanushenko}, {Manchado}, {Mao}, {Maraston}, {Marques-Chaves}, {Masseron},
  {Masters}, {McBride}, {McDermid}, {McGrath}, {McGreer}, {Medina Pe{\~n}a},
  {Melendez}, {Merloni}, {Merrifield}, {Meszaros}, {Meza}, {Minchev},
  {Minniti}, {Miyaji}, {More}, {Mulchaey}, {M{\"u}ller-S{\'a}nchez}, {Muna},
  {Munoz}, {Myers}, {Nair}, {Nandra}, {Correa do Nascimento}, {Negrete},
  {Ness}, {Newman}, {Nichol}, {Nidever}, {Nitschelm}, {Ntelis}, {O'Connell},
  {Oelkers}, {Oravetz}, {Oravetz}, {Pace}, {Padilla}, {Palanque-Delabrouille},
  {Alonso Palicio}, {Pan}, {Parejko}, {Parikh}, {P{\^a}ris}, {Park}, {Patten},
  {Peirani}, {Pellejero-Ibanez}, {Penny}, {Percival}, {Perez-Fournon},
  {Petitjean}, {Pieri}, {Pinsonneault}, {Pisani}, {Poleski}, {Prada},
  {Prakash}, {Queiroz}, {Raddick}, {Raichoor}, {Barboza Rembold}, {Richstein},
  {Riffel}, {Riffel}, {Rix}, {Robin}, {Rockosi}, {Rodr{\'\i}guez-Torres},
  {Roman-Lopes}, {Rom{\'a}n-Z{\'u}{\~n}iga}, {Rosado}, {Ross}, {Rossi}, {Ruan},
  {Ruggeri}, {Rykoff}, {Salazar-Albornoz}, {Salvato}, {S{\'a}nchez}, {Aguado},
  {S{\'a}nchez-Gallego}, {Santana}, {Santiago}, {Sayres}, {Schiavon}, {da Silva
  Schimoia}, {Schlafly}, {Schlegel}, {Schneider}, {Schultheis}, {Schuster},
  {Schwope}, {Seo}, {Shao}, {Shen}, {Shetrone}, {Shull}, {Simon}, {Skinner},
  {Skrutskie}, {Slosar}, {Smith}, {Sobeck}, {Sobreira}, {Somers}, {Souto},
  {Stark}, {Stassun}, {Stauffer}, {Steinmetz}, {Storchi-Bergmann},
  {Streblyanska}, {Stringfellow}, {Su{\'a}rez}, {Sun}, {Suzuki}, {Szigeti},
  {Taghizadeh-Popp}, {Tang}, {Tao}, {Tayar}, {Tembe}, {Teske}, {Thakar},
  {Thomas}, {Thompson}, {Tinker}, {Tissera}, {Tojeiro}, {Hernandez Toledo}, {de
  la Torre}, {Tremonti}, {Troup}, {Valenzuela}, {Martinez Valpuesta},
  {Vargas-Gonz{\'a}lez}, {Vargas-Maga{\~n}a}, {Vazquez}, {Villanova}, {Vivek},
  {Vogt}, {Wake}, {Walterbos}, {Wang}, {Weaver}, {Weijmans}, {Weinberg},
  {Westfall}, {Whelan}, {Wild}, {Wilson}, {Wood-Vasey}, {Wylezalek}, {Xiao},
  {Yan}, {Yang}, {Ybarra}, {Y{\`e}che}, {Zakamska}, {Zamora}, {Zarrouk},
  {Zasowski}, {Zhang}, {Zhao}, {Zheng}, {Zheng}, {Zhou}, {Zhou}, {Zhu},
  {Zoccali}, \& {Zou}}]{blanton2017}
{Blanton}, M.~R., {Bershady}, M.~A., {Abolfathi}, B., {et~al.} 2017, \aj, 154,
  28, \dodoi{10.3847/1538-3881/aa7567}

\bibitem[{Bowen \& Vaughan(1973)}]{bowen1973}
Bowen, I.~S., \& Vaughan, A.~H. 1973, Appl. Opt., 12, 1430,
  \dodoi{10.1364/AO.12.001430}

\bibitem[{{Bressan} {et~al.}(2012){Bressan}, {Marigo}, {Girardi}, {Salasnich},
  {Dal Cero}, {Rubele}, \& {Nanni}}]{bressan2012}
{Bressan}, A., {Marigo}, P., {Girardi}, L., {et~al.} 2012, \mnras, 427, 127,
  \dodoi{10.1111/j.1365-2966.2012.21948.x}

\bibitem[{{Cappellari} \& {Copin}(2003)}]{cappellari2003}
{Cappellari}, M., \& {Copin}, Y. 2003, \mnras, 342, 345,
  \dodoi{10.1046/j.1365-8711.2003.06541.x}

\bibitem[{{Carrera} {et~al.}(2017){Carrera}, {Conn}, {No{\"e}l}, {Read}, \&
  {L{\'o}pez S{\'a}nchez}}]{carrera2017}
{Carrera}, R., {Conn}, B.~C., {No{\"e}l}, N. E.~D., {Read}, J.~I., \&
  {L{\'o}pez S{\'a}nchez}, {\'A}.~R. 2017, \mnras, 471, 4571,
  \dodoi{10.1093/mnras/stx1932}

\bibitem[{{Chen} {et~al.}(2015){Chen}, {Bressan}, {Girardi}, {Marigo}, {Kong},
  \& {Lanza}}]{chen2015}
{Chen}, Y., {Bressan}, A., {Girardi}, L., {et~al.} 2015, \mnras, 452, 1068,
  \dodoi{10.1093/mnras/stv1281}

\bibitem[{{Chen} {et~al.}(2014){Chen}, {Girardi}, {Bressan}, {Marigo},
  {Barbieri}, \& {Kong}}]{chen2014}
{Chen}, Y., {Girardi}, L., {Bressan}, A., {et~al.} 2014, \mnras, 444, 2525,
  \dodoi{10.1093/mnras/stu1605}

\bibitem[{{Choi} {et~al.}(2022){Choi}, {Olsen}, {Besla}, {van der Marel},
  {Zivick}, {Kallivayalil}, \& {Nidever}}]{choi2022}
{Choi}, Y., {Olsen}, K. A.~G., {Besla}, G., {et~al.} 2022, \apj, 927, 153,
  \dodoi{10.3847/1538-4357/ac4e90}

\bibitem[{{Choi} {et~al.}(2018){Choi}, {Nidever}, {Olsen}, {Blum}, {Besla},
  {Zaritsky}, {van der Marel}, {Bell}, {Gallart}, {Cioni}, {Johnson}, {Vivas},
  {Saha}, {de Boer}, {No{\"e}l}, {Monachesi}, {Massana}, {Conn},
  {Martinez-Delgado}, {Mu{\~n}oz}, \& {Stringfellow}}]{choi2018a}
{Choi}, Y., {Nidever}, D.~L., {Olsen}, K., {et~al.} 2018, \apj, 866, 90,
  \dodoi{10.3847/1538-4357/aae083}

\bibitem[{{Clark} {et~al.}(2023){Clark}, {Roman-Duval}, {Gordon}, {Bot},
  {Smith}, \& {Hagen}}]{clark2023}
{Clark}, C. J.~R., {Roman-Duval}, J.~C., {Gordon}, K.~D., {et~al.} 2023, \apj,
  946, 42, \dodoi{10.3847/1538-4357/acbb66}

\bibitem[{{Cullinane} {et~al.}(2023){Cullinane}, {Mackey}, {Da Costa},
  {Koposov}, \& {Erkal}}]{cullinane2023}
{Cullinane}, L.~R., {Mackey}, A.~D., {Da Costa}, G.~S., {Koposov}, S.~E., \&
  {Erkal}, D. 2023, \mnras, 518, L25, \dodoi{10.1093/mnrasl/slac129}

\bibitem[{{Danforth} {et~al.}(2002){Danforth}, {Howk}, {Fullerton}, {Blair}, \&
  {Sembach}}]{danforth2002}
{Danforth}, C.~W., {Howk}, J.~C., {Fullerton}, A.~W., {Blair}, W.~P., \&
  {Sembach}, K.~R. 2002, \apjs, 139, 81, \dodoi{10.1086/338239}

\bibitem[{{de Grijs} \& {Bono}(2015)}]{degrijs2015}
{de Grijs}, R., \& {Bono}, G. 2015, \aj, 149, 179,
  \dodoi{10.1088/0004-6256/149/6/179}

\bibitem[{{De Leo} {et~al.}(2020){De Leo}, {Carrera}, {No{\"e}l}, {Read},
  {Erkal}, \& {Gallart}}]{deleo2020}
{De Leo}, M., {Carrera}, R., {No{\"e}l}, N. E.~D., {et~al.} 2020, \mnras, 495,
  98, \dodoi{10.1093/mnras/staa1122}

\bibitem[{{Di Teodoro} {et~al.}(2019){Di Teodoro}, {McClure-Griffiths},
  {Jameson}, {D{\'e}nes}, {Dickey}, {Stanimirovi{\'c}}, {Staveley-Smith},
  {Anderson}, {Bunton}, {Chippendale}, {Lee-Waddell}, {MacLeod}, \&
  {Voronkov}}]{diteodoro2019}
{Di Teodoro}, E.~M., {McClure-Griffiths}, N.~M., {Jameson}, K.~E., {et~al.}
  2019, \mnras, 483, 392, \dodoi{10.1093/mnras/sty3095}

\bibitem[{{Dias} {et~al.}(2022){Dias}, {Parisi}, {Angelo}, {Maia}, {Oliveira},
  {Souza}, {Kerber}, {Santos}, {P{\'e}rez-Villegas}, {Sanmartim}, {Quint},
  {Fraga}, {Barbuy}, {Bica}, {Santrich}, {Hernandez-Jimenez}, {Geisler},
  {Minniti}, {De B{\'o}rtoli}, {Bassino}, \& {Rocha}}]{dias_2022}
{Dias}, B., {Parisi}, M.~C., {Angelo}, M., {et~al.} 2022, \mnras, 512, 4334,
  \dodoi{10.1093/mnras/stac259}

\bibitem[{{Diaz} \& {Bekki}(2012)}]{diaz2012}
{Diaz}, J.~D., \& {Bekki}, K. 2012, \apj, 750, 36,
  \dodoi{10.1088/0004-637X/750/1/36}

\bibitem[{{Dobbie} {et~al.}(2014){Dobbie}, {Cole}, {Subramaniam}, \&
  {Keller}}]{dobbie2014}
{Dobbie}, P.~D., {Cole}, A.~A., {Subramaniam}, A., \& {Keller}, S. 2014,
  Monthly Notices of the Royal Astronomical Society, 442, 1663,
  \dodoi{10.1093/mnras/stu910}

\bibitem[{{El Youssoufi} {et~al.}(2023){El Youssoufi}, {Cioni}, {Kacharov},
  {Bell}, {Matjevi{\'c}}, {Bekki}, {de Grijs}, {Ivanov}, \& {van
  Loon}}]{elyoussoufi2023}
{El Youssoufi}, D., {Cioni}, M.-R.~L., {Kacharov}, N., {et~al.} 2023, \mnras,
  523, 347, \dodoi{10.1093/mnras/stad1339}

\bibitem[{{Erkal} {et~al.}(2019){Erkal}, {Belokurov}, {Laporte}, {Koposov},
  {Li}, {Grillmair}, {Kallivayalil}, {Price-Whelan}, {Evans}, {Hawkins},
  {Hendel}, {Mateu}, {Navarro}, {del Pino}, {Slater}, {Sohn}, \& {Orphan Aspen
  Treasury Collaboration}}]{erkal2019}
{Erkal}, D., {Belokurov}, V., {Laporte}, C.~F.~P., {et~al.} 2019, \mnras, 487,
  2685, \dodoi{10.1093/mnras/stz1371}

\bibitem[{{Evans} \& {Howarth}(2008)}]{evans2008}
{Evans}, C.~J., \& {Howarth}, I.~D. 2008, \mnras, 386, 826,
  \dodoi{10.1111/j.1365-2966.2008.13012.x}

\bibitem[{{For} {et~al.}(2014){For}, {Staveley-Smith}, {Matthews}, \&
  {McClure-Griffiths}}]{for2014}
{For}, B.~Q., {Staveley-Smith}, L., {Matthews}, D., \& {McClure-Griffiths},
  N.~M. 2014, \apj, 792, 43, \dodoi{10.1088/0004-637X/792/1/43}

\bibitem[{{For} {et~al.}(2016){For}, {Staveley-Smith}, {McClure-Griffiths},
  {Westmeier}, \& {Bekki}}]{for2016}
{For}, B.~Q., {Staveley-Smith}, L., {McClure-Griffiths}, N.~M., {Westmeier},
  T., \& {Bekki}, K. 2016, \mnras, 461, 892, \dodoi{10.1093/mnras/stw1364}

\bibitem[{{Foreman-Mackey} {et~al.}(2013){Foreman-Mackey}, {Hogg}, {Lang}, \&
  {Goodman}}]{foremanmackey2013}
{Foreman-Mackey}, D., {Hogg}, D.~W., {Lang}, D., \& {Goodman}, J. 2013, \pasp,
  125, 306, \dodoi{10.1086/670067}

\bibitem[{{Gaia Collaboration} {et~al.}(2016){Gaia Collaboration}, {Prusti},
  {de Bruijne}, {Brown}, {Vallenari}, {Babusiaux}, {Bailer-Jones}, {Bastian},
  {Biermann}, {Evans}, {Eyer}, {Jansen}, {Jordi}, {Klioner}, {Lammers},
  {Lindegren}, {Luri}, {Mignard}, {Milligan}, {Panem}, {Poinsignon},
  {Pourbaix}, {Randich}, {Sarri}, {Sartoretti}, {Siddiqui}, {Soubiran},
  {Valette}, {van Leeuwen}, {Walton}, {Aerts}, {Arenou}, {Cropper}, {Drimmel},
  {H{\o}g}, {Katz}, {Lattanzi}, {O'Mullane}, {Grebel}, {Holland}, {Huc},
  {Passot}, {Bramante}, {Cacciari}, {Casta{\~n}eda}, {Chaoul}, {Cheek}, {De
  Angeli}, {Fabricius}, {Guerra}, {Hern{\'a}ndez}, {Jean-Antoine-Piccolo},
  {Masana}, {Messineo}, {Mowlavi}, {Nienartowicz}, {Ord{\'o}{\~n}ez-Blanco},
  {Panuzzo}, {Portell}, {Richards}, {Riello}, {Seabroke}, {Tanga},
  {Th{\'e}venin}, {Torra}, {Els}, {Gracia-Abril}, {Comoretto},
  {Garcia-Reinaldos}, {Lock}, {Mercier}, {Altmann}, {Andrae}, {Astraatmadja},
  {Bellas-Velidis}, {Benson}, {Berthier}, {Blomme}, {Busso}, {Carry},
  {Cellino}, {Clementini}, {Cowell}, {Creevey}, {Cuypers}, {Davidson}, {De
  Ridder}, {de Torres}, {Delchambre}, {Dell'Oro}, {Ducourant}, {Fr{\'e}mat},
  {Garc{\'\i}a-Torres}, {Gosset}, {Halbwachs}, {Hambly}, {Harrison}, {Hauser},
  {Hestroffer}, {Hodgkin}, {Huckle}, {Hutton}, {Jasniewicz}, {Jordan},
  {Kontizas}, {Korn}, {Lanzafame}, {Manteiga}, {Moitinho}, {Muinonen},
  {Osinde}, {Pancino}, {Pauwels}, {Petit}, {Recio-Blanco}, {Robin}, {Sarro},
  {Siopis}, {Smith}, {Smith}, {Sozzetti}, {Thuillot}, {van Reeven}, {Viala},
  {Abbas}, {Abreu Aramburu}, {Accart}, {Aguado}, {Allan}, {Allasia},
  {Altavilla}, {{\'A}lvarez}, {Alves}, {Anderson}, {Andrei}, {Anglada Varela},
  {Antiche}, {Antoja}, {Ant{\'o}n}, {Arcay}, {Atzei}, {Ayache}, {Bach},
  {Baker}, {Balaguer-N{\'u}{\~n}ez}, {Barache}, {Barata}, {Barbier}, {Barblan},
  {Baroni}, {Barrado y Navascu{\'e}s}, {Barros}, {Barstow}, {Becciani},
  {Bellazzini}, {Bellei}, {Bello Garc{\'\i}a}, {Belokurov}, {Bendjoya},
  {Berihuete}, {Bianchi}, {Bienaym{\'e}}, {Billebaud}, {Blagorodnova},
  {Blanco-Cuaresma}, {Boch}, {Bombrun}, {Borrachero}, {Bouquillon}, {Bourda},
  {Bouy}, {Bragaglia}, {Breddels}, {Brouillet}, {Br{\"u}semeister},
  {Bucciarelli}, {Budnik}, {Burgess}, {Burgon}, {Burlacu}, {Busonero}, {Buzzi},
  {Caffau}, {Cambras}, {Campbell}, {Cancelliere}, {Cantat-Gaudin}, {Carlucci},
  {Carrasco}, {Castellani}, {Charlot}, {Charnas}, {Charvet}, {Chassat},
  {Chiavassa}, {Clotet}, {Cocozza}, {Collins}, {Collins}, {Costigan}, {Crifo},
  {Cross}, {Crosta}, {Crowley}, {Dafonte}, {Damerdji}, {Dapergolas}, {David},
  {David}, {De Cat}, {de Felice}, {de Laverny}, {De Luise}, {De March}, {de
  Martino}, {de Souza}, {Debosscher}, {del Pozo}, {Delbo}, {Delgado},
  {Delgado}, {di Marco}, {Di Matteo}, {Diakite}, {Distefano}, {Dolding}, {Dos
  Anjos}, {Drazinos}, {Dur{\'a}n}, {Dzigan}, {Ecale}, {Edvardsson}, {Enke},
  {Erdmann}, {Escolar}, {Espina}, {Evans}, {Eynard Bontemps}, {Fabre},
  {Fabrizio}, {Faigler}, {Falc{\~a}o}, {Farr{\`a}s Casas}, {Faye}, {Federici},
  {Fedorets}, {Fern{\'a}ndez-Hern{\'a}ndez}, {Fernique}, {Fienga}, {Figueras},
  {Filippi}, {Findeisen}, {Fonti}, {Fouesneau}, {Fraile}, {Fraser}, {Fuchs},
  {Furnell}, {Gai}, {Galleti}, {Galluccio}, {Garabato}, {Garc{\'\i}a-Sedano},
  {Gar{\'e}}, {Garofalo}, {Garralda}, {Gavras}, {Gerssen}, {Geyer}, {Gilmore},
  {Girona}, {Giuffrida}, {Gomes}, {Gonz{\'a}lez-Marcos},
  {Gonz{\'a}lez-N{\'u}{\~n}ez}, {Gonz{\'a}lez-Vidal}, {Granvik}, {Guerrier},
  {Guillout}, {Guiraud}, {G{\'u}rpide}, {Guti{\'e}rrez-S{\'a}nchez}, {Guy},
  {Haigron}, {Hatzidimitriou}, {Haywood}, {Heiter}, {Helmi}, {Hobbs},
  {Hofmann}, {Holl}, {Holland}, {Hunt}, {Hypki}, {Icardi}, {Irwin}, {Jevardat
  de Fombelle}, {Jofr{\'e}}, {Jonker}, {Jorissen}, {Julbe}, {Karampelas},
  {Kochoska}, {Kohley}, {Kolenberg}, {Kontizas}, {Koposov}, {Kordopatis},
  {Koubsky}, {Kowalczyk}, {Krone-Martins}, {Kudryashova}, {Kull}, {Bachchan},
  {Lacoste-Seris}, {Lanza}, {Lavigne}, {Le Poncin-Lafitte}, {Lebreton},
  {Lebzelter}, {Leccia}, {Leclerc}, {Lecoeur-Taibi}, {Lemaitre}, {Lenhardt},
  {Leroux}, {Liao}, {Licata}, {Lindstr{\o}m}, {Lister}, {Livanou}, {Lobel},
  {L{\"o}ffler}, {L{\'o}pez}, {Lopez-Lozano}, {Lorenz}, {Loureiro},
  {MacDonald}, {Magalh{\~a}es Fernandes}, {Managau}, {Mann}, {Mantelet},
  {Marchal}, {Marchant}, {Marconi}, {Marie}, {Marinoni}, {Marrese},
  {Marschalk{\'o}}, {Marshall}, {Mart{\'\i}n-Fleitas}, {Martino}, {Mary},
  {Matijevi{\v{c}}}, {Mazeh}, {McMillan}, {Messina}, {Mestre}, {Michalik},
  {Millar}, {Miranda}, {Molina}, {Molinaro}, {Molinaro}, {Moln{\'a}r},
  {Moniez}, {Montegriffo}, {Monteiro}, {Mor}, {Mora}, {Morbidelli}, {Morel},
  {Morgenthaler}, {Morley}, {Morris}, {Mulone}, {Muraveva}, {Musella},
  {Narbonne}, {Nelemans}, {Nicastro}, {Noval}, {Ord{\'e}novic},
  {Ordieres-Mer{\'e}}, {Osborne}, {Pagani}, {Pagano}, {Pailler}, {Palacin},
  {Palaversa}, {Parsons}, {Paulsen}, {Pecoraro}, {Pedrosa}, {Pentik{\"a}inen},
  {Pereira}, {Pichon}, {Piersimoni}, {Pineau}, {Plachy}, {Plum}, {Poujoulet},
  {Pr{\v{s}}a}, {Pulone}, {Ragaini}, {Rago}, {Rambaux}, {Ramos-Lerate},
  {Ranalli}, {Rauw}, {Read}, {Regibo}, {Renk}, {Reyl{\'e}}, {Ribeiro},
  {Rimoldini}, {Ripepi}, {Riva}, {Rixon}, {Roelens}, {Romero-G{\'o}mez},
  {Rowell}, {Royer}, {Rudolph}, {Ruiz-Dern}, {Sadowski}, {Sagrist{\`a}
  Sell{\'e}s}, {Sahlmann}, {Salgado}, {Salguero}, {Sarasso}, {Savietto},
  {Schnorhk}, {Schultheis}, {Sciacca}, {Segol}, {Segovia}, {Segransan},
  {Serpell}, {Shih}, {Smareglia}, {Smart}, {Smith}, {Solano}, {Solitro},
  {Sordo}, {Soria Nieto}, {Souchay}, {Spagna}, {Spoto}, {Stampa}, {Steele},
  {Steidelm{\"u}ller}, {Stephenson}, {Stoev}, {Suess}, {S{\"u}veges}, {Surdej},
  {Szabados}, {Szegedi-Elek}, {Tapiador}, {Taris}, {Tauran}, {Taylor},
  {Teixeira}, {Terrett}, {Tingley}, {Trager}, {Turon}, {Ulla}, {Utrilla},
  {Valentini}, {van Elteren}, {Van Hemelryck}, {van Leeuwen}, {Varadi},
  {Vecchiato}, {Veljanoski}, {Via}, {Vicente}, {Vogt}, {Voss}, {Votruba},
  {Voutsinas}, {Walmsley}, {Weiler}, {Weingrill}, {Werner}, {Wevers},
  {Whitehead}, {Wyrzykowski}, {Yoldas}, {{\v{Z}}erjal}, {Zucker}, {Zurbach},
  {Zwitter}, {Alecu}, {Allen}, {Allende Prieto}, {Amorim},
  {Anglada-Escud{\'e}}, {Arsenijevic}, {Azaz}, {Balm}, {Beck}, {Bernstein},
  {Bigot}, {Bijaoui}, {Blasco}, {Bonfigli}, {Bono}, {Boudreault}, {Bressan},
  {Brown}, {Brunet}, {Bunclark}, {Buonanno}, {Butkevich}, {Carret}, {Carrion},
  {Chemin}, {Ch{\'e}reau}, {Corcione}, {Darmigny}, {de Boer}, {de Teodoro}, {de
  Zeeuw}, {Delle Luche}, {Domingues}, {Dubath}, {Fodor}, {Fr{\'e}zouls},
  {Fries}, {Fustes}, {Fyfe}, {Gallardo}, {Gallegos}, {Gardiol}, {Gebran},
  {Gomboc}, {G{\'o}mez}, {Grux}, {Gueguen}, {Heyrovsky}, {Hoar}, {Iannicola},
  {Isasi Parache}, {Janotto}, {Joliet}, {Jonckheere}, {Keil}, {Kim},
  {Klagyivik}, {Klar}, {Knude}, {Kochukhov}, {Kolka}, {Kos}, {Kutka}, {Lainey},
  {LeBouquin}, {Liu}, {Loreggia}, {Makarov}, {Marseille}, {Martayan},
  {Martinez-Rubi}, {Massart}, {Meynadier}, {Mignot}, {Munari}, {Nguyen},
  {Nordlander}, {Ocvirk}, {O'Flaherty}, {Olias Sanz}, {Ortiz}, {Osorio},
  {Oszkiewicz}, {Ouzounis}, {Palmer}, {Park}, {Pasquato}, {Peltzer}, {Peralta},
  {P{\'e}turaud}, {Pieniluoma}, {Pigozzi}, {Poels}, {Prat}, {Prod'homme},
  {Raison}, {Rebordao}, {Risquez}, {Rocca-Volmerange}, {Rosen}, {Ruiz-Fuertes},
  {Russo}, {Sembay}, {Serraller Vizcaino}, {Short}, {Siebert}, {Silva},
  {Sinachopoulos}, {Slezak}, {Soffel}, {Sosnowska}, {Strai{\v{z}}ys}, {ter
  Linden}, {Terrell}, {Theil}, {Tiede}, {Troisi}, {Tsalmantza}, {Tur},
  {Vaccari}, {Vachier}, {Valles}, {Van Hamme}, {Veltz}, {Virtanen}, {Wallut},
  {Wichmann}, {Wilkinson}, {Ziaeepour}, \& {Zschocke}}]{gaia2016}
{Gaia Collaboration}, {Prusti}, T., {de Bruijne}, J.~H.~J., {et~al.} 2016,
  \aap, 595, A1, \dodoi{10.1051/0004-6361/201629272}

\bibitem[{{Gaia Collaboration} {et~al.}(2018{\natexlab{a}}){Gaia
  Collaboration}, {Helmi}, {van Leeuwen}, {McMillan}, {Massari}, {Antoja},
  {Robin}, {Lindegren}, {Bastian}, {Arenou}, {Babusiaux}, {Biermann},
  {Breddels}, {Hobbs}, {Jordi}, {Pancino}, {Reyl{\'e}}, {Veljanoski}, {Brown},
  {Vallenari}, {Prusti}, {de Bruijne}, {Bailer-Jones}, {Evans}, {Eyer},
  {Jansen}, {Klioner}, {Lammers}, {Luri}, {Mignard}, {Panem}, {Pourbaix},
  {Randich}, {Sartoretti}, {Siddiqui}, {Soubiran}, {Walton}, {Cropper},
  {Drimmel}, {Katz}, {Lattanzi}, {Bakker}, {Cacciari}, {Casta{\~n}eda},
  {Chaoul}, {Cheek}, {De Angeli}, {Fabricius}, {Guerra}, {Holl}, {Masana},
  {Messineo}, {Mowlavi}, {Nienartowicz}, {Panuzzo}, {Portell}, {Riello},
  {Seabroke}, {Tanga}, {Th{\'e}venin}, {Gracia-Abril}, {Comoretto},
  {Garcia-Reinaldos}, {Teyssier}, {Altmann}, {Andrae}, {Audard},
  {Bellas-Velidis}, {Benson}, {Berthier}, {Blomme}, {Burgess}, {Busso},
  {Carry}, {Cellino}, {Clementini}, {Clotet}, {Creevey}, {Davidson}, {De
  Ridder}, {Delchambre}, {Dell'Oro}, {Ducourant},
  {Fern{\'a}ndez-Hern{\'a}ndez}, {Fouesneau}, {Fr{\'e}mat}, {Galluccio},
  {Garc{\'\i}a-Torres}, {Gonz{\'a}lez-N{\'u}{\~n}ez}, {Gonz{\'a}lez-Vidal},
  {Gosset}, {Guy}, {Halbwachs}, {Hambly}, {Harrison}, {Hern{\'a}ndez},
  {Hestroffer}, {Hodgkin}, {Hutton}, {Jasniewicz}, {Jean-Antoine-Piccolo},
  {Jordan}, {Korn}, {Krone-Martins}, {Lanzafame}, {Lebzelter}, {L{\"o}ffler},
  {Manteiga}, {Marrese}, {Mart{\'\i}n-Fleitas}, {Moitinho}, {Mora}, {Muinonen},
  {Osinde}, {Pauwels}, {Petit}, {Recio-Blanco}, {Richards}, {Rimoldini},
  {Sarro}, {Siopis}, {Smith}, {Sozzetti}, {S{\"u}veges}, {Torra}, {van Reeven},
  {Abbas}, {Abreu Aramburu}, {Accart}, {Aerts}, {Altavilla}, {{\'A}lvarez},
  {Alvarez}, {Alves}, {Anderson}, {Andrei}, {Anglada Varela}, {Antiche},
  {Arcay}, {Astraatmadja}, {Bach}, {Baker}, {Balaguer-N{\'u}{\~n}ez}, {Balm},
  {Barache}, {Barata}, {Barbato}, {Barblan}, {Barklem}, {Barrado}, {Barros},
  {Barstow}, {Bartholom{\'e} Mu{\~n}oz}, {Bassilana}, {Becciani}, {Bellazzini},
  {Berihuete}, {Bertone}, {Bianchi}, {Bienaym{\'e}}, {Blanco-Cuaresma}, {Boch},
  {Boeche}, {Bombrun}, {Borrachero}, {Bossini}, {Bouquillon}, {Bourda},
  {Bragaglia}, {Bramante}, {Bressan}, {Brouillet}, {Br{\"u}semeister},
  {Brugaletta}, {Bucciarelli}, {Burlacu}, {Busonero}, {Butkevich}, {Buzzi},
  {Caffau}, {Cancelliere}, {Cannizzaro}, {Cantat-Gaudin}, {Carballo},
  {Carlucci}, {Carrasco}, {Casamiquela}, {Castellani}, {Castro-Ginard},
  {Charlot}, {Chemin}, {Chiavassa}, {Cocozza}, {Costigan}, {Cowell}, {Crifo},
  {Crosta}, {Crowley}, {Cuypers}, {Dafonte}, {Damerdji}, {Dapergolas}, {David},
  {David}, {de Laverny}, {De Luise}, {De March}, {de Martino}, {de Souza}, {de
  Torres}, {Debosscher}, {del Pozo}, {Delbo}, {Delgado}, {Delgado}, {Di
  Matteo}, {Diakite}, {Diener}, {Distefano}, {Dolding}, {Drazinos},
  {Dur{\'a}n}, {Edvardsson}, {Enke}, {Eriksson}, {Esquej}, {Eynard Bontemps},
  {Fabre}, {Fabrizio}, {Faigler}, {Falc{\~a}o}, {Farr{\`a}s Casas}, {Federici},
  {Fedorets}, {Fernique}, {Figueras}, {Filippi}, {Findeisen}, {Fonti},
  {Fraile}, {Fraser}, {Fr{\'e}zouls}, {Gai}, {Galleti}, {Garabato},
  {Garc{\'\i}a-Sedano}, {Garofalo}, {Garralda}, {Gavel}, {Gavras}, {Gerssen},
  {Geyer}, {Giacobbe}, {Gilmore}, {Girona}, {Giuffrida}, {Glass}, {Gomes},
  {Granvik}, {Gueguen}, {Guerrier}, {Guiraud}, {Guti{\'e}rrez-S{\'a}nchez},
  {Hofmann}, {Holland}, {Huckle}, {Hypki}, {Icardi}, {Jan{\ss}en}, {Jevardat de
  Fombelle}, {Jonker}, {Juh{\'a}sz}, {Julbe}, {Karampelas}, {Kewley}, {Klar},
  {Kochoska}, {Kohley}, {Kolenberg}, {Kontizas}, {Kontizas}, {Koposov},
  {Kordopatis}, {Kostrzewa-Rutkowska}, {Koubsky}, {Lambert}, {Lanza}, {Lasne},
  {Lavigne}, {Le Fustec}, {Le Poncin-Lafitte}, {Lebreton}, {Leccia}, {Leclerc},
  {Lecoeur-Taibi}, {Lenhardt}, {Leroux}, {Liao}, {Licata}, {Lindstr{\o}m},
  {Lister}, {Livanou}, {Lobel}, {L{\'o}pez}, {Managau}, {Mann}, {Mantelet},
  {Marchal}, {Marchant}, {Marconi}, {Marinoni}, {Marschalk{\'o}}, {Marshall},
  {Martino}, {Marton}, {Mary}, {Matijevi{\v{c}}}, {Mazeh}, {Messina},
  {Michalik}, {Millar}, {Molina}, {Molinaro}, {Moln{\'a}r}, {Montegriffo},
  {Mor}, {Morbidelli}, {Morel}, {Morris}, {Mulone}, {Muraveva}, {Musella},
  {Nelemans}, {Nicastro}, {Noval}, {O'Mullane}, {Ord{\'e}novic},
  {Ord{\'o}{\~n}ez-Blanco}, {Osborne}, {Pagani}, {Pagano}, {Pailler},
  {Palacin}, {Palaversa}, {Panahi}, {Pawlak}, {Piersimoni}, {Pineau}, {Plachy},
  {Plum}, {Poggio}, {Poujoulet}, {Pr{\v{s}}a}, {Pulone}, {Racero}, {Ragaini},
  {Rambaux}, {Ramos-Lerate}, {Regibo}, {Riclet}, {Ripepi}, {Riva}, {Rivard},
  {Rixon}, {Roegiers}, {Roelens}, {Romero-G{\'o}mez}, {Rowell}, {Royer},
  {Ruiz-Dern}, {Sadowski}, {Sagrist{\`a} Sell{\'e}s}, {Sahlmann}, {Salgado},
  {Salguero}, {Sanna}, {Santana-Ros}, {Sarasso}, {Savietto}, {Schultheis},
  {Sciacca}, {Segol}, {Segovia}, {S{\'e}gransan}, {Shih}, {Siltala}, {Silva},
  {Smart}, {Smith}, {Solano}, {Solitro}, {Sordo}, {Soria Nieto}, {Souchay},
  {Spagna}, {Spoto}, {Stampa}, {Steele}, {Steidelm{\"u}ller}, {Stephenson},
  {Stoev}, {Suess}, {Surdej}, {Szabados}, {Szegedi-Elek}, {Tapiador}, {Taris},
  {Tauran}, {Taylor}, {Teixeira}, {Terrett}, {Teyssand ier}, {Thuillot},
  {Titarenko}, {Torra Clotet}, {Turon}, {Ulla}, {Utrilla}, {Uzzi}, {Vaillant},
  {Valentini}, {Valette}, {van Elteren}, {Van Hemelryck}, {van Leeuwen},
  {Vaschetto}, {Vecchiato}, {Viala}, {Vicente}, {Vogt}, {von Essen}, {Voss},
  {Votruba}, {Voutsinas}, {Walmsley}, {Weiler}, {Wertz}, {Wevems},
  {Wyrzykowski}, {Yoldas}, {{\v{Z}}erjal}, {Ziaeepour}, {Zorec}, {Zschocke},
  {Zucker}, {Zurbach}, \& {Zwitter}}]{gaiacollaboration2018}
{Gaia Collaboration}, {Helmi}, A., {van Leeuwen}, F., {et~al.}
  2018{\natexlab{a}}, \aap, 616, A12, \dodoi{10.1051/0004-6361/201832698}

\bibitem[{{Gaia Collaboration} {et~al.}(2018{\natexlab{b}}){Gaia
  Collaboration}, {Helmi}, {van Leeuwen}, {McMillan}, {Massari}, {Antoja},
  {Robin}, {Lindegren}, {Bastian}, {Arenou}, {Babusiaux}, {Biermann},
  {Breddels}, {Hobbs}, {Jordi}, {Pancino}, {Reyl{\'e}}, {Veljanoski}, {Brown},
  {Vallenari}, {Prusti}, {de Bruijne}, {Bailer-Jones}, {Evans}, {Eyer},
  {Jansen}, {Klioner}, {Lammers}, {Luri}, {Mignard}, {Panem}, {Pourbaix},
  {Randich}, {Sartoretti}, {Siddiqui}, {Soubiran}, {Walton}, {Cropper},
  {Drimmel}, {Katz}, {Lattanzi}, {Bakker}, {Cacciari}, {Casta{\~n}eda},
  {Chaoul}, {Cheek}, {De Angeli}, {Fabricius}, {Guerra}, {Holl}, {Masana},
  {Messineo}, {Mowlavi}, {Nienartowicz}, {Panuzzo}, {Portell}, {Riello},
  {Seabroke}, {Tanga}, {Th{\'e}venin}, {Gracia-Abril}, {Comoretto},
  {Garcia-Reinaldos}, {Teyssier}, {Altmann}, {Andrae}, {Audard},
  {Bellas-Velidis}, {Benson}, {Berthier}, {Blomme}, {Burgess}, {Busso},
  {Carry}, {Cellino}, {Clementini}, {Clotet}, {Creevey}, {Davidson}, {De
  Ridder}, {Delchambre}, {Dell'Oro}, {Ducourant},
  {Fern{\'a}ndez-Hern{\'a}ndez}, {Fouesneau}, {Fr{\'e}mat}, {Galluccio},
  {Garc{\'\i}a-Torres}, {Gonz{\'a}lez-N{\'u}{\~n}ez}, {Gonz{\'a}lez-Vidal},
  {Gosset}, {Guy}, {Halbwachs}, {Hambly}, {Harrison}, {Hern{\'a}ndez},
  {Hestroffer}, {Hodgkin}, {Hutton}, {Jasniewicz}, {Jean-Antoine-Piccolo},
  {Jordan}, {Korn}, {Krone-Martins}, {Lanzafame}, {Lebzelter}, {L{\"o}ffler},
  {Manteiga}, {Marrese}, {Mart{\'\i}n-Fleitas}, {Moitinho}, {Mora}, {Muinonen},
  {Osinde}, {Pauwels}, {Petit}, {Recio-Blanco}, {Richards}, {Rimoldini},
  {Sarro}, {Siopis}, {Smith}, {Sozzetti}, {S{\"u}veges}, {Torra}, {van Reeven},
  {Abbas}, {Abreu Aramburu}, {Accart}, {Aerts}, {Altavilla}, {{\'A}lvarez},
  {Alvarez}, {Alves}, {Anderson}, {Andrei}, {Anglada Varela}, {Antiche},
  {Arcay}, {Astraatmadja}, {Bach}, {Baker}, {Balaguer-N{\'u}{\~n}ez}, {Balm},
  {Barache}, {Barata}, {Barbato}, {Barblan}, {Barklem}, {Barrado}, {Barros},
  {Barstow}, {Bartholom{\'e} Mu{\~n}oz}, {Bassilana}, {Becciani}, {Bellazzini},
  {Berihuete}, {Bertone}, {Bianchi}, {Bienaym{\'e}}, {Blanco-Cuaresma}, {Boch},
  {Boeche}, {Bombrun}, {Borrachero}, {Bossini}, {Bouquillon}, {Bourda},
  {Bragaglia}, {Bramante}, {Bressan}, {Brouillet}, {Br{\"u}semeister},
  {Brugaletta}, {Bucciarelli}, {Burlacu}, {Busonero}, {Butkevich}, {Buzzi},
  {Caffau}, {Cancelliere}, {Cannizzaro}, {Cantat-Gaudin}, {Carballo},
  {Carlucci}, {Carrasco}, {Casamiquela}, {Castellani}, {Castro-Ginard},
  {Charlot}, {Chemin}, {Chiavassa}, {Cocozza}, {Costigan}, {Cowell}, {Crifo},
  {Crosta}, {Crowley}, {Cuypers}, {Dafonte}, {Damerdji}, {Dapergolas}, {David},
  {David}, {de Laverny}, {De Luise}, {De March}, {de Martino}, {de Souza}, {de
  Torres}, {Debosscher}, {del Pozo}, {Delbo}, {Delgado}, {Delgado}, {Di
  Matteo}, {Diakite}, {Diener}, {Distefano}, {Dolding}, {Drazinos},
  {Dur{\'a}n}, {Edvardsson}, {Enke}, {Eriksson}, {Esquej}, {Eynard Bontemps},
  {Fabre}, {Fabrizio}, {Faigler}, {Falc{\~a}o}, {Farr{\`a}s Casas}, {Federici},
  {Fedorets}, {Fernique}, {Figueras}, {Filippi}, {Findeisen}, {Fonti},
  {Fraile}, {Fraser}, {Fr{\'e}zouls}, {Gai}, {Galleti}, {Garabato},
  {Garc{\'\i}a-Sedano}, {Garofalo}, {Garralda}, {Gavel}, {Gavras}, {Gerssen},
  {Geyer}, {Giacobbe}, {Gilmore}, {Girona}, {Giuffrida}, {Glass}, {Gomes},
  {Granvik}, {Gueguen}, {Guerrier}, {Guiraud}, {Guti{\'e}rrez-S{\'a}nchez},
  {Hofmann}, {Holland}, {Huckle}, {Hypki}, {Icardi}, {Jan{\ss}en}, {Jevardat de
  Fombelle}, {Jonker}, {Juh{\'a}sz}, {Julbe}, {Karampelas}, {Kewley}, {Klar},
  {Kochoska}, {Kohley}, {Kolenberg}, {Kontizas}, {Kontizas}, {Koposov},
  {Kordopatis}, {Kostrzewa-Rutkowska}, {Koubsky}, {Lambert}, {Lanza}, {Lasne},
  {Lavigne}, {Le Fustec}, {Le Poncin-Lafitte}, {Lebreton}, {Leccia}, {Leclerc},
  {Lecoeur-Taibi}, {Lenhardt}, {Leroux}, {Liao}, {Licata}, {Lindstr{\o}m},
  {Lister}, {Livanou}, {Lobel}, {L{\'o}pez}, {Managau}, {Mann}, {Mantelet},
  {Marchal}, {Marchant}, {Marconi}, {Marinoni}, {Marschalk{\'o}}, {Marshall},
  {Martino}, {Marton}, {Mary}, {Matijevi{\v{c}}}, {Mazeh}, {Messina},
  {Michalik}, {Millar}, {Molina}, {Molinaro}, {Moln{\'a}r}, {Montegriffo},
  {Mor}, {Morbidelli}, {Morel}, {Morris}, {Mulone}, {Muraveva}, {Musella},
  {Nelemans}, {Nicastro}, {Noval}, {O'Mullane}, {Ord{\'e}novic},
  {Ord{\'o}{\~n}ez-Blanco}, {Osborne}, {Pagani}, {Pagano}, {Pailler},
  {Palacin}, {Palaversa}, {Panahi}, {Pawlak}, {Piersimoni}, {Pineau}, {Plachy},
  {Plum}, {Poggio}, {Poujoulet}, {Pr{\v{s}}a}, {Pulone}, {Racero}, {Ragaini},
  {Rambaux}, {Ramos-Lerate}, {Regibo}, {Riclet}, {Ripepi}, {Riva}, {Rivard},
  {Rixon}, {Roegiers}, {Roelens}, {Romero-G{\'o}mez}, {Rowell}, {Royer},
  {Ruiz-Dern}, {Sadowski}, {Sagrist{\`a} Sell{\'e}s}, {Sahlmann}, {Salgado},
  {Salguero}, {Sanna}, {Santana-Ros}, {Sarasso}, {Savietto}, {Schultheis},
  {Sciacca}, {Segol}, {Segovia}, {S{\'e}gransan}, {Shih}, {Siltala}, {Silva},
  {Smart}, {Smith}, {Solano}, {Solitro}, {Sordo}, {Soria Nieto}, {Souchay},
  {Spagna}, {Spoto}, {Stampa}, {Steele}, {Steidelm{\"u}ller}, {Stephenson},
  {Stoev}, {Suess}, {Surdej}, {Szabados}, {Szegedi-Elek}, {Tapiador}, {Taris},
  {Tauran}, {Taylor}, {Teixeira}, {Terrett}, {Teyssandier}, {Thuillot},
  {Titarenko}, {Torra Clotet}, {Turon}, {Ulla}, {Utrilla}, {Uzzi}, {Vaillant},
  {Valentini}, {Valette}, {van Elteren}, {Van Hemelryck}, {van Leeuwen},
  {Vaschetto}, {Vecchiato}, {Viala}, {Vicente}, {Vogt}, {von Essen}, {Voss},
  {Votruba}, {Voutsinas}, {Walmsley}, {Weiler}, {Wertz}, {Wevems},
  {Wyrzykowski}, {Yoldas}, {{\v{Z}}erjal}, {Ziaeepour}, {Zorec}, {Zschocke},
  {Zucker}, {Zurbach}, \& {Zwitter}}]{gaia2018}
---. 2018{\natexlab{b}}, \aap, 616, A12, \dodoi{10.1051/0004-6361/201832698}

\bibitem[{{Gaia Collaboration} {et~al.}(2021){Gaia Collaboration}, {Luri},
  {Chemin}, {Clementini}, {Delgado}, {McMillan}, {Romero-G{\'o}mez},
  {Balbinot}, {Castro-Ginard}, {Mor}, {Ripepi}, {Sarro}, {Cioni}, {Fabricius},
  {Garofalo}, {Helmi}, {Muraveva}, {Brown}, {Vallenari}, {Prusti}, {de
  Bruijne}, {Babusiaux}, {Biermann}, {Creevey}, {Evans}, {Eyer}, {Hutton},
  {Jansen}, {Jordi}, {Klioner}, {Lammers}, {Lindegren}, {Mignard}, {Panem},
  {Pourbaix}, {Randich}, {Sartoretti}, {Soubiran}, {Walton}, {Arenou},
  {Bailer-Jones}, {Bastian}, {Cropper}, {Drimmel}, {Katz}, {Lattanzi}, {van
  Leeuwen}, {Bakker}, {Casta{\~n}eda}, {De Angeli}, {Ducourant}, {Fouesneau},
  {Fr{\'e}mat}, {Guerra}, {Guerrier}, {Guiraud}, {Jean-Antoine Piccolo},
  {Masana}, {Messineo}, {Mowlavi}, {Nicolas}, {Nienartowicz}, {Pailler},
  {Panuzzo}, {Riclet}, {Roux}, {Seabroke}, {Sordo}, {Tanga}, {Th{\'e}venin},
  {Gracia-Abril}, {Portell}, {Teyssier}, {Altmann}, {Andrae}, {Bellas-Velidis},
  {Benson}, {Berthier}, {Blomme}, {Brugaletta}, {Burgess}, {Busso}, {Carry},
  {Cellino}, {Cheek}, {Damerdji}, {Davidson}, {Delchambre}, {Dell'Oro},
  {Fern{\'a}ndez-Hern{\'a}ndez}, {Galluccio}, {Garc{\'\i}a-Lario},
  {Garcia-Reinaldos}, {Gonz{\'a}lez-N{\'u}{\~n}ez}, {Gosset}, {Haigron},
  {Halbwachs}, {Hambly}, {Harrison}, {Hatzidimitriou}, {Heiter},
  {Hern{\'a}ndez}, {Hestroffer}, {Hodgkin}, {Holl}, {Jan{\ss}en}, {Jevardat de
  Fombelle}, {Jordan}, {Krone-Martins}, {Lanzafame}, {L{\"o}ffler}, {Lorca},
  {Manteiga}, {Marchal}, {Marrese}, {Moitinho}, {Mora}, {Muinonen}, {Osborne},
  {Pancino}, {Pauwels}, {Recio-Blanco}, {Richards}, {Riello}, {Rimoldini},
  {Robin}, {Roegiers}, {Rybizki}, {Siopis}, {Smith}, {Sozzetti}, {Ulla},
  {Utrilla}, {van Leeuwen}, {van Reeven}, {Abbas}, {Abreu Aramburu}, {Accart},
  {Aerts}, {Aguado}, {Ajaj}, {Altavilla}, {{\'A}lvarez}, {{\'A}lvarez
  Cid-Fuentes}, {Alves}, {Anderson}, {Anglada Varela}, {Antoja}, {Audard},
  {Baines}, {Baker}, {Balaguer-N{\'u}{\~n}ez}, {Balog}, {Barache}, {Barbato},
  {Barros}, {Barstow}, {Bartolom{\'e}}, {Bassilana}, {Bauchet},
  {Baudesson-Stella}, {Becciani}, {Bellazzini}, {Bernet}, {Bertone}, {Bianchi},
  {Blanco-Cuaresma}, {Boch}, {Bombrun}, {Bossini}, {Bouquillon}, {Bragaglia},
  {Bramante}, {Breedt}, {Bressan}, {Brouillet}, {Bucciarelli}, {Burlacu},
  {Busonero}, {Butkevich}, {Buzzi}, {Caffau}, {Cancelliere}, {C{\'a}novas},
  {Cantat-Gaudin}, {Carballo}, {Carlucci}, {Carnerero}, {Carrasco},
  {Casamiquela}, {Castellani}, {Castro Sampol}, {Chaoul}, {Charlot},
  {Chiavassa}, {Comoretto}, {Cooper}, {Cornez}, {Cowell}, {Crifo}, {Crosta},
  {Crowley}, {Dafonte}, {Dapergolas}, {David}, {David}, {de Laverny}, {De
  Luise}, {De March}, {De Ridder}, {de Souza}, {de Teodoro}, {de Torres}, {del
  Peloso}, {del Pozo}, {Delgado}, {Delisle}, {Di Matteo}, {Diakite}, {Diener},
  {Distefano}, {Dolding}, {Eappachen}, {Enke}, {Esquej}, {Fabre}, {Fabrizio},
  {Faigler}, {Fedorets}, {Fernique}, {Fienga}, {Figueras}, {Fouron},
  {Fragkoudi}, {Fraile}, {Franke}, {Gai}, {Garabato}, {Garcia-Gutierrez},
  {Garc{\'\i}a-Torres}, {Gavras}, {Gerlach}, {Geyer}, {Giacobbe}, {Gilmore},
  {Girona}, {Giuffrida}, {Gomez}, {Gonzalez-Santamaria}, {Gonz{\'a}lez-Vidal},
  {Granvik}, {Guti{\'e}rrez-S{\'a}nchez}, {Guy}, {Hauser}, {Haywood},
  {Hidalgo}, {Hilger}, {H{\l}adczuk}, {Hobbs}, {Holland}, {Huckle},
  {Jasniewicz}, {Jonker}, {Juaristi Campillo}, {Julbe}, {Karbevska},
  {Kervella}, {Khanna}, {Kochoska}, {Kontizas}, {Kordopatis}, {Korn},
  {Kostrzewa-Rutkowska}, {Kruszy{\'n}ska}, {Lambert}, {Lanza}, {Lasne}, {Le
  Campion}, {Le Fustec}, {Lebreton}, {Lebzelter}, {Leccia}, {Leclerc},
  {Lecoeur-Taibi}, {Liao}, {Licata}, {Lindstr{\o}m}, {Lister}, {Livanou},
  {Lobel}, {Madrero Pardo}, {Managau}, {Mann}, {Marchant}, {Marconi}, {Marcos
  Santos}, {Marinoni}, {Marocco}, {Marshall}, {Martin Polo},
  {Mart{\'\i}n-Fleitas}, {Masip}, {Massari}, {Mastrobuono-Battisti}, {Mazeh},
  {Messina}, {Michalik}, {Millar}, {Mints}, {Molina}, {Molinaro}, {Moln{\'a}r},
  {Montegriffo}, {Morbidelli}, {Morel}, {Morris}, {Mulone}, {Munoz}, {Murphy},
  {Musella}, {Noval}, {Ord{\'e}novic}, {Orr{\`u}}, {Osinde}, {Pagani},
  {Pagano}, {Palaversa}, {Palicio}, {Panahi}, {Pawlak}, {Pe{\~n}alosa
  Esteller}, {Penttil{\"a}}, {Piersimoni}, {Pineau}, {Plachy}, {Plum},
  {Poggio}, {Poretti}, {Poujoulet}, {Pr{\v{s}}a}, {Pulone}, {Racero},
  {Ragaini}, {Rainer}, {Raiteri}, {Rambaux}, {Ramos}, {Ramos-Lerate}, {Re
  Fiorentin}, {Regibo}, {Reyl{\'e}}, {Riva}, {Rixon}, {Robichon}, {Robin},
  {Roelens}, {Rohrbasser}, {Rowell}, {Royer}, {Rybicki}, {Sadowski},
  {Sagrist{\`a} Sell{\'e}s}, {Sahlmann}, {Salgado}, {Salguero}, {Samaras},
  {Gimenez}, {Sanna}, {Santove{\~n}a}, {Sarasso}, {Schultheis}, {Sciacca},
  {Segol}, {Segovia}, {S{\'e}gransan}, {Semeux}, {Siddiqui}, {Siebert},
  {Siltala}, {Slezak}, {Smart}, {Solano}, {Solitro}, {Souami}, {Souchay},
  {Spagna}, {Spoto}, {Steele}, {Steidelm{\"u}ller}, {Stephenson},
  {S{\"u}veges}, {Szabados}, {Szegedi-Elek}, {Taris}, {Tauran}, {Taylor},
  {Teixeira}, {Thuillot}, {Tonello}, {Torra}, {Torra}, {Turon}, {Unger},
  {Vaillant}, {van Dillen}, {Vanel}, {Vecchiato}, {Viala}, {Vicente},
  {Voutsinas}, {Weiler}, {Wevers}, {Wyrzykowski}, {Yoldas}, {Yvard}, {Zhao},
  {Zorec}, {Zucker}, {Zurbach}, \& {Zwitter}}]{gaia2021}
{Gaia Collaboration}, {Luri}, X., {Chemin}, L., {et~al.} 2021, \aap, 649, A7,
  \dodoi{10.1051/0004-6361/202039588}

\bibitem[{{Gaia Collaboration} {et~al.}(2023){Gaia Collaboration}, {Vallenari},
  {Brown}, {Prusti}, {de Bruijne}, {Arenou}, {Babusiaux}, {Biermann},
  {Creevey}, {Ducourant}, {Evans}, {Eyer}, {Guerra}, {Hutton}, {Jordi},
  {Klioner}, {Lammers}, {Lindegren}, {Luri}, {Mignard}, {Panem}, {Pourbaix},
  {Randich}, {Sartoretti}, {Soubiran}, {Tanga}, {Walton}, {Bailer-Jones},
  {Bastian}, {Drimmel}, {Jansen}, {Katz}, {Lattanzi}, {van Leeuwen}, {Bakker},
  {Cacciari}, {Casta{\~n}eda}, {De Angeli}, {Fabricius}, {Fouesneau},
  {Fr{\'e}mat}, {Galluccio}, {Guerrier}, {Heiter}, {Masana}, {Messineo},
  {Mowlavi}, {Nicolas}, {Nienartowicz}, {Pailler}, {Panuzzo}, {Riclet}, {Roux},
  {Seabroke}, {Sordo}, {Th{\'e}venin}, {Gracia-Abril}, {Portell}, {Teyssier},
  {Altmann}, {Andrae}, {Audard}, {Bellas-Velidis}, {Benson}, {Berthier},
  {Blomme}, {Burgess}, {Busonero}, {Busso}, {C{\'a}novas}, {Carry}, {Cellino},
  {Cheek}, {Clementini}, {Damerdji}, {Davidson}, {de Teodoro}, {Nu{\~n}ez
  Campos}, {Delchambre}, {Dell'Oro}, {Esquej}, {Fern{\'a}ndez-Hern{\'a}ndez},
  {Fraile}, {Garabato}, {Garc{\'\i}a-Lario}, {Gosset}, {Haigron}, {Halbwachs},
  {Hambly}, {Harrison}, {Hern{\'a}ndez}, {Hestroffer}, {Hodgkin}, {Holl},
  {Jan{\ss}en}, {Jevardat de Fombelle}, {Jordan}, {Krone-Martins}, {Lanzafame},
  {L{\"o}ffler}, {Marchal}, {Marrese}, {Moitinho}, {Muinonen}, {Osborne},
  {Pancino}, {Pauwels}, {Recio-Blanco}, {Reyl{\'e}}, {Riello}, {Rimoldini},
  {Roegiers}, {Rybizki}, {Sarro}, {Siopis}, {Smith}, {Sozzetti}, {Utrilla},
  {van Leeuwen}, {Abbas}, {{\'A}brah{\'a}m}, {Abreu Aramburu}, {Aerts},
  {Aguado}, {Ajaj}, {Aldea-Montero}, {Altavilla}, {{\'A}lvarez}, {Alves},
  {Anders}, {Anderson}, {Anglada Varela}, {Antoja}, {Baines}, {Baker},
  {Balaguer-N{\'u}{\~n}ez}, {Balbinot}, {Balog}, {Barache}, {Barbato},
  {Barros}, {Barstow}, {Bartolom{\'e}}, {Bassilana}, {Bauchet}, {Becciani},
  {Bellazzini}, {Berihuete}, {Bernet}, {Bertone}, {Bianchi}, {Binnenfeld},
  {Blanco-Cuaresma}, {Blazere}, {Boch}, {Bombrun}, {Bossini}, {Bouquillon},
  {Bragaglia}, {Bramante}, {Breedt}, {Bressan}, {Brouillet}, {Brugaletta},
  {Bucciarelli}, {Burlacu}, {Butkevich}, {Buzzi}, {Caffau}, {Cancelliere},
  {Cantat-Gaudin}, {Carballo}, {Carlucci}, {Carnerero}, {Carrasco},
  {Casamiquela}, {Castellani}, {Castro-Ginard}, {Chaoul}, {Charlot}, {Chemin},
  {Chiaramida}, {Chiavassa}, {Chornay}, {Comoretto}, {Contursi}, {Cooper},
  {Cornez}, {Cowell}, {Crifo}, {Cropper}, {Crosta}, {Crowley}, {Dafonte},
  {Dapergolas}, {David}, {David}, {de Laverny}, {De Luise}, {De March}, {De
  Ridder}, {de Souza}, {de Torres}, {del Peloso}, {del Pozo}, {Delbo},
  {Delgado}, {Delisle}, {Demouchy}, {Dharmawardena}, {Di Matteo}, {Diakite},
  {Diener}, {Distefano}, {Dolding}, {Edvardsson}, {Enke}, {Fabre}, {Fabrizio},
  {Faigler}, {Fedorets}, {Fernique}, {Fienga}, {Figueras}, {Fournier},
  {Fouron}, {Fragkoudi}, {Gai}, {Garcia-Gutierrez}, {Garcia-Reinaldos},
  {Garc{\'\i}a-Torres}, {Garofalo}, {Gavel}, {Gavras}, {Gerlach}, {Geyer},
  {Giacobbe}, {Gilmore}, {Girona}, {Giuffrida}, {Gomel}, {Gomez},
  {Gonz{\'a}lez-N{\'u}{\~n}ez}, {Gonz{\'a}lez-Santamar{\'\i}a},
  {Gonz{\'a}lez-Vidal}, {Granvik}, {Guillout}, {Guiraud},
  {Guti{\'e}rrez-S{\'a}nchez}, {Guy}, {Hatzidimitriou}, {Hauser}, {Haywood},
  {Helmer}, {Helmi}, {Sarmiento}, {Hidalgo}, {Hilger}, {H{\l}adczuk}, {Hobbs},
  {Holland}, {Huckle}, {Jardine}, {Jasniewicz}, {Jean-Antoine Piccolo},
  {Jim{\'e}nez-Arranz}, {Jorissen}, {Juaristi Campillo}, {Julbe}, {Karbevska},
  {Kervella}, {Khanna}, {Kontizas}, {Kordopatis}, {Korn}, {K{\'o}sp{\'a}l},
  {Kostrzewa-Rutkowska}, {Kruszy{\'n}ska}, {Kun}, {Laizeau}, {Lambert},
  {Lanza}, {Lasne}, {Le Campion}, {Lebreton}, {Lebzelter}, {Leccia}, {Leclerc},
  {Lecoeur-Taibi}, {Liao}, {Licata}, {Lindstr{\o}m}, {Lister}, {Livanou},
  {Lobel}, {Lorca}, {Loup}, {Madrero Pardo}, {Magdaleno Romeo}, {Managau},
  {Mann}, {Manteiga}, {Marchant}, {Marconi}, {Marcos}, {Marcos Santos},
  {Mar{\'\i}n Pina}, {Marinoni}, {Marocco}, {Marshall}, {Martin Polo},
  {Mart{\'\i}n-Fleitas}, {Marton}, {Mary}, {Masip}, {Massari},
  {Mastrobuono-Battisti}, {Mazeh}, {McMillan}, {Messina}, {Michalik}, {Millar},
  {Mints}, {Molina}, {Molinaro}, {Moln{\'a}r}, {Monari}, {Mongui{\'o}},
  {Montegriffo}, {Montero}, {Mor}, {Mora}, {Morbidelli}, {Morel}, {Morris},
  {Muraveva}, {Murphy}, {Musella}, {Nagy}, {Noval}, {Oca{\~n}a}, {Ogden},
  {Ordenovic}, {Osinde}, {Pagani}, {Pagano}, {Palaversa}, {Palicio},
  {Pallas-Quintela}, {Panahi}, {Payne-Wardenaar}, {Pe{\~n}alosa Esteller},
  {Penttil{\"a}}, {Pichon}, {Piersimoni}, {Pineau}, {Plachy}, {Plum}, {Poggio},
  {Pr{\v{s}}a}, {Pulone}, {Racero}, {Ragaini}, {Rainer}, {Raiteri}, {Rambaux},
  {Ramos}, {Ramos-Lerate}, {Re Fiorentin}, {Regibo}, {Richards}, {Rios Diaz},
  {Ripepi}, {Riva}, {Rix}, {Rixon}, {Robichon}, {Robin}, {Robin}, {Roelens},
  {Rogues}, {Rohrbasser}, {Romero-G{\'o}mez}, {Rowell}, {Royer}, {Ruz Mieres},
  {Rybicki}, {Sadowski}, {S{\'a}ez N{\'u}{\~n}ez}, {Sagrist{\`a} Sell{\'e}s},
  {Sahlmann}, {Salguero}, {Samaras}, {Sanchez Gimenez}, {Sanna},
  {Santove{\~n}a}, {Sarasso}, {Schultheis}, {Sciacca}, {Segol}, {Segovia},
  {S{\'e}gransan}, {Semeux}, {Shahaf}, {Siddiqui}, {Siebert}, {Siltala},
  {Silvelo}, {Slezak}, {Slezak}, {Smart}, {Snaith}, {Solano}, {Solitro},
  {Souami}, {Souchay}, {Spagna}, {Spina}, {Spoto}, {Steele},
  {Steidelm{\"u}ller}, {Stephenson}, {S{\"u}veges}, {Surdej}, {Szabados},
  {Szegedi-Elek}, {Taris}, {Taylor}, {Teixeira}, {Tolomei}, {Tonello}, {Torra},
  {Torra}, {Torralba Elipe}, {Trabucchi}, {Tsounis}, {Turon}, {Ulla}, {Unger},
  {Vaillant}, {van Dillen}, {van Reeven}, {Vanel}, {Vecchiato}, {Viala},
  {Vicente}, {Voutsinas}, {Weiler}, {Wevers}, {Wyrzykowski}, {Yoldas}, {Yvard},
  {Zhao}, {Zorec}, {Zucker}, \& {Zwitter}}]{gaia2023}
{Gaia Collaboration}, {Vallenari}, A., {Brown}, A.~G.~A., {et~al.} 2023, \aap,
  674, A1, \dodoi{10.1051/0004-6361/202243940}

\bibitem[{{Garc{\'\i}a P{\'e}rez} {et~al.}(2016){Garc{\'\i}a P{\'e}rez},
  {Allende Prieto}, {Holtzman}, {Shetrone}, {M{\'e}sz{\'a}ros}, {Bizyaev},
  {Carrera}, {Cunha}, {Garc{\'\i}a-Hern{\'a}ndez}, {Johnson}, {Majewski},
  {Nidever}, {Schiavon}, {Shane}, {Smith}, {Sobeck}, {Troup}, {Zamora},
  {Weinberg}, {Bovy}, {Eisenstein}, {Feuillet}, {Frinchaboy}, {Hayden},
  {Hearty}, {Nguyen}, {O'Connell}, {Pinsonneault}, {Wilson}, \&
  {Zasowski}}]{garcia-perez2016}
{Garc{\'\i}a P{\'e}rez}, A.~E., {Allende Prieto}, C., {Holtzman}, J.~A.,
  {et~al.} 2016, \aj, 151, 144, \dodoi{10.3847/0004-6256/151/6/144}

\bibitem[{{Gerrard} {et~al.}(2023){Gerrard}, {Federrath}, {Pingel},
  {McClure-Griffiths}, {Marchal}, {Joncas}, {Clark}, {Stanimirovi{\'c}}, {Lee},
  {van Loon}, {Dickey}, {D{\'e}nes}, {Ma}, {Dempsey}, \& {Lynn}}]{gerrard2023}
{Gerrard}, I.~A., {Federrath}, C., {Pingel}, N.~M., {et~al.} 2023, arXiv
  e-prints, arXiv:2309.10755.
\newblock \doarXiv{2309.10755}

\bibitem[{{Gordon} {et~al.}(2011){Gordon}, {Meixner}, {Meade}, {Whitney},
  {Engelbracht}, {Bot}, {Boyer}, {Lawton}, {Sewi{\l}o}, {Babler}, {Bernard},
  {Bracker}, {Block}, {Blum}, {Bolatto}, {Bonanos}, {Harris}, {Hora},
  {Indebetouw}, {Misselt}, {Reach}, {Shiao}, {Tielens}, {Carlson},
  {Churchwell}, {Clayton}, {Chen}, {Cohen}, {Fukui}, {Gorjian}, {Hony},
  {Israel}, {Kawamura}, {Kemper}, {Leroy}, {Li}, {Madden}, {Marble},
  {McDonald}, {Mizuno}, {Mizuno}, {Muller}, {Oliveira}, {Olsen}, {Onishi},
  {Paladini}, {Paradis}, {Points}, {Robitaille}, {Rubin}, {Sandstrom}, {Sato},
  {Shibai}, {Simon}, {Smith}, {Srinivasan}, {Vijh}, {Van Dyk}, {van Loon}, \&
  {Zaritsky}}]{gordon2011}
{Gordon}, K.~D., {Meixner}, M., {Meade}, M.~R., {et~al.} 2011, \aj, 142, 102,
  \dodoi{10.1088/0004-6256/142/4/102}

\bibitem[{{Gro{\ss}schedl} {et~al.}(2021){Gro{\ss}schedl}, {Alves}, {Meingast},
  \& {Herbst-Kiss}}]{grossschedl2021}
{Gro{\ss}schedl}, J.~E., {Alves}, J., {Meingast}, S., \& {Herbst-Kiss}, G.
  2021, \aap, 647, A91, \dodoi{10.1051/0004-6361/202038913}

\bibitem[{{Gunn} {et~al.}(2006){Gunn}, {Siegmund}, {Mannery}, {Owen}, {Hull},
  {Leger}, {Carey}, {Knapp}, {York}, {Boroski}, {Kent}, {Lupton}, {Rockosi},
  {Evans}, {Waddell}, {Anderson}, {Annis}, {Barentine}, {Bartoszek}, {Bastian},
  {Bracker}, {Brewington}, {Briegel}, {Brinkmann}, {Brown}, {Carr},
  {Czarapata}, {Drennan}, {Dombeck}, {Federwitz}, {Gillespie}, {Gonzales},
  {Hansen}, {Harvanek}, {Hayes}, {Jordan}, {Kinney}, {Klaene}, {Kleinman},
  {Kron}, {Kresinski}, {Lee}, {Limmongkol}, {Lindenmeyer}, {Long}, {Loomis},
  {McGehee}, {Mantsch}, {Neilsen}, {Neswold}, {Newman}, {Nitta}, {Peoples},
  {Pier}, {Prieto}, {Prosapio}, {Rivetta}, {Schneider}, {Snedden}, \&
  {Wang}}]{gunn2006}
{Gunn}, J.~E., {Siegmund}, W.~A., {Mannery}, E.~J., {et~al.} 2006, \aj, 131,
  2332, \dodoi{10.1086/500975}

\bibitem[{Harris {et~al.}(2020)Harris, Millman, van~der Walt, Gommers,
  Virtanen, Cournapeau, Wieser, Taylor, Berg, Smith, Kern, Picus, Hoyer, van
  Kerkwijk, Brett, Haldane, del R{'{\i}}o, Wiebe, Peterson,
  G{'{e}}rard-Marchant, Sheppard, Reddy, Weckesser, Abbasi, Gohlke, \&
  Oliphant}]{harris2020array}
Harris, C.~R., Millman, K.~J., van~der Walt, S.~J., {et~al.} 2020, Nature, 585,
  357, \dodoi{10.1038/s41586-020-2649-2}

\bibitem[{{Harris} \& {Zaritsky}(2006)}]{harris2006}
{Harris}, J., \& {Zaritsky}, D. 2006, \aj, 131, 2514, \dodoi{10.1086/500974}

\bibitem[{{Hasselquist} {et~al.}(2021){Hasselquist}, {Hayes}, {Lian},
  {Weinberg}, {Zasowski}, {Horta}, {Beaton}, {Feuillet}, {Garro}, {Gallart},
  {Smith}, {Holtzman}, {Minniti}, {Lacerna}, {Shetrone}, {J{\"o}nsson},
  {Cioni}, {Fillingham}, {Cunha}, {O'Connell}, {Fern{\'a}ndez-Trincado},
  {Mu{\~n}oz}, {Schiavon}, {Almeida}, {Anguiano}, {Beers}, {Bizyaev},
  {Brownstein}, {Cohen}, {Frinchaboy}, {Garc{\'\i}a-Hern{\'a}ndez}, {Geisler},
  {Lane}, {Majewski}, {Nidever}, {Nitschelm}, {Povick}, {Price-Whelan},
  {Roman-Lopes}, {Rosado}, {Sobeck}, {Stringfellow}, {Valenzuela}, {Villanova},
  \& {Vincenzo}}]{hasselquist2021}
{Hasselquist}, S., {Hayes}, C.~R., {Lian}, J., {et~al.} 2021, \apj, 923, 172,
  \dodoi{10.3847/1538-4357/ac25f9}

\bibitem[{{Hatzidimitriou} \& {Hawkins}(1989)}]{hatzidimitriou1989}
{Hatzidimitriou}, D., \& {Hawkins}, M.~R.~S. 1989, \mnras, 241, 667,
  \dodoi{10.1093/mnras/241.4.667}

\bibitem[{{Hindman}(1964)}]{hindman1964}
{Hindman}, J.~V. 1964, \nat, 202, 377, \dodoi{10.1038/202377b0}

\bibitem[{{Hindman}(1967)}]{hindman1967}
---. 1967, Australian Journal of Physics, 20, 147, \dodoi{10.1071/PH670147}

\bibitem[{{Hotan} {et~al.}(2021){Hotan}, {Bunton}, {Chippendale}, {Whiting},
  {Tuthill}, {Moss}, {McConnell}, {Amy}, {Huynh}, {Allison}, {Anderson},
  {Bannister}, {Bastholm}, {Beresford}, {Bock}, {Bolton}, {Chapman}, {Chow},
  {Collier}, {Cooray}, {Cornwell}, {Diamond}, {Edwards}, {Feain}, {Franzen},
  {George}, {Gupta}, {Hampson}, {Harvey-Smith}, {Hayman}, {Heywood}, {Jacka},
  {Jackson}, {Jackson}, {Jeganathan}, {Johnston}, {Kesteven}, {Kleiner},
  {Koribalski}, {Lee-Waddell}, {Lenc}, {Lensson}, {Mackay}, {Mahony},
  {McClure-Griffiths}, {McConigley}, {Mirtschin}, {Ng}, {Norris}, {Pearce},
  {Phillips}, {Pilawa}, {Raja}, {Reynolds}, {Roberts}, {Roxby}, {Sadler},
  {Shields}, {Schinckel}, {Serra}, {Shaw}, {Sweetnam}, {Troup}, {Tzioumis},
  {Voronkov}, \& {Westmeier}}]{hotan2021}
{Hotan}, A.~W., {Bunton}, J.~D., {Chippendale}, A.~P., {et~al.} 2021, \pasa,
  38, e009, \dodoi{10.1017/pasa.2021.1}

\bibitem[{{Hubeny} {et~al.}(2021){Hubeny}, {Allende Prieto}, {Osorio}, \&
  {Lanz}}]{Hubeny2021}
{Hubeny}, I., {Allende Prieto}, C., {Osorio}, Y., \& {Lanz}, T. 2021, arXiv
  e-prints, arXiv:2104.02829, \dodoi{10.48550/arXiv.2104.02829}

\bibitem[{Hunter(2007)}]{Hunter:2007}
Hunter, J.~D. 2007, Computing In Science \& Engineering, 9, 90

\bibitem[{{Israel} {et~al.}(1993){Israel}, {Johansson}, {Lequeux}, {Booth},
  {Nyman}, {Crane}, {Rubio}, {de Graauw}, {Kutner}, {Gredel}, {Boulanger},
  {Garay}, \& {Westerlund}}]{israel1993}
{Israel}, F.~P., {Johansson}, L.~E.~B., {Lequeux}, J., {et~al.} 1993, \aap,
  276, 25

\bibitem[{{Jacyszyn-Dobrzeniecka} {et~al.}(2016){Jacyszyn-Dobrzeniecka},
  {Skowron}, {Mr{\'o}z}, {Skowron}, {Soszy{\'n}ski}, {Udalski}, {Pietrukowicz},
  {Koz{\l}owski}, {Wyrzykowski}, {Poleski}, {Pawlak}, {Szyma{\'n}ski}, \&
  {Ulaczyk}}]{jacyszyn2016}
{Jacyszyn-Dobrzeniecka}, A.~M., {Skowron}, D.~M., {Mr{\'o}z}, P., {et~al.}
  2016, \actaa, 66, 149.
\newblock \doarXiv{1602.09141}

\bibitem[{{Jacyszyn-Dobrzeniecka} {et~al.}(2017){Jacyszyn-Dobrzeniecka},
  {Skowron}, {Mr{\'o}z}, {Soszy{\'n}ski}, {Udalski}, {Pietrukowicz}, {Skowron},
  {Poleski}, {Koz{\l}owski}, {Wyrzykowski}, {Pawlak}, {Szyma{\'n}ski}, \&
  {Ulaczyk}}]{Jacyszyn-Dobrzeniecka2017}
---. 2017, \actaa, 67, 1, \dodoi{10.32023/0001-5237/67.1.1}

\bibitem[{{Jameson} {et~al.}(2016){Jameson}, {Bolatto}, {Leroy}, {Meixner},
  {Roman-Duval}, {Gordon}, {Hughes}, {Israel}, {Rubio}, {Indebetouw}, {Madden},
  {Bot}, {Hony}, {Cormier}, {Pellegrini}, {Galametz}, \&
  {Sonneborn}}]{jameson2016}
{Jameson}, K.~E., {Bolatto}, A.~D., {Leroy}, A.~K., {et~al.} 2016, \apj, 825,
  12, \dodoi{10.3847/0004-637X/825/1/12}

\bibitem[{{Johnson}(1961)}]{johnson1961}
{Johnson}, H.~M. 1961, \pasp, 73, 20, \dodoi{10.1086/127613}

\bibitem[{{J{\"o}nsson} {et~al.}(2020){J{\"o}nsson}, {Holtzman}, {Allende
  Prieto}, {Cunha}, {Garc{\'\i}a-Hern{\'a}ndez}, {Hasselquist}, {Masseron},
  {Osorio}, {Shetrone}, {Smith}, {Stringfellow}, {Bizyaev}, {Edvardsson},
  {Majewski}, {M{\'e}sz{\'a}ros}, {Souto}, {Zamora}, {Beaton}, {Bovy}, {Donor},
  {Pinsonneault}, {Poovelil}, \& {Sobeck}}]{jonsson2020}
{J{\"o}nsson}, H., {Holtzman}, J.~A., {Allende Prieto}, C., {et~al.} 2020, \aj,
  160, 120, \dodoi{10.3847/1538-3881/aba592}

\bibitem[{{Kalberla} \& {Haud}(2015)}]{kalberla2015}
{Kalberla}, P.~M.~W., \& {Haud}, U. 2015, \aap, 578, A78,
  \dodoi{10.1051/0004-6361/201525859}

\bibitem[{{Kalberla} {et~al.}(2010){Kalberla}, {McClure-Griffiths}, {Pisano},
  {Calabretta}, {Ford}, {Lockman}, {Staveley-Smith}, {Kerp}, {Winkel},
  {Murphy}, \& {Newton-McGee}}]{kalberla2010}
{Kalberla}, P.~M.~W., {McClure-Griffiths}, N.~M., {Pisano}, D.~J., {et~al.}
  2010, \aap, 521, A17, \dodoi{10.1051/0004-6361/200913979}

\bibitem[{{Kallivayalil} {et~al.}(2006{\natexlab{a}}){Kallivayalil}, {van der
  Marel}, \& {Alcock}}]{kallivayalil2006b}
{Kallivayalil}, N., {van der Marel}, R.~P., \& {Alcock}, C. 2006{\natexlab{a}},
  \apj, 652, 1213, \dodoi{10.1086/508014}

\bibitem[{{Kallivayalil} {et~al.}(2006{\natexlab{b}}){Kallivayalil}, {van der
  Marel}, {Alcock}, {Axelrod}, {Cook}, {Drake}, \& {Geha}}]{kallivayalil2006a}
{Kallivayalil}, N., {van der Marel}, R.~P., {Alcock}, C., {et~al.}
  2006{\natexlab{b}}, \apj, 638, 772, \dodoi{10.1086/498972}

\bibitem[{{Kerr} {et~al.}(1954){Kerr}, {Hindman}, \& {Robinson}}]{kerr1954}
{Kerr}, F.~J., {Hindman}, J.~F., \& {Robinson}, B.~J. 1954, Australian Journal
  of Physics, 7, 297, \dodoi{10.1071/PH540297}

\bibitem[{{Leroy} {et~al.}(2007){Leroy}, {Bolatto}, {Stanimirovi{\'c}},
  {Mizuno}, {Israel}, \& {Bot}}]{leroy2007}
{Leroy}, A., {Bolatto}, A., {Stanimirovi{\'c}}, S., {et~al.} 2007, \apj, 658,
  1027, \dodoi{10.1086/511150}

\bibitem[{{Lindner} {et~al.}(2015){Lindner}, {Vera-Ciro}, {Murray},
  {Stanimirovi{\'c}}, {Babler}, {Heiles}, {Hennebelle}, {Goss}, \&
  {Dickey}}]{lindner2015}
{Lindner}, R.~R., {Vera-Ciro}, C., {Murray}, C.~E., {et~al.} 2015, \aj, 149,
  138, \dodoi{10.1088/0004-6256/149/4/138}

\bibitem[{{Lucchini} {et~al.}(2021){Lucchini}, {D'Onghia}, \&
  {Fox}}]{lucchini2021}
{Lucchini}, S., {D'Onghia}, E., \& {Fox}, A.~J. 2021, \apjl, 921, L36,
  \dodoi{10.3847/2041-8213/ac3338}

\bibitem[{{Ma} {et~al.}(2023){Ma}, {McClure-Griffiths}, {Clark}, {Gibson}, {van
  Loon}, {Soler}, {Putman}, {Dickey}, {Lee}, {Jameson}, {Uscanga}, {Dempsey},
  {D{\'e}nes}, {Lynn}, \& {Pingel}}]{ma2023}
{Ma}, Y.~K., {McClure-Griffiths}, N.~M., {Clark}, S.~E., {et~al.} 2023, \mnras,
  521, 60, \dodoi{10.1093/mnras/stad462}

\bibitem[{{Mackey} {et~al.}(2016){Mackey}, {Koposov}, {Erkal}, {Belokurov}, {Da
  Costa}, \& {G{\'o}mez}}]{mackey2016}
{Mackey}, A.~D., {Koposov}, S.~E., {Erkal}, D., {et~al.} 2016, \mnras, 459,
  239, \dodoi{10.1093/mnras/stw497}

\bibitem[{{Majewski} {et~al.}(2011){Majewski}, {Zasowski}, \&
  {Nidever}}]{majewski2011}
{Majewski}, S.~R., {Zasowski}, G., \& {Nidever}, D.~L. 2011, \apj, 739, 25,
  \dodoi{10.1088/0004-637X/739/1/25}

\bibitem[{{Majewski} {et~al.}(2017){Majewski}, {Schiavon}, {Frinchaboy},
  {Allende Prieto}, {Barkhouser}, {Bizyaev}, {Blank}, {Brunner}, {Burton},
  {Carrera}, {Chojnowski}, {Cunha}, {Epstein}, {Fitzgerald}, {Garc{\'\i}a
  P{\'e}rez}, {Hearty}, {Henderson}, {Holtzman}, {Johnson}, {Lam}, {Lawler},
  {Maseman}, {M{\'e}sz{\'a}ros}, {Nelson}, {Nguyen}, {Nidever}, {Pinsonneault},
  {Shetrone}, {Smee}, {Smith}, {Stolberg}, {Skrutskie}, {Walker}, {Wilson},
  {Zasowski}, {Anders}, {Basu}, {Beland}, {Blanton}, {Bovy}, {Brownstein},
  {Carlberg}, {Chaplin}, {Chiappini}, {Eisenstein}, {Elsworth}, {Feuillet},
  {Fleming}, {Galbraith-Frew}, {Garc{\'\i}a}, {Garc{\'\i}a-Hern{\'a}ndez},
  {Gillespie}, {Girardi}, {Gunn}, {Hasselquist}, {Hayden}, {Hekker}, {Ivans},
  {Kinemuchi}, {Klaene}, {Mahadevan}, {Mathur}, {Mosser}, {Muna}, {Munn},
  {Nichol}, {O'Connell}, {Parejko}, {Robin}, {Rocha-Pinto}, {Schultheis},
  {Serenelli}, {Shane}, {Silva Aguirre}, {Sobeck}, {Thompson}, {Troup},
  {Weinberg}, \& {Zamora}}]{majewski2017}
{Majewski}, S.~R., {Schiavon}, R.~P., {Frinchaboy}, P.~M., {et~al.} 2017, \aj,
  154, 94, \dodoi{10.3847/1538-3881/aa784d}

\bibitem[{{Marigo} {et~al.}(2017){Marigo}, {Girardi}, {Bressan}, {Rosenfield},
  {Aringer}, {Chen}, {Dussin}, {Nanni}, {Pastorelli}, {Rodrigues}, {Trabucchi},
  {Bladh}, {Dalcanton}, {Groenewegen}, {Montalb{\'a}n}, \& {Wood}}]{marigo2017}
{Marigo}, P., {Girardi}, L., {Bressan}, A., {et~al.} 2017, \apj, 835, 77,
  \dodoi{10.3847/1538-4357/835/1/77}

\bibitem[{{Mart{\'\i}nez-Delgado} {et~al.}(2019){Mart{\'\i}nez-Delgado},
  {Vivas}, {Grebel}, {Gallart}, {Pieres}, {Bell}, {Zivick}, {Lemasle}, {Clifton
  Johnson}, {Carballo-Bello}, {No{\"e}l}, {Cioni}, {Choi}, {Besla}, {Schmidt},
  {Zaritsky}, {Gruendl}, {Seibert}, {Nidever}, {Monteagudo}, {Monelli}, {Hubl},
  {van der Marel}, {Ballesteros}, {Stringfellow}, {Walker}, {Blum}, {Bell},
  {Conn}, {Olsen}, {Martin}, {Chu}, {Inno}, {Boer}, {Kallivayalil}, {De Leo},
  {Beletsky}, {Neyer}, \& {Mu{\~n}oz}}]{martinez2019}
{Mart{\'\i}nez-Delgado}, D., {Vivas}, A.~K., {Grebel}, E.~K., {et~al.} 2019,
  \aap, 631, A98, \dodoi{10.1051/0004-6361/201936021}

\bibitem[{{Massana} {et~al.}(2022){Massana}, {Ruiz-Lara}, {No{\"e}l},
  {Gallart}, {Nidever}, {Choi}, {Sakowska}, {Besla}, {Olsen}, {Monelli},
  {Dorta}, {Stringfellow}, {Cassisi}, {Bernard}, {Zaritsky}, {Cioni},
  {Monachesi}, {van der Marel}, {de Boer}, \& {Walker}}]{massana2022}
{Massana}, P., {Ruiz-Lara}, T., {No{\"e}l}, N.~E.~D., {et~al.} 2022, \mnras,
  513, L40, \dodoi{10.1093/mnrasl/slac030}

\bibitem[{{Mathewson} {et~al.}(1974){Mathewson}, {Cleary}, \&
  {Murray}}]{mathewson1974}
{Mathewson}, D.~S., {Cleary}, M.~N., \& {Murray}, J.~D. 1974, \apj, 190, 291,
  \dodoi{10.1086/152875}

\bibitem[{{Mathewson} \& {Ford}(1984)}]{mathewson1984}
{Mathewson}, D.~S., \& {Ford}, V.~L. 1984, in IAU Symposium, Vol. 1983,
  Structure and Evolution of the Magellanic Clouds, ed. S.~{van den Bergh} \&
  K.~S.~D. {de Boer}, 125--136

\bibitem[{{Mathewson} {et~al.}(1988){Mathewson}, {Ford}, \&
  {Visvanathan}}]{mathewson1988}
{Mathewson}, D.~S., {Ford}, V.~L., \& {Visvanathan}, N. 1988, \apj, 333, 617,
  \dodoi{10.1086/166772}

\bibitem[{{McClure-Griffiths} {et~al.}(2009){McClure-Griffiths}, {Pisano},
  {Calabretta}, {Ford}, {Lockman}, {Staveley-Smith}, {Kalberla}, {Bailin},
  {Dedes}, {Janowiecki}, {Gibson}, {Murphy}, {Nakanishi}, \&
  {Newton-McGee}}]{mcg2009}
{McClure-Griffiths}, N.~M., {Pisano}, D.~J., {Calabretta}, M.~R., {et~al.}
  2009, \apjs, 181, 398, \dodoi{10.1088/0067-0049/181/2/398}

\bibitem[{{McClure-Griffiths} {et~al.}(2018){McClure-Griffiths}, {D{\'e}nes},
  {Dickey}, {Stanimirovi{\'c}}, {Staveley-Smith}, {Jameson}, {Di Teodoro},
  {Allison}, {Collier}, {Chippendale}, {Franzen}, {G{\"u}rkan}, {Heald},
  {Hotan}, {Kleiner}, {Lee-Waddell}, {McConnell}, {Popping}, {Rhee}, {Riseley},
  {Voronkov}, \& {Whiting}}]{mcg2018}
{McClure-Griffiths}, N.~M., {D{\'e}nes}, H., {Dickey}, J.~M., {et~al.} 2018,
  Nature Astronomy, 2, 901, \dodoi{10.1038/s41550-018-0608-8}

\bibitem[{{Mizuno} {et~al.}(2001){Mizuno}, {Rubio}, {Mizuno}, {Yamaguchi},
  {Onishi}, \& {Fukui}}]{mizuno2001}
{Mizuno}, N., {Rubio}, M., {Mizuno}, A., {et~al.} 2001, \pasj, 53, L45,
  \dodoi{10.1093/pasj/53.6.L45}

\bibitem[{{Mucciarelli} {et~al.}(2023){Mucciarelli}, {Minelli}, {Bellazzini},
  {Lardo}, {Romano}, {Origlia}, \& {Ferraro}}]{mucciarelli2023}
{Mucciarelli}, A., {Minelli}, A., {Bellazzini}, M., {et~al.} 2023, \aap, 671,
  A124, \dodoi{10.1051/0004-6361/202245133}

\bibitem[{{Muller} \& {Bekki}(2007)}]{muller2007}
{Muller}, E., \& {Bekki}, K. 2007, \mnras, 381, L11,
  \dodoi{10.1111/j.1745-3933.2007.00356.x}

\bibitem[{{Muller} {et~al.}(2004){Muller}, {Stanimirovi{\'c}}, {Rosolowsky}, \&
  {Staveley-Smith}}]{muller2004}
{Muller}, E., {Stanimirovi{\'c}}, S., {Rosolowsky}, E., \& {Staveley-Smith}, L.
  2004, \apj, 616, 845, \dodoi{10.1086/425154}

\bibitem[{{Muller} {et~al.}(2010){Muller}, {Ott}, {Hughes}, {Pineda}, {Wong},
  {Mizuno}, {Kawamura}, {Mizuno}, {Fukui}, {Onishi}, \& {Rubio}}]{muller2010}
{Muller}, E., {Ott}, J., {Hughes}, A., {et~al.} 2010, \apj, 712, 1248,
  \dodoi{10.1088/0004-637X/712/2/1248}

\bibitem[{{Murai} \& {Fujimoto}(1980)}]{murai1980}
{Murai}, T., \& {Fujimoto}, M. 1980, \pasj, 32, 581

\bibitem[{{Murray} {et~al.}(2019){Murray}, {Peek}, {Di Teodoro},
  {McClure-Griffiths}, {Dickey}, \& {D{\'e}nes}}]{murray2019}
{Murray}, C.~E., {Peek}, J.~E.~G., {Di Teodoro}, E.~M., {et~al.} 2019, \apj,
  887, 267, \dodoi{10.3847/1538-4357/ab510f}

\bibitem[{{Nidever} {et~al.}(2010){Nidever}, {Majewski}, {Butler Burton}, \&
  {Nigra}}]{nidever2010}
{Nidever}, D.~L., {Majewski}, S.~R., {Butler Burton}, W., \& {Nigra}, L. 2010,
  \apj, 723, 1618, \dodoi{10.1088/0004-637X/723/2/1618}

\bibitem[{{Nidever} {et~al.}(2013){Nidever}, {Monachesi}, {Bell}, {Majewski},
  {Mu{\~n}oz}, \& {Beaton}}]{nidever2013}
{Nidever}, D.~L., {Monachesi}, A., {Bell}, E.~F., {et~al.} 2013, \apj, 779,
  145, \dodoi{10.1088/0004-637X/779/2/145}

\bibitem[{{Nidever} {et~al.}(2015){Nidever}, {Holtzman}, {Allende Prieto},
  {Beland}, {Bender}, {Bizyaev}, {Burton}, {Desphande}, {Fleming}, {Garc{\'\i}a
  P{\'e}rez}, {Hearty}, {Majewski}, {M{\'e}sz{\'a}ros}, {Muna}, {Nguyen},
  {Schiavon}, {Shetrone}, {Skrutskie}, {Sobeck}, \& {Wilson}}]{nidever2015}
{Nidever}, D.~L., {Holtzman}, J.~A., {Allende Prieto}, C., {et~al.} 2015, \aj,
  150, 173, \dodoi{10.1088/0004-6256/150/6/173}

\bibitem[{{Nidever} {et~al.}(2020){Nidever}, {Hasselquist}, {Hayes}, {Hawkins},
  {Povick}, {Majewski}, {Smith}, {Anguiano}, {Stringfellow}, {Sobeck}, {Cunha},
  {Beers}, {Bestenlehner}, {Cohen}, {Garcia-Hernandez}, {J{\"o}nsson},
  {Nitschelm}, {Shetrone}, {Lacerna}, {Allende Prieto}, {Beaton}, {Dell'Agli},
  {Fern{\'a}ndez-Trincado}, {Feuillet}, {Gallart}, {Hearty}, {Holtzman},
  {Manchado}, {Mu{\~n}oz}, {O'Connell}, \& {Rosado}}]{nidever2020}
{Nidever}, D.~L., {Hasselquist}, S., {Hayes}, C.~R., {et~al.} 2020, \apj, 895,
  88, \dodoi{10.3847/1538-4357/ab7305}

\bibitem[{{Niederhofer} {et~al.}(2018){Niederhofer}, {Cioni}, {Rubele},
  {Schmidt}, {Bekki}, {de Grijs}, {Emerson}, {Ivanov}, {Marconi}, {Oliveira},
  {Petr-Gotzens}, {Ripepi}, {van Loon}, \& {Zaggia}}]{niederhofer2018}
{Niederhofer}, F., {Cioni}, M. R.~L., {Rubele}, S., {et~al.} 2018, \aap, 613,
  L8, \dodoi{10.1051/0004-6361/201833144}

\bibitem[{{Niederhofer} {et~al.}(2021){Niederhofer}, {Cioni}, {Rubele},
  {Schmidt}, {Diaz}, {Matijev{\u{i}}c}, {Bekki}, {Bell}, {de Grijs}, {El
  Youssoufi}, {Ivanov}, {Oliveira}, {Ripepi}, {Subramanian}, {Sun}, \& {van
  Loon}}]{niederhofer2021}
{Niederhofer}, F., {Cioni}, M.-R.~L., {Rubele}, S., {et~al.} 2021, \mnras, 502,
  2859, \dodoi{10.1093/mnras/stab206}

\bibitem[{{Olsen} {et~al.}(2015){Olsen}, {Blum}, {Smart}, {Zaritsky}, {Boyer},
  {Gordon}, \& {Massey}}]{olsen2015}
{Olsen}, K.~A.~G., {Blum}, R.~D., {Smart}, B., {et~al.} 2015, in Astronomical
  Society of the Pacific Conference Series, Vol. 491, Fifty Years of Wide Field
  Studies in the Southern Hemisphere: Resolved Stellar Populations of the
  Galactic Bulge and Magellanic Clouds, ed. S.~{Points} \& A.~{Kunder}, 257

\bibitem[{{Omkumar} {et~al.}(2021){Omkumar}, {Subramanian}, {Niederhofer},
  {Diaz}, {Cioni}, {El Youssoufi}, {Bekki}, {de Grijs}, \& {van
  Loon}}]{omkumar2021}
{Omkumar}, A.~O., {Subramanian}, S., {Niederhofer}, F., {et~al.} 2021, \mnras,
  500, 2757, \dodoi{10.1093/mnras/staa3085}

\bibitem[{{Osorio} {et~al.}(2020){Osorio}, {Allende Prieto}, {Hubeny},
  {M{\'e}sz{\'a}ros}, \& {Shetrone}}]{Osorio2020}
{Osorio}, Y., {Allende Prieto}, C., {Hubeny}, I., {M{\'e}sz{\'a}ros}, S., \&
  {Shetrone}, M. 2020, \aap, 637, A80, \dodoi{10.1051/0004-6361/201937054}

\bibitem[{{Pardy} {et~al.}(2018){Pardy}, {D'Onghia}, \& {Fox}}]{pardy2018}
{Pardy}, S.~A., {D'Onghia}, E., \& {Fox}, A.~J. 2018, \apj, 857, 101,
  \dodoi{10.3847/1538-4357/aab95b}

\bibitem[{{Pastorelli} {et~al.}(2019){Pastorelli}, {Marigo}, {Girardi}, {Chen},
  {Rubele}, {Trabucchi}, {Aringer}, {Bladh}, {Bressan}, {Montalb{\'a}n},
  {Boyer}, {Dalcanton}, {Eriksson}, {Groenewegen}, {H{\"o}fner}, {Lebzelter},
  {Nanni}, {Rosenfield}, {Wood}, \& {Cioni}}]{pastorelli2019}
{Pastorelli}, G., {Marigo}, P., {Girardi}, L., {et~al.} 2019, \mnras, 485,
  5666, \dodoi{10.1093/mnras/stz725}

\bibitem[{{Pastorelli} {et~al.}(2020){Pastorelli}, {Marigo}, {Girardi},
  {Aringer}, {Chen}, {Rubele}, {Trabucchi}, {Bladh}, {Boyer}, {Bressan},
  {Dalcanton}, {Groenewegen}, {Lebzelter}, {Mowlavi}, {Chubb}, {Cioni}, {de
  Grijs}, {Ivanov}, {Nanni}, {van Loon}, \& {Zaggia}}]{pastorelli2020}
---. 2020, \mnras, 498, 3283, \dodoi{10.1093/mnras/staa2565}

\bibitem[{{Patel} {et~al.}(2020){Patel}, {Kallivayalil}, {Garavito-Camargo},
  {Besla}, {Weisz}, {van der Marel}, {Boylan-Kolchin}, {Pawlowski}, \&
  {G{\'o}mez}}]{patel2020}
{Patel}, E., {Kallivayalil}, N., {Garavito-Camargo}, N., {et~al.} 2020, \apj,
  893, 121, \dodoi{10.3847/1538-4357/ab7b75}

\bibitem[{{Pe{\~n}arrubia} {et~al.}(2016){Pe{\~n}arrubia}, {G{\'o}mez},
  {Besla}, {Erkal}, \& {Ma}}]{penarrubia2016}
{Pe{\~n}arrubia}, J., {G{\'o}mez}, F.~A., {Besla}, G., {Erkal}, D., \& {Ma},
  Y.-Z. 2016, \mnras, 456, L54, \dodoi{10.1093/mnrasl/slv160}

\bibitem[{{Piatek} {et~al.}(2008){Piatek}, {Pryor}, \&
  {Olszewski}}]{piatek2008}
{Piatek}, S., {Pryor}, C., \& {Olszewski}, E.~W. 2008, \aj, 135, 1024,
  \dodoi{10.1088/0004-6256/135/3/1024}

\bibitem[{{Pieres} {et~al.}(2017){Pieres}, {Santiago}, {Drlica-Wagner},
  {Bechtol}, {Marel}, {Besla}, {Martin}, {Belokurov}, {Gallart},
  {Martinez-Delgado}, {Marshall}, {N{\"o}el}, {Majewski}, {Cioni}, {Li},
  {Hartley}, {Luque}, {Conn}, {Walker}, {Balbinot}, {Stringfellow}, {Olsen},
  {Nidever}, {da Costa}, {Ogando}, {Maia}, {Neto}, {Abbott}, {Abdalla},
  {Allam}, {Annis}, {Benoit-L{\'e}vy}, {Rosell}, {Kind}, {Carretero}, {Cunha},
  {D'Andrea}, {Desai}, {Diehl}, {Doel}, {Flaugher}, {Fosalba},
  {Garc{\'\i}a-Bellido}, {Gruen}, {Gruendl}, {Gschwend}, {Gutierrez},
  {Honscheid}, {James}, {Kuehn}, {Kuropatkin}, {Menanteau}, {Miquel}, {Plazas},
  {Romer}, {Sako}, {Sanchez}, {Scarpine}, {Schubnell}, {Sevilla-Noarbe},
  {Smith}, {Soares-Santos}, {Sobreira}, {Suchyta}, {Swanson}, {Tarle},
  {Tucker}, \& {Wester}}]{pieres2017}
{Pieres}, A., {Santiago}, B.~X., {Drlica-Wagner}, A., {et~al.} 2017, \mnras,
  468, 1349, \dodoi{10.1093/mnras/stx507}

\bibitem[{{Pingel} {et~al.}(2022){Pingel}, {Dempsey}, {McClure-Griffiths},
  {Dickey}, {Jameson}, {Arce}, {Anglada}, {Bland-Hawthorn}, {Breen},
  {Buckland-Willis}, {Clark}, {Dawson}, {D{\'e}nes}, {Di Teodoro}, {For},
  {Foster}, {G{\'o}mez}, {Imai}, {Joncas}, {Kim}, {Lee}, {Lynn}, {Leahy}, {Ma},
  {Marchal}, {McConnell}, {Miville-Desch{\`e}nes}, {Moss}, {Murray}, {Nidever},
  {Peek}, {Stanimirovi{\'c}}, {Staveley-Smith}, {Tepper-Garcia}, {Tremblay},
  {Uscanga}, {van Loon}, {V{\'a}zquez-Semadeni}, {Allison}, {Anderson}, {Ball},
  {Bell}, {Bock}, {Bunton}, {Cooray}, {Cornwell}, {Koribalski}, {Gupta},
  {Hayman}, {Harvey-Smith}, {Lee-Waddell}, {Ng}, {Phillips}, {Voronkov},
  {Westmeier}, \& {Whiting}}]{pingel2022}
{Pingel}, N.~M., {Dempsey}, J., {McClure-Griffiths}, N.~M., {et~al.} 2022,
  \pasa, 39, e005, \dodoi{10.1017/pasa.2021.59}

\bibitem[{{Putman} {et~al.}(1998){Putman}, {Gibson}, {Staveley-Smith}, {Banks},
  {Barnes}, {Bhatal}, {Disney}, {Ekers}, {Freeman}, {Haynes}, {Henning},
  {Jerjen}, {Kilborn}, {Koribalski}, {Knezek}, {Malin}, {Mould}, {Oosterloo},
  {Price}, {Ryder}, {Sadler}, {Stewart}, {Stootman}, {Vaile}, {Webster}, \&
  {Wright}}]{putman1998}
{Putman}, M.~E., {Gibson}, B.~K., {Staveley-Smith}, L., {et~al.} 1998, \nat,
  394, 752, \dodoi{10.1038/29466}

\bibitem[{{Riener} {et~al.}(2019){Riener}, {Kainulainen}, {Henshaw}, {Orkisz},
  {Murray}, \& {Beuther}}]{riener2019}
{Riener}, M., {Kainulainen}, J., {Henshaw}, J.~D., {et~al.} 2019, \aap, 628,
  A78, \dodoi{10.1051/0004-6361/201935519}

\bibitem[{{Ripepi} {et~al.}(2017){Ripepi}, {Cioni}, {Moretti}, {Marconi},
  {Bekki}, {Clementini}, {de Grijs}, {Emerson}, {Groenewegen}, {Ivanov},
  {Molinaro}, {Muraveva}, {Oliveira}, {Piatti}, {Subramanian}, \& {van
  Loon}}]{ripepi2017}
{Ripepi}, V., {Cioni}, M.-R.~L., {Moretti}, M.~I., {et~al.} 2017, \mnras, 472,
  808, \dodoi{10.1093/mnras/stx2096}

\bibitem[{{Rubio} {et~al.}(1991){Rubio}, {Garay}, {Montani}, \&
  {Thaddeus}}]{rubio1991}
{Rubio}, M., {Garay}, G., {Montani}, J., \& {Thaddeus}, P. 1991, \apj, 368,
  173, \dodoi{10.1086/169680}

\bibitem[{{Rubio} {et~al.}(1996){Rubio}, {Lequeux}, {Boulanger}, {Booth},
  {Garay}, {de Graauw}, {Israel}, {Johansson}, {Kutner}, \&
  {Nyman}}]{rubio1996}
{Rubio}, M., {Lequeux}, J., {Boulanger}, F., {et~al.} 1996, \aaps, 118, 263

\bibitem[{{Russell} \& {Dopita}(1992)}]{russell1992}
{Russell}, S.~C., \& {Dopita}, M.~A. 1992, \apj, 384, 508,
  \dodoi{10.1086/170893}

\bibitem[{{Salda{\~n}o} {et~al.}(2022){Salda{\~n}o}, {Rubio}, {Bolatto},
  {Verdugo}, {Jameson}, \& {Leroy}}]{saldano2022}
{Salda{\~n}o}, H.~P., {Rubio}, M., {Bolatto}, A.~D., {et~al.} 2022, arXiv
  e-prints, arXiv:2211.07792, \dodoi{10.48550/arXiv.2211.07792}

\bibitem[{{Santana} {et~al.}(2021){Santana}, {Beaton}, {Covey}, {O'Connell},
  {Longa-Pe{\~n}a}, {Cohen}, {Fern{\'a}ndez-Trincado}, {Hayes}, {Zasowski},
  {Sobeck}, {Majewski}, {Chojnowski}, {De Lee}, {Oelkers}, {Stringfellow},
  {Almeida}, {Anguiano}, {Donor}, {Frinchaboy}, {Hasselquist}, {Johnson},
  {Kollmeier}, {Nidever}, {Price-Whelan}, {Rojas-Arriagada}, {Schultheis},
  {Shetrone}, {Simon}, {Aerts}, {Borissova}, {Drout}, {Geisler}, {Law},
  {Medina}, {Minniti}, {Monachesi}, {Mu{\~n}oz}, {Poleski}, {Roman-Lopes},
  {Schlaufman}, {Stutz}, {Teske}, {Tkachenko}, {Van Saders}, {Weinberger}, \&
  {Zoccali}}]{santana2021}
{Santana}, F.~A., {Beaton}, R.~L., {Covey}, K.~R., {et~al.} 2021, \aj, 162,
  303, \dodoi{10.3847/1538-3881/ac2cbc}

\bibitem[{{Schruba} {et~al.}(2012){Schruba}, {Leroy}, {Walter}, {Bigiel},
  {Brinks}, {de Blok}, {Kramer}, {Rosolowsky}, {Sandstrom}, {Schuster},
  {Usero}, {Weiss}, \& {Wiesemeyer}}]{schruba2012}
{Schruba}, A., {Leroy}, A.~K., {Walter}, F., {et~al.} 2012, \aj, 143, 138,
  \dodoi{10.1088/0004-6256/143/6/138}

\bibitem[{{Scowcroft} {et~al.}(2016){Scowcroft}, {Freedman}, {Madore},
  {Monson}, {Persson}, {Rich}, {Seibert}, \& {Rigby}}]{scowcroft2016}
{Scowcroft}, V., {Freedman}, W.~L., {Madore}, B.~F., {et~al.} 2016, \apj, 816,
  49, \dodoi{10.3847/0004-637X/816/2/49}

\bibitem[{{Skrutskie} {et~al.}(2006){Skrutskie}, {Cutri}, {Stiening},
  {Weinberg}, {Schneider}, {Carpenter}, {Beichman}, {Capps}, {Chester},
  {Elias}, {Huchra}, {Liebert}, {Lonsdale}, {Monet}, {Price}, {Seitzer},
  {Jarrett}, {Kirkpatrick}, {Gizis}, {Howard}, {Evans}, {Fowler}, {Fullmer},
  {Hurt}, {Light}, {Kopan}, {Marsh}, {McCallon}, {Tam}, {Van Dyk}, \&
  {Wheelock}}]{skrutskie2006}
{Skrutskie}, M.~F., {Cutri}, R.~M., {Stiening}, R., {et~al.} 2006, \aj, 131,
  1163, \dodoi{10.1086/498708}

\bibitem[{{Stanimirovi{\'c}} {et~al.}(1999){Stanimirovi{\'c}},
  {Staveley-Smith}, {Dickey}, {Sault}, \& {Snowden}}]{stanimirovic1999}
{Stanimirovi{\'c}}, S., {Staveley-Smith}, L., {Dickey}, J.~M., {Sault}, R.~J.,
  \& {Snowden}, S.~L. 1999, \mnras, 302, 417,
  \dodoi{10.1046/j.1365-8711.1999.02013.x}

\bibitem[{{Stanimirovi{\'c}} {et~al.}(2004){Stanimirovi{\'c}},
  {Staveley-Smith}, \& {Jones}}]{stanimirovic2004}
{Stanimirovi{\'c}}, S., {Staveley-Smith}, L., \& {Jones}, P.~A. 2004, \apj,
  604, 176, \dodoi{10.1086/381869}

\bibitem[{{Staveley-Smith} {et~al.}(1997){Staveley-Smith}, {Sault},
  {Hatzidimitriou}, {Kesteven}, \& {McConnell}}]{staveleysmith1997}
{Staveley-Smith}, L., {Sault}, R.~J., {Hatzidimitriou}, D., {Kesteven}, M.~J.,
  \& {McConnell}, D. 1997, \mnras, 289, 225, \dodoi{10.1093/mnras/289.2.225}

\bibitem[{{Subramanian} \& {Subramaniam}(2012)}]{subramanian2012}
{Subramanian}, S., \& {Subramaniam}, A. 2012, \apj, 744, 128,
  \dodoi{10.1088/0004-637X/744/2/128}

\bibitem[{{Subramanian} {et~al.}(2017){Subramanian}, {Rubele}, {Sun},
  {Girardi}, {de Grijs}, {van Loon}, {Cioni}, {Piatti}, {Bekki}, {Emerson},
  {Ivanov}, {Kerber}, {Marconi}, {Ripepi}, \& {Tatton}}]{subramanian2017}
{Subramanian}, S., {Rubele}, S., {Sun}, N.-C., {et~al.} 2017, \mnras, 467,
  2980, \dodoi{10.1093/mnras/stx205}

\bibitem[{{Tang} {et~al.}(2014){Tang}, {Bressan}, {Rosenfield}, {Slemer},
  {Marigo}, {Girardi}, \& {Bianchi}}]{tang2014}
{Tang}, J., {Bressan}, A., {Rosenfield}, P., {et~al.} 2014, \mnras, 445, 4287,
  \dodoi{10.1093/mnras/stu2029}

\bibitem[{{Tatton} {et~al.}(2021){Tatton}, {van Loon}, {Cioni}, {Bekki},
  {Bell}, {Choudhury}, {de Grijs}, {Groenewegen}, {Ivanov}, {Marconi},
  {Oliveira}, {Ripepi}, {Rubele}, {Subramanian}, \& {Sun}}]{tatton2021}
{Tatton}, B.~L., {van Loon}, J.~T., {Cioni}, M. R.~L., {et~al.} 2021, \mnras,
  504, 2983, \dodoi{10.1093/mnras/staa3857}

\bibitem[{{Tokuda} {et~al.}(2021){Tokuda}, {Kondo}, {Ohno}, {Konishi}, {Sano},
  {Tsuge}, {Zahorecz}, {Goto}, {Neelamkodan}, {Wong}, {Sewi{\l}o}, {Fukushima},
  {Takekoshi}, {Muraoka}, {Kawamura}, {Tachihara}, {Fukui}, \&
  {Onishi}}]{tokuda2021}
{Tokuda}, K., {Kondo}, H., {Ohno}, T., {et~al.} 2021, \apj, 922, 171,
  \dodoi{10.3847/1538-4357/ac1ff4}

\bibitem[{{Toomre} \& {Toomre}(1972)}]{toomre1972}
{Toomre}, A., \& {Toomre}, J. 1972, \apj, 178, 623, \dodoi{10.1086/151823}

\bibitem[{{van de Ven} {et~al.}(2006){van de Ven}, {van den Bosch}, {Verolme},
  \& {de Zeeuw}}]{vandeven2006}
{van de Ven}, G., {van den Bosch}, R.~C.~E., {Verolme}, E.~K., \& {de Zeeuw},
  P.~T. 2006, \aap, 445, 513, \dodoi{10.1051/0004-6361:20053061}

\bibitem[{{van der Marel} \& {Kallivayalil}(2014)}]{vandermarel2014}
{van der Marel}, R.~P., \& {Kallivayalil}, N. 2014, \apj, 781, 121,
  \dodoi{10.1088/0004-637X/781/2/121}

\bibitem[{{Vasiliev}(2024)}]{vasiliev2024}
{Vasiliev}, E. 2024, \mnras, 527, 437, \dodoi{10.1093/mnras/stad2612}

\bibitem[{{Virtanen} {et~al.}(2020){Virtanen}, {Gommers}, {Oliphant},
  {Haberland}, {Reddy}, {Cournapeau}, {Burovski}, {Peterson}, {Weckesser},
  {Bright}, {van der Walt}, {Brett}, {Wilson}, {Jarrod Millman}, {Mayorov},
  {Nelson}, {Jones}, {Kern}, {Larson}, {Carey}, {Polat}, {Feng}, {Moore}, {Vand
  erPlas}, {Laxalde}, {Perktold}, {Cimrman}, {Henriksen}, {Quintero}, {Harris},
  {Archibald}, {Ribeiro}, {Pedregosa}, {van Mulbregt}, \&
  {Contributors}}]{Virtanen_2020}
{Virtanen}, P., {Gommers}, R., {Oliphant}, T.~E., {et~al.} 2020, Nature
  Methods, 17, 261, \dodoi{https://doi.org/10.1038/s41592-019-0686-2}

\bibitem[{{Wannier} {et~al.}(1972){Wannier}, {Wrixon}, \&
  {Wilson}}]{wannier1972}
{Wannier}, P., {Wrixon}, G.~T., \& {Wilson}, R.~W. 1972, \aap, 18, 224

\bibitem[{{Welty} {et~al.}(2012){Welty}, {Xue}, \& {Wong}}]{welty2012}
{Welty}, D.~E., {Xue}, R., \& {Wong}, T. 2012, \apj, 745, 173,
  \dodoi{10.1088/0004-637X/745/2/173}

\bibitem[{{Wilson} {et~al.}(2019){Wilson}, {Hearty}, {Skrutskie}, {Majewski},
  {Holtzman}, {Eisenstein}, {Gunn}, {Blank}, {Henderson}, {Smee}, {Nelson},
  {Nidever}, {Arns}, {Barkhouser}, {Barr}, {Beland}, {Bershady}, {Blanton},
  {Brunner}, {Burton}, {Carey}, {Carr}, {Colque}, {Crane}, {Damke}, {Davidson},
  {Dean}, {Di Mille}, {Don}, {Ebelke}, {Evans}, {Fitzgerald}, {Gillespie},
  {Hall}, {Harding}, {Harding}, {Hammond}, {Hancock}, {Harrison}, {Hope},
  {Horne}, {Karakla}, {Lam}, {Leger}, {MacDonald}, {Maseman}, {Matsunari},
  {Melton}, {Mitcheltree}, {O'Brien}, {O'Connell}, {Patten}, {Richardson},
  {Rieke}, {Rieke}, {Roman-Lopes}, {Schiavon}, {Sobeck}, {Stolberg}, {Stoll},
  {Tembe}, {Trujillo}, {Uomoto}, {Vernieri}, {Walker}, {Weinberg}, {Young},
  {Anthony-Brumfield}, {Bizyaev}, {Breslauer}, {De Lee}, {Downey}, {Halverson},
  {Huehnerhoff}, {Klaene}, {Leon}, {Long}, {Mahadevan}, {Malanushenko},
  {Nguyen}, {Owen}, {S{\'a}nchez-Gallego}, {Sayres}, {Shane}, {Shectman},
  {Shetrone}, {Skinner}, {Stauffer}, \& {Zhao}}]{wilson2019}
{Wilson}, J.~C., {Hearty}, F.~R., {Skrutskie}, M.~F., {et~al.} 2019, \pasp,
  131, 055001, \dodoi{10.1088/1538-3873/ab0075}

\bibitem[{{Yanchulova Merica-Jones} {et~al.}(2017){Yanchulova Merica-Jones},
  {Sandstrom}, {Johnson}, {Dalcanton}, {Dolphin}, {Gordon}, {Roman-Duval},
  {Weisz}, \& {Williams}}]{yanchulova2017}
{Yanchulova Merica-Jones}, P., {Sandstrom}, K.~M., {Johnson}, L.~C., {et~al.}
  2017, \apj, 847, 102, \dodoi{10.3847/1538-4357/aa8a67}

\bibitem[{{Zamora} {et~al.}(2015){Zamora}, {Garc{\'\i}a-Hern{\'a}ndez},
  {Allende Prieto}, {Carrera}, {Koesterke}, {Edvardsson}, {Castelli}, {Plez},
  {Bizyaev}, {Cunha}, {Garc{\'\i}a P{\'e}rez}, {Gustafsson}, {Holtzman},
  {Lawler}, {Majewski}, {Manchado}, {M{\'e}sz{\'a}ros}, {Shane}, {Shetrone},
  {Smith}, \& {Zasowski}}]{Zamora2015}
{Zamora}, O., {Garc{\'\i}a-Hern{\'a}ndez}, D.~A., {Allende Prieto}, C.,
  {et~al.} 2015, \aj, 149, 181, \dodoi{10.1088/0004-6256/149/6/181}

\bibitem[{{Zasowski} {et~al.}(2017){Zasowski}, {Cohen}, {Chojnowski},
  {Santana}, {Oelkers}, {Andrews}, {Beaton}, {Bender}, {Bird}, {Bovy},
  {Carlberg}, {Covey}, {Cunha}, {Dell'Agli}, {Fleming}, {Frinchaboy},
  {Garc{\'\i}a-Hern{\'a}ndez}, {Harding}, {Holtzman}, {Johnson}, {Kollmeier},
  {Majewski}, {M{\'e}sz{\'a}ros}, {Munn}, {Mu{\~n}oz}, {Ness}, {Nidever},
  {Poleski}, {Rom{\'a}n-Z{\'u}{\~n}iga}, {Shetrone}, {Simon}, {Smith},
  {Sobeck}, {Stringfellow}, {Szigeti{\'a}ros}, {Tayar}, \&
  {Troup}}]{zasowski2017}
{Zasowski}, G., {Cohen}, R.~E., {Chojnowski}, S.~D., {et~al.} 2017, \aj, 154,
  198, \dodoi{10.3847/1538-3881/aa8df9}

\bibitem[{{Zhang} {et~al.}(2023){Zhang}, {Green}, \& {Rix}}]{zhang2023}
{Zhang}, X., {Green}, G.~M., \& {Rix}, H.-W. 2023, arXiv e-prints,
  arXiv:2303.03420, \dodoi{10.48550/arXiv.2303.03420}

\bibitem[{{Zivick} {et~al.}(2021){Zivick}, {Kallivayalil}, \& {van der
  Marel}}]{zivick2021}
{Zivick}, P., {Kallivayalil}, N., \& {van der Marel}, R.~P. 2021, \apj, 910,
  36, \dodoi{10.3847/1538-4357/abe1bb}

\bibitem[{{Zivick} {et~al.}(2018){Zivick}, {Kallivayalil}, {van der Marel},
  {Besla}, {Linden}, {Koz{\l}owski}, {Fritz}, {Kochanek}, {Anderson}, {Sohn},
  {Geha}, \& {Alcock}}]{zivick2018}
{Zivick}, P., {Kallivayalil}, N., {van der Marel}, R.~P., {et~al.} 2018, \apj,
  864, 55, \dodoi{10.3847/1538-4357/aad4b0}

\bibitem[{{Zivick} {et~al.}(2019){Zivick}, {Kallivayalil}, {Besla}, {Sohn},
  {van der Marel}, {del Pino}, {Linden}, {Fritz}, \& {Anderson}}]{zivick2019}
{Zivick}, P., {Kallivayalil}, N., {Besla}, G., {et~al.} 2019, \apj, 874, 78,
  \dodoi{10.3847/1538-4357/ab0554}

\end{thebibliography}
\bibliographystyle{aasjournal}

\end{document}